\documentclass[preprint,12pt]{elsarticle}
\usepackage{graphicx} 
\usepackage[T1]{fontenc}
\usepackage[usenames]{color} 
\usepackage{ulem}
\usepackage{amsmath,amssymb,theorem} 
\usepackage{bm}
\usepackage{dcolumn}
\usepackage{hhline}
\usepackage{float}
\usepackage[caption=false,font=large,labelfont={bf,sf,large}]{subfig}

\journal{Journal of Alloys and Compounds}

\begin{document}

\begin{frontmatter}

\title{Domain formation in the deposition of thin films of two-component mixtures}

\author{Tung B. T. To}
\ead{tungto@if.uff.br}
\author{F\'abio D. A. Aar\~ao Reis}
\ead{reis@if.uff.br}
\address{Instituto de F\'{i}sica, Universidade Federal Fluminense,\\
Avenida Litor\^{a}nea s/n, 24210-340 Niter\'{o}i, RJ, Brazil}

\begin{abstract}

We perform a kinetic Monte Carlo simulation study of a model of thin film deposition of a
two-component mixture in which the activation energy for diffusion of an adatom is additive
over its nearest neighbors and in which the interactions between adatoms of the same
species are stronger than those between adatoms of different species.
The film morphology is investigated for broad ranges of values of the ratios between terrace
diffusivity and atomic flux, of the probabilities of detachment from lateral neighbors,
and of probabilities of crossing terrace edges.
First, we consider a symmetric case in which the interactions of adatoms of the same species
are the same for the two components.
For low and intermediate temperatures, we show the formation of narrow domains,
whose widths are $3$-$30$ lattice constants, which meander in the layers parallel
to the substrate, and are connected through very long distances.
The domain width depends on the diffusivity-flux ratio of the same species,
but it is weakly affected by interactions of different species,
so rough estimates of that width can be obtained by using tabulated properties of
films with a single component.
At high temperatures, the separation of domains is enhanced, but their long distance
connectivity is lost, which shows that the formation of the long meandering domains
is restricted to a certain range of temperature and flux.
Similar morphological features are obtained when the intra-species interactions are different
and the adatom diffusion coefficients on terraces of the same species differ up to two orders of magnitude.
An approximate scaling relation for the domain width is obtained, but the species with the largest
mobility constrains the temperature range in which the long connected domains are found.

\end{abstract}


\end{frontmatter}

\section{Introduction}
\label{intro}

The separation of domains in thin films with different components such as metals, semiconductors, or
organic molecules is important for their technological applications because the domain
organization may provide special mechanical, chemical, optical, or electronic properties to the films
and may be used to design nanostructures with particular shapes.
For instance, phase separation is helpful for production of nanowires
after selective removal of one of the components of binary alloy films
\citep{adams1993,fukutani2007,thogersen2018}
and domain formation is required for photovoltaic applications of hybrid organic films
\citep{banerjeePRL2013,lorchJCP2017}.

The production of films with the desired morphologies at microscale or nanoscale is known
to depend on the thorough control of the deposition conditions, which
motivated several models of deposition of binary alloy films in the last decades.
In one of the pioneering models, the growth of Al-Ge films was represented by the sequential addition
of a small number of layers followed by the relaxation according to an Ising Hamiltonian \cite{adams1993}.
These features are experimentally observed in mixtures of metallic \citep{adams1992,fukutani2007} and
semiconductor \citep{Kuech2016} components.
The possibility of lateral and vertical modulation of concentration, which depends on the growth conditions,
was shown in a phase-field model in Ref. \protect\cite{lu2012}.
This type of model also describes the deposition of films at a coarse-grained
level, with applications to mixtures of Cu and Mo \citep{ankit2019} and of C and Ni
\citep{kairaitis2014,kairaitis2018}.
There are also several models describing the atomic scale processes in the deposition of alloy films
\citep{drossel1996,leonard1997,he2006}, in many cases including applications to specific materials such
as Ge-Mn \citep{mouton2014}, Ni-Ti \citep{zhu2014}, Ga-As-Bi \citep{rodriguez2016},
CoPt/SiO${}_2$ \citep{luCGD2017},
PbTe-CdTe \citep{minkowski2016,minkowski2018}, Ni-SrTiO${}_{\text 3}$ \citep{hennes2018},
and Si-Al \citep{thogersen2018}, Ag-Cu \citep{elofsson2016}, and Ag-Au \citep{elofsson2018}.
In models with relaxation processes restricted to the film surface, an important result is the
description of compositional modulations in the vertical and/or in the lateral directions
at nanoscale, as observed in experiments
\citep{mouton2014,zhu2014,tian2014,minkowski2016,minkowski2018,hennes2018,thogersen2018,kairaitis2018,tait2018}.

The atomistic models for deposition of films with binary mixtures represent the competition of
atomic/molecular flux and the diffusion of the adsorbed atoms/molecules.
They extend the framework previously developed for modeling thin
film deposition with a single chemical species \citep{michely,etb}.
One of the most successful models for metal and semiconductor epitaxy is that of
Clarke and Vvedensky (CV) \citep{cv}, in which the rate of adatom hops depends on an energy barrier
that characterizes the current site neighborhood; for details, see e.g. Ref. \cite{etb}.
For this reason, the CV-type model was extended to the description of deposition of binary alloy monolayers
and submonolayers \citep{kotrla2000,dumont2008,einax2007,einax2009,einax,han2011,han2014}
and of thin films \citep{tao2008}.

In the present work, we use kinetic Monte Carlo (KMC) simulations of a CV-type model
to study the structure of thin films with two components in which
the formation of domains of the same species is favored.
The model parameters correspond to broad ranges of values of diffusion coefficients of adsorbed atoms,
binding energies, and atomic fluxes, in order to explore possible properties of several materials
and a variety of deposition conditions.
We first consider the case of two species with similar dynamics, i.e. similar activation energy
of terrace diffusion and similar binding energy.
In a certain range of temperature and flux, it shows the formation of very large domains that
meander across the film, connecting distant points, but have very narrow widths.
We also consider the case of two species with different surface dynamics and show that the
species with the largest mobility determines the conditions in which that domain morphology can be observed.
Approximate relations for the domain widths are obtained in terms of
the model parameters, which may help to obtain the desired nanoscale patterns
in real films of two-component mixtures.

The rest of this paper is organized as follows.
In Sec. \ref{basic}, we present the deposition model, the simulation methods, and
the main quantitites to be measured.
Sec. \ref{results} presents the results for two species with similar surface dynamics and
for two species with different surface dynamics in separate subsections, where the
interpretations and consequences of the results are discussed.
Sec. \ref{conclusion} summarizes our results and presents our conclusions.

\section{Model and methods}
\label{basic}

\subsection{Deposition model}
\label{models}

The model is defined in a simple cubic lattice in which the edge of a site is taken as the unit length;
thus, the values of lengths shown throughout this work are dimensionless.
The substrate is initially flat, located at $z=0$, and has lateral size $L$.
Periodic boundary conditions are considered in $x$ and $y$ directions.
The two species of atoms/molecules are termed A and B particles; each deposited
particle occupies one lattice site, independently of being A or B.
We impose the solid-on-solid condition to the deposit, i.e. overhangs are not allowed.
The set of particles with the same $\left( x,y\right)$ positions is termed a column of the deposit.

Deposition occurs with a flux of particles in the $z$ direction towards the substrate.
The number of incident particles per column per unit time is $F$ (unit ${\text s}^{-1}$).
The probabilities of deposition of A and B particles are the same, so the films
have an equimolar mixture.
Adatom desorption is not considered in the model.
We consider that mobile particles are only those at the top of each column.

Growth models that account for activated adatom diffusion of a single species
usually consider that the hopping rate of the adatom can be written in the form
$k = \nu\exp{\left[ -E_{act}/\left( k_B T\right)\right]}$, where
$\nu$ is a frequency (which depends on the temperature in the CV model \citep{cv}),
$E_{act}$ is an activation energy,
$k_B$ is the Boltzmann constant, and $T$ is the temperature \citep{etb}.
In the simplest cases, $E_{act}$ has an additive form in terms of nearest neighbor (NN) interaction
energies at the current position:
$E_{act}=E_s+nE_b$, where $E_s$ denotes the energy for diffusion on terraces
and $E_b$ denotes the bond energy with a lateral NN.
The terrace diffusion coefficient is $D=\nu\exp{\left[ -E_s/\left( k_B T\right)\right]}$.
More realistic models also consider an additional energy barrier for crossing terrace edges
\citep{ES}, which leads to a probability for the hop to be executed to a lower or upper terrace.

In modeling two-component mixtures in the submonolayer regime, this approach was generalized to
consider that $E_s$ depends on the species of the mobile adatom and
$E_b$ is different for A-A, B-B, and A-B interactions \citep{einax2007}.
In the case of thin films studied here, we have to consider that $E_s$ depends on the species
of the mobile adatom and on the species in terrace below that adatom.

These features can be summarized in a general expression for the rate at which a particle
$p=$\{A,B\} hops:
\begin{equation}
k_{pqn_An_B}(T)=D_{pq}  (\epsilon_{pA})^{n_A} (\epsilon_{pB})^{n_B} ,
\label{hoppingrate}
\end{equation}
where:

\par\noindent  $q=$\{A,B,S\} denotes the site below $p$, which includes the possibility of substrate sites;

\par\noindent  $n_A$ and $n_B$ are the number of lateral neighbors of species A and B, respectively;

\par\noindent $D_{pq}= \left( {2k_BT}/{h}\right)
\exp{\left[ -\left(E_{pq}^{\left( ter\right)}\right)/\left( k_BT\right)\right]}$ 
is the diffusion coefficient of particle $p$ on a terrace with a particle $q$ immediately below it
(i.e. the hopping rate without lateral neighbors, $n_A=n_B=0$),
and $E_{pq}^{\left( ter\right)}>0$ is the activation energy for this process;

\par\noindent
$\epsilon_{pA}=\exp{\left[ -\left( E_{pA}^{\left( lat\right)}\right) /\left( k_BT\right)\right]}$ and
$\epsilon_{pB}=\exp{\left[- \left( E_{pB}^{\left( lat\right)}\right) /\left(k_BT\right)\right]}$,
where $E_{pA}^{\left( lat\right)}$ and $E_{pB}^{\left( lat\right)}$ are bond energies with
lateral neighbors A and B, respectively.
Hereafter, the parameters $\epsilon_{pq}$ are termed detachment probabilities because their
effects are to reduce the hopping rates when the adatoms have lateral neighbors.

\begin{figure}[!h]
\begin{center}
\includegraphics[width=0.7\textwidth]{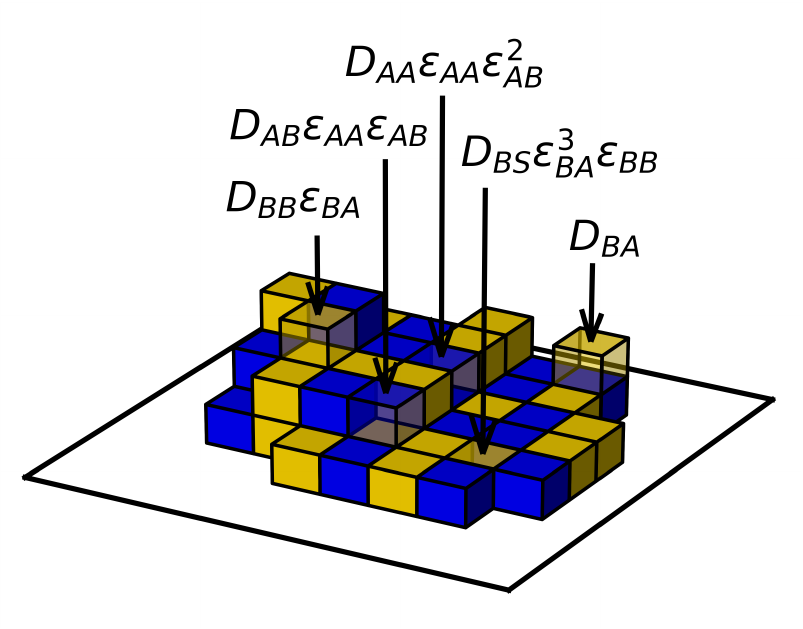}
\caption{
A small part of a deposit with A (blue) and B (yellow) particles and the
hopping rates of some surface particles.
}
\label{rates}
\end{center}
\end{figure}

In our model, we consider $E_{pq}^{\left( ter\right)}=E_{qp}^{\left( ter\right)}$ and $E_{pq}^{\left( lat\right)}=E_{qp}^{\left( lat\right)}$, which gives $D_{AB}=D_{BA}$ and $\epsilon_{AB}=\epsilon_{BA}$. Fig. \ref{rates} shows the hopping rates of mobile particles in some configurations.

When a mobile particle hops, it moves to the top of a nearest neighbor column,
which is randomly chosen among four directions, $\left( \pm x,\pm y\right)$.
Upward and downward hops are allowed.

An additional energy barrier is considered when a particle attempts to hop to a layer at a
different height $z$, which is the Ehrlich-Schw\"{o}ebel barrier \citep{ES}.
Here this mechanism is represented by a single probability $P$
for any particle to hop to an upper or lower layer; with probability $1-P$,
the hop attempt is not executed.
In all films studied here, local height differences are not large; thus, long
distance hops in the vertical direction are not observed
(if such hops were possible, then the probability of hopping to a different terrace should account
for the diffusion along vertical walls and for the probable return to the original position;
see Ref. \protect\cite{lealJPCM}).

Here we consider cases in which the diffusion on terraces of a different species is faster than
that on terraces of the same species, i.e. $D_{AB}> D_{AA}$ and $D_{AB}> D_{BB}$.
This represents a weaker interaction between atoms of different species when compared to
interactions of atoms of the same species.
We also assume that interactions with the substrate are weak and set $D_{AS}=D_{BS}=D_{AB}$.
Thus, $D_{AB}$ is the parameter representing the largest hopping rates of the system,
while the smaller values of $D_{AA}$ and $D_{BB}$ favor the formation of domains of the same species
at the film surface.

In the CV model with a single species with terrace diffusion coefficient $D$,
the main parameter is the dimensionless diffusion-to-deposition ratio $R=D/F$.
Here, we extend that reasoning with the definition of three ratios:
\begin{equation}
R_{AA} = \frac{D_{AA}}{F} \qquad , \qquad R_{BB} = \frac{D_{BB}}{F} \qquad , \qquad 
R_{AB} = \frac{D_{AB}}{F} .
\label{defR}
\end{equation}
The remaining dimensionless parameters of the model in the present work are the detachment
probabilities $\epsilon_{AA}$, $\epsilon_{BB}$, and $\epsilon_{AB}$, and the jump probability $P$.
These variables are more useful for the interpretation of scaling regimes than the sets of
activation energies and temperature, as shown in works with a single atomic species 
\citep{etb,cv2015}.

\subsection{Simulation methods}
\label{simulations}

We perform simulations on lattices with $L=1024$, which are sufficiently large to represent many
different microscopic environments in a single deposit.
For most metals and semiconductors, this length corresponds to $0.2$-$0.3\mu$m.
For each parameter set, we generated $10$ deposits and observed small fluctuations in
the average quantities; significant differences are observed only in the sizes of the largest clusters
of A and B particles at large temperatures, but their orders of magnitude are the same in the different
samples.
The films typically have $30$ deposited layers, which corresponds to thicknesses of order
$\sim 10$nm for metals or semiconductors.

The ratios $R_{AA}$ and $R_{BB}$ range between ${10}^4$ and ${10}^7$, and the values of $R_{AB}$
are $10$-$100$ times larger than the largest single-species ratio ($R_{AA}$ or $R_{BB}$).
The values of $\epsilon_{AA}$ and $\epsilon_{BB}$ range between ${10}^{-5}$ and ${10}^{-2}$.
$\epsilon_{AB}$ is chosen in the range ${10}^{-2}$-${10}^{-1}$ to represent weak bonds
between atoms of different species.
The jump probability $P$ is chosen in the range ${10}^{-3}$-${10}^{-1}$.

Since the diffusion coefficients $D_{pq}$ and the detachment probabilities $\epsilon_{pq}$
increase with temperature, it is interesting to compare results in which those parameters
are simultaneously raised.
For this reason, we performed several simulations with the conditions
$R_{AB}/R_{AA}=\epsilon_{AB}/\epsilon_{AA}$, $R_{AB}/R_{BB}=\epsilon_{AB}/\epsilon_{BB}$,
which correspond to
$E_{AB}^{\left( ter\right)}-E_{AB}^{\left( lat\right)} =
E_{AA}^{\left( ter\right)}-E_{AA}^{\left( lat\right)} =
E_{BB}^{\left( ter\right)}-E_{BB}^{\left( lat\right)}$.
However, we also performed simulations with parameters that are not constrained by these relations.

The simulations are implemented with an extension of the algorithm for a single species, which is
described in detail in Ref. \cite{lam1997} and was formerly used in submonolayer growth simulations
in our group \citep{submonorev}.

The $L^2$ surface particles have their positions ($x,y$) grouped into ninety lists $X_{pqn_An_B}$
according to the species ($p=A,B$), the type of the particle below it ($q=A,B,S$),
the number of particles $A$ at nearest-neighbor sites ($n_A=0, \dots, 4$)
and the number of particles $B$ at nearest-neighbor sites ($n_B=0, \dots, 4$), with the latter numbers
constrained to $0 \leq n_A+n_B \leq 4$.
The position of a surface particle in a list $X_{pqn_An_B}$ is stored in an inverted-list matrix
$M_1(x,y)$.
In addition, a matrix $M_2(x,y,z)$ stores the particles ($A$ or $B$) and empty sites,
with $M_2(x,y,0)=S$, and is used for rapid access to the configuration of the neighborhood of a
mobile particle.

At each step of the simulation, the rates of all possible events (namely, deposition of A,
deposition of B, and hop of one of the $L^2$ surface particles) are calculated and their sum is
denoted as $\Sigma$.
The probability of each deposition event is the ratio between its rate and $\Sigma$.
Since all particles in each list $X_{pqn_An_B}$ have the same hopping rate, the probability of that
list is the product of the number of particles in the list and the hopping rate divided by $\Sigma$.
The event to be executed is then chosen according to those probabilities.
In the case of choosing a list, one of its particles is randomly chosen to hop, in a direction
which is also randomly chosen among four possibilities ($\pm x$, $\pm y$); in the case of a hop
to a different layer, the hop is executed with probability $P$, otherwise nothing occurs.

After a simulation step, the time is incremented by $1/\Sigma$ minus the natural logarithm of a 
randomly chosen number in the interval $\left( 0,1\right]$; the latter contribution has very small
effect on the total deposition time.

\subsection{Basic quantities}
\label{quantities}

The fluctuations of the surfaces of the growing films are monitored with the calculation of the
roughness, which is defined as
\begin{equation}
W(L,t)\equiv \left< {\left[ \overline{{\left( h - \overline{h}\right) }^2}  \right] }^{1/2} \right>  ,
\label{defw}
\end{equation}
where the overbars indicate spatial averages and the angular brackets indicate configurational averages.
Since $W$ accounts for height fluctuations in the whole film surface, it is frequently called global
roughness; the local asperity of the surface is sometimes quantified by the local surface roughness
\cite{chamereis2004}.

Visual inspection of cross sections of the deposited films showed meandering domains of both species.
To characterize these domains, we calculate characteristic lengths and areas in horizontal
cross sections at heights $z=5$, $10$, $15$, and $20$.
The distributions of domain area $S$ of both species, $P_A\left( S\right)$ and $P_B\left( S\right)$,
are calculated using the Hoshen-Kopelman algorithm for identification of connected clusters \cite{hoshen}.

For the calculation of domain widths, in each line of constant $x$ or $y$, we measure the
size of all segments of consecutive sites of the same species.
This is equivalent to the definition of a spatial persistence length in one-dimensional interface
problems \citep{majumdarPRL2001}.
The average values of those segment sizes are defined as the domain widths $\lambda_A$ and $\lambda_B$;
in cases where A and B interactions are similar, the average domain width is 
$\lambda = \left( \lambda_A +\lambda_B \right) /2$.

\section{Results}
\label{results}

\subsection{Two species with the same surface mobility}
\label{symmetric}

Here we analyze data for $R_{AA}=R_{BB}$ and $\epsilon_{AA}=\epsilon_{BB}$,
which is hereafter termed symmetric case.
It represents mixtures of two species that have similar values of activation energy of
surface diffusion and of bond energy.

\subsubsection{Low temperatures}
\label{lowtempsym}

Here we consider $R_{AA}={10}^4$, which is typical of low temperatures.
For atoms with $E_{AA}^{\left( ter\right)}\approx 0.5eV$ and a flux of $\sim 1$ monolayer per second,
this is obtained close to room temperature.
For some metals with very small activation barriers of surface diffusion (e.g.
$E_{AA}^{\left( ter\right)}\sim 0.2eV$ \citep{etb}), this corresponds to much lower temperatures.
The values of $\epsilon_{AA}$ are also expected to be small in these conditions;
for instance, with a low value $E_{AA}^{\left( lat\right)}\sim 0.2eV$ and at room temperature, we have
$\epsilon_{AA}\sim {10}^{-3}$.

\begin{figure}
\begin{center}
    \subfloat[]{\includegraphics[width=0.35\textwidth]{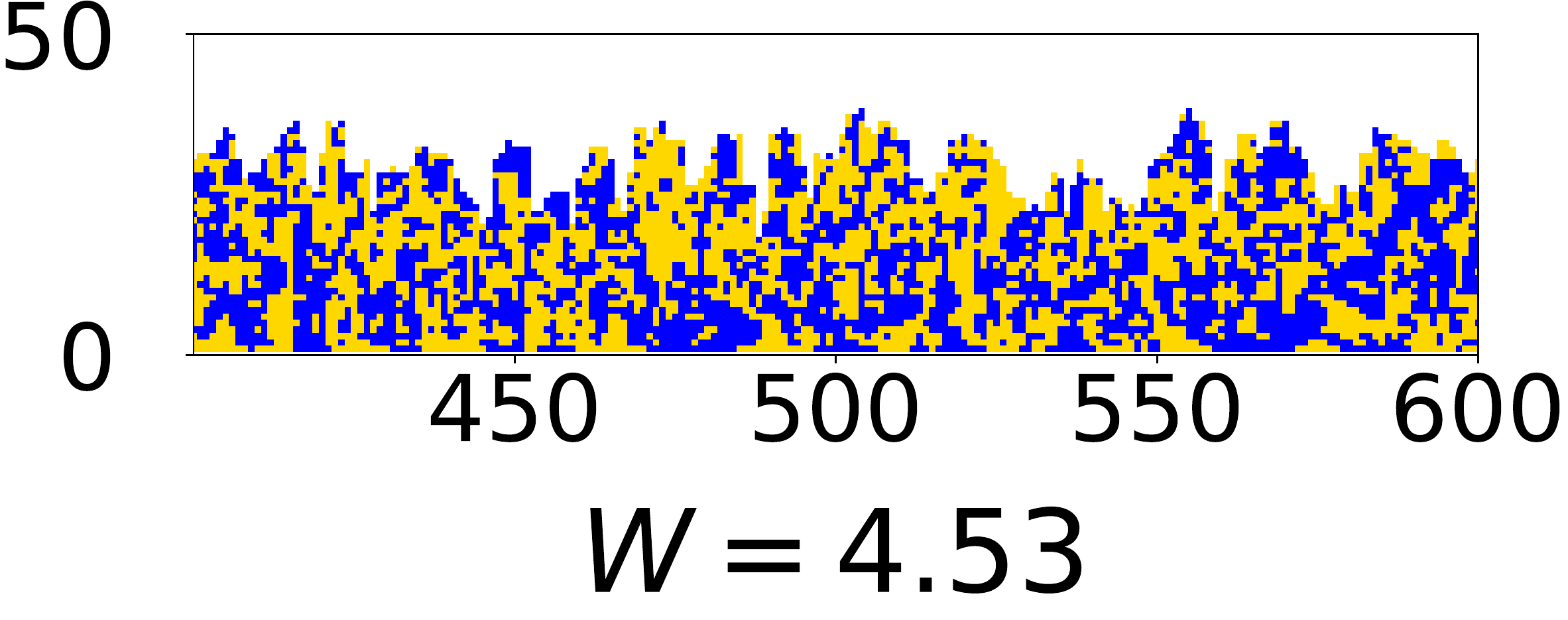}} \hspace{1cm}
    \subfloat[]{\includegraphics[width=0.35\textwidth]{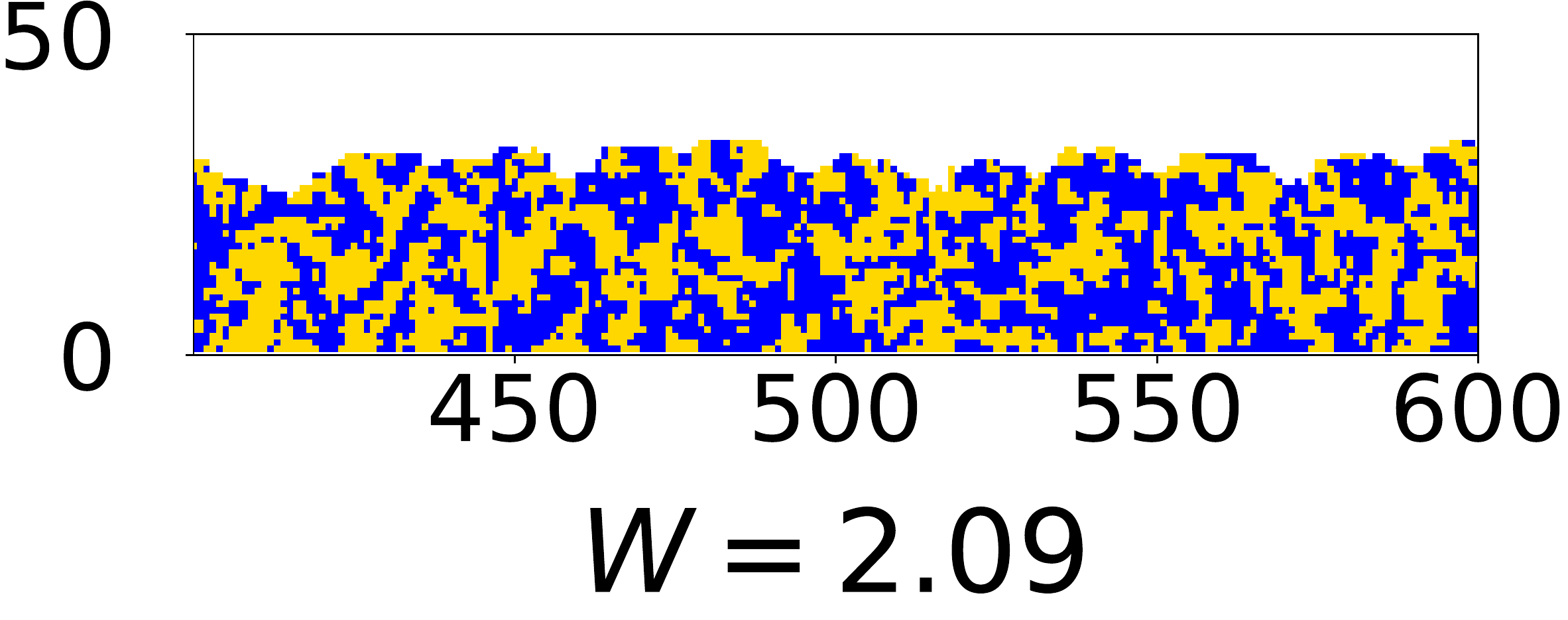}} \\
    \subfloat[]{\includegraphics[width=0.35\textwidth]{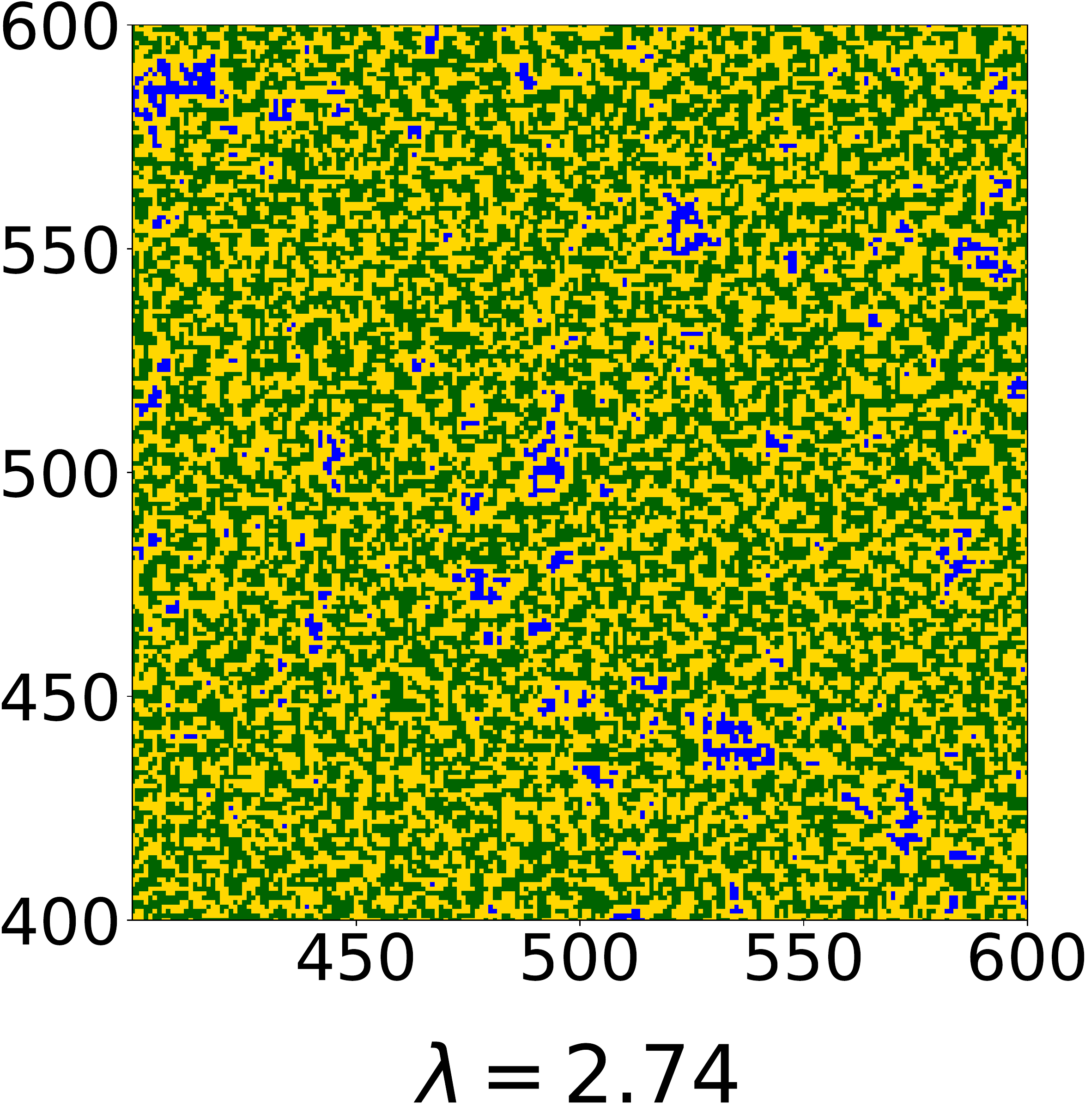}} \hspace{1cm}
    \subfloat[]{\includegraphics[width=0.35\textwidth]{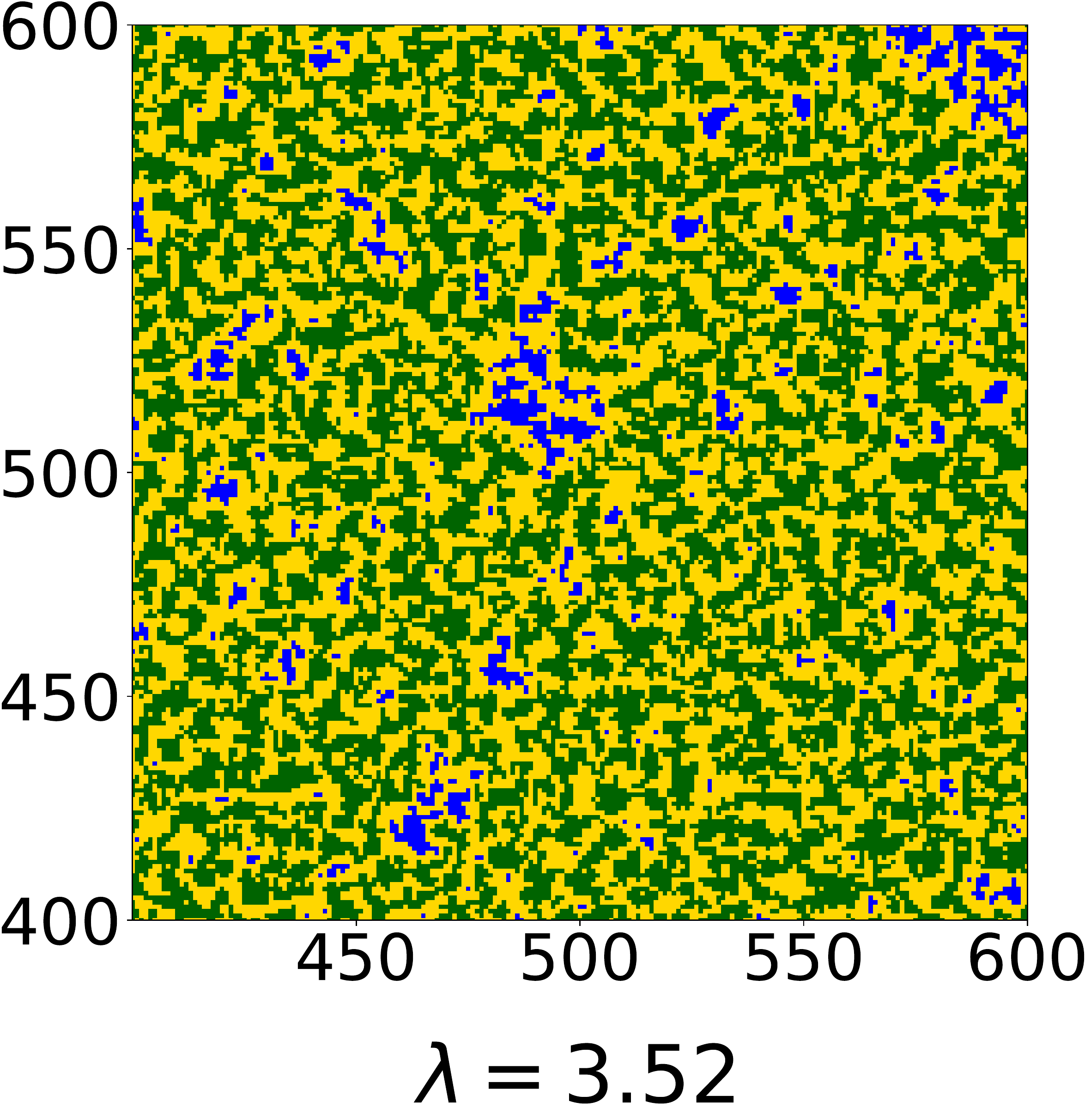}} \\
	\subfloat[]{\includegraphics[width=0.35\textwidth]{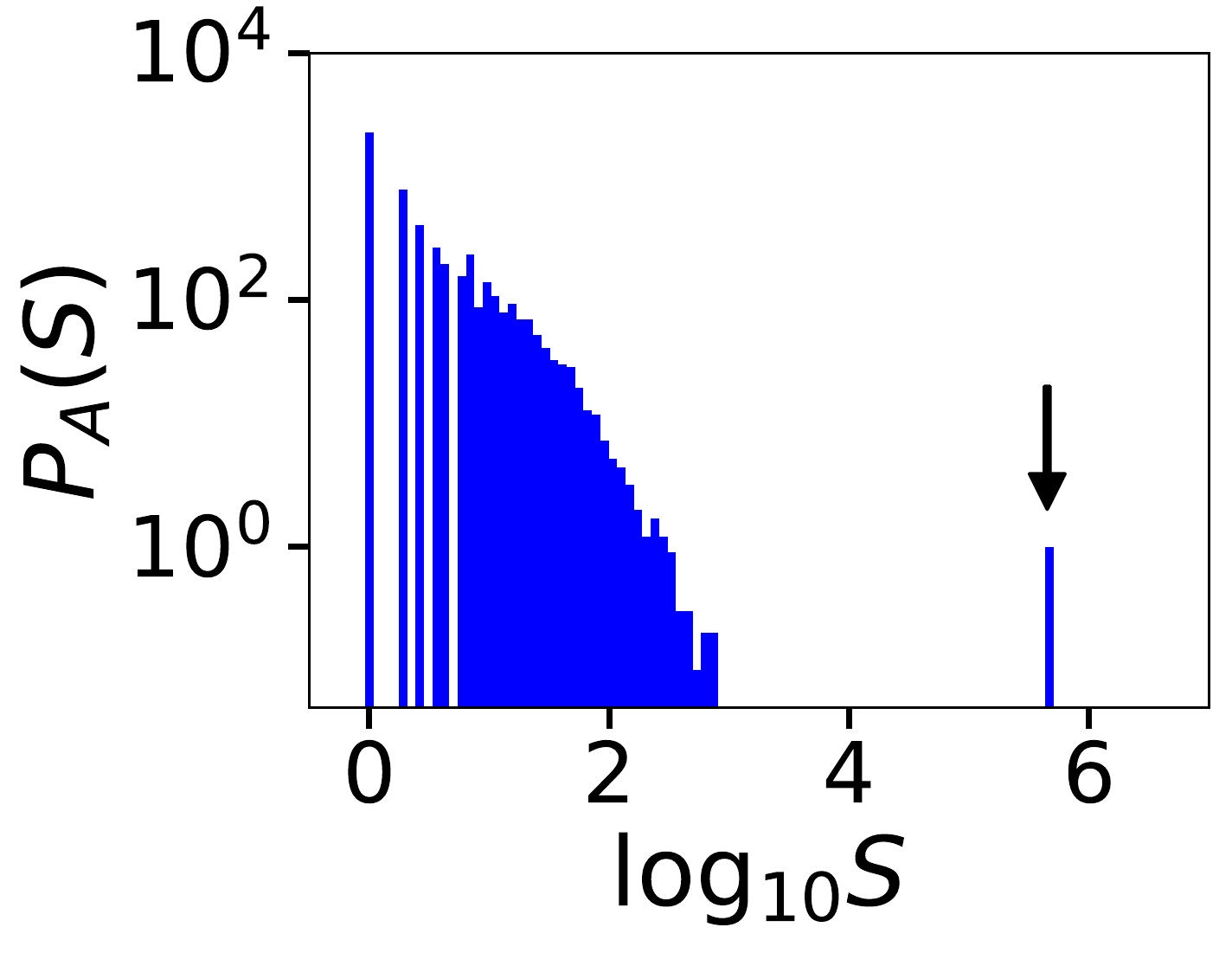}} \hspace{1cm}
    \subfloat[]{\includegraphics[width=0.35\textwidth]{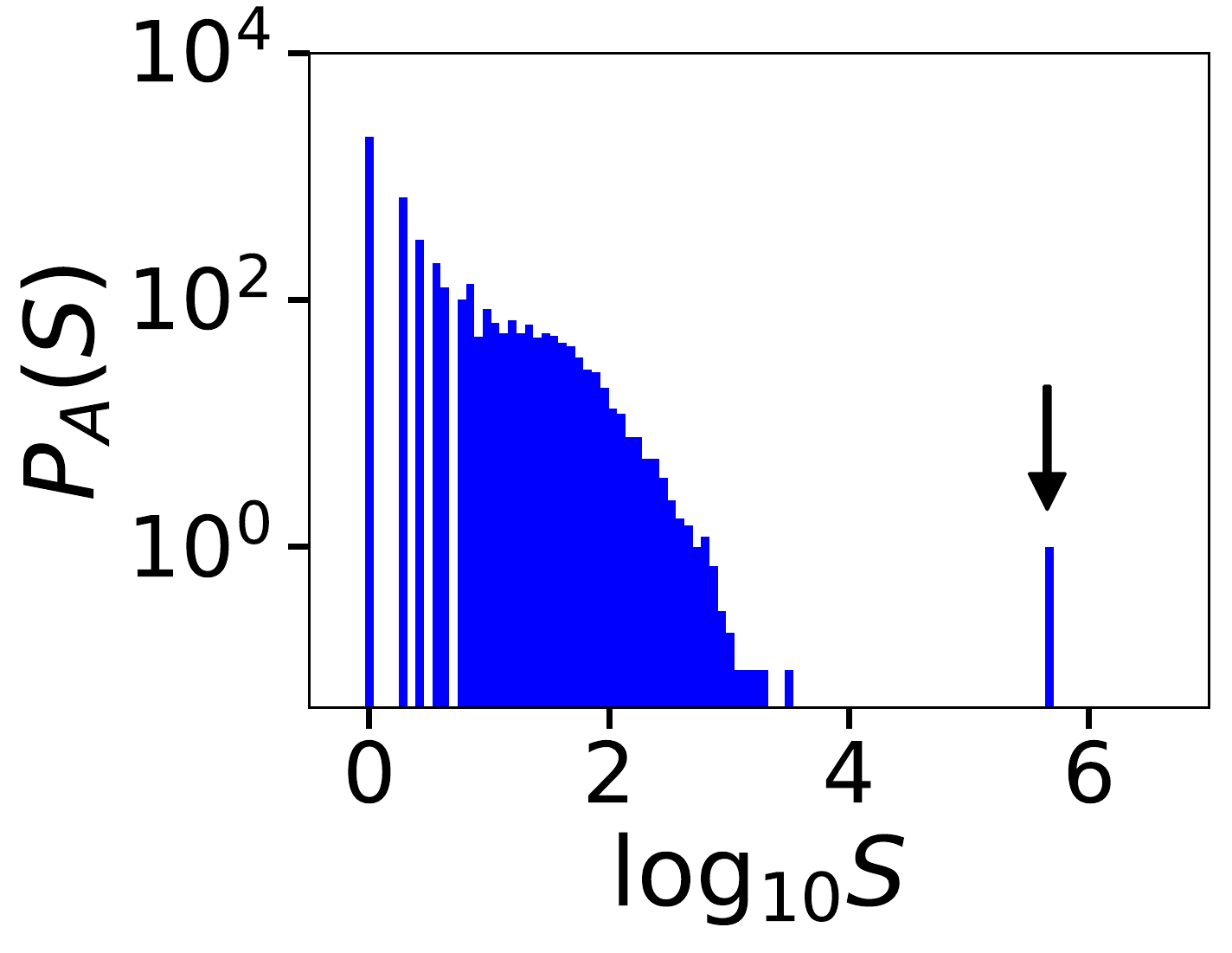}}
\caption{
(a) Vertical cross sections, (c) horizontal cross sections at $z=10$, and
(e) domain size distribution at $z=10$ of films deposited with
$R_{AA}={10}^4$, $\epsilon_{AA}={10}^{-3}$, $R_{AB}={10}^6$, $\epsilon_{AB}=0.1$, and $P={10}^{-3}$.
(b) Vertical cross sections, (d) horizontal cross sections at $z=10$, and
(f) domain size distribution at $z=10$ of films deposited with
$R_{AA}={10}^4$, $\epsilon_{AA}={10}^{-2}$, $R_{AB}={10}^5$, $\epsilon_{AB}=0.1$, and $P={10}^{-1}$.
A particles are in blue, B in yellow, and the largest domain of A is in green in (c) and (d).
The surface roughness is indicated in (a) and (b), and
the average domain sizes are indicated in (c) and (d).
In (e) and (f), the arrows indicate the largest domain sizes.
}
\label{sym4}
\end{center}
\end{figure}

Fig. \ref{sym4}(a) shows vertical cross sections of films with
$R_{AB}={10}^6$, $\epsilon_{AB}=0.1$, $\epsilon_{AA}={10}^{-3}$, and $P={10}^{-3}$.
Fig. \ref{sym4}(b) shows vertical cross sections of films with
$R_{AB}={10}^5$, $\epsilon_{AB}=0.1$, $\epsilon_{AA}={10}^{-2}$, and $P={10}^{-1}$.
The values of the surface roughness $W$, which are shown below those images,
illustrate the general trend of $W$ to increase as $P$ decreases.
This parallels the observations in films with a single species \citep{etb,lealJPCM}.
In both cases, we observe the formation of disordered narrow domains;
some of those domains connect the substrate and the top surface.

Figs. \ref{sym4}(c) and \ref{sym4}(d) show horizontal cross sections at height $z=10$
of films deposited with the same parameters of Figs. \ref{sym4}(a) and \ref{sym4}(b), respectively.
The average domain width $\lambda$ is indicated below those plots and quantitatively confirm
that the domains are very narrow.
Figs. \ref{sym4}(e) and \ref{sym4}(f) show the corresponding distributions of domain size
at height $z=10$ (recall that A and B are equivalent in this case).
The distributions obtained at $z=15$ were similar.

These distributions have a decreasing part with average $\langle S \rangle\sim {10}^2$,
which corresponds to small domains of A and B with randomly distributed sizes.
However, they also have peaks at a size $S\sim {10}^6$, which is indicated by arrows in
Figs. \ref{sym4}(e) and \ref{sym4}(f) and appears in all samples simulated in these low
temperature conditions.
This indicates that the films have large domains of A and B crossing it in the horizontal directions;
the particles belonging to the largest A domain in Figs. \ref{sym4}(c) and \ref{sym4}(d) are highlighted.
That domain meanders across the sample with a very irregular shape and a very narrow and fluctuating
width, which is confirmed by the small value $\lambda$.

We also deposited films with $R_{AA}={10}^4$ and other five different sets of parameters.
The results are similar to those shown in Figs. \ref{sym4}(a)-(f): the films have large domains,
with total sizes of order $S\sim {10}^6$, but the average widths $\lambda$ vary between
$2.7$ and $4.2$, which corresponds to sizes $\sim 1$nm for metals and semiconductors.
No significant effect of $R_{AB}$, $\epsilon_{AB}$, or $\epsilon_{AA}$ is observed.

It is important to observe that the formation of long connected domains in a square grid,
which is the case of the horizontal sections of Figs. \ref{sym4}(b) and \ref{sym4}(d),
is not possible if the distribution of the two species is random because a
50-50 mixture is below the critical percolation point \citep{hoshen}.
Thus, the observed connectivity represents a long-range correlation of atomic positions.

\subsubsection{Effects of the diffusion coefficient on terraces of a different species}
\label{DABsym}

Here we analyze the effects of the diffusion coefficient $D_{AB}$ of a particle on a terrace
of a different species, i.e. A on a terrace of B or vice-versa.
This is possible by varying $R_{AB}$ while the other model parameters are kept fixed.
Note that $D_{AB}$ is set equal to $D_{AS}$,
which controls the formation of islands in the submonolayer regime because they
predominantly grow by capturing freely moving adatoms \cite{etb}.
We consider cases of non-negligible barriers for crossing terrace edges: $P\leq {10}^{-2}$.

\begin{figure}
\begin{center}
    \subfloat{\includegraphics[width=0.35\textwidth]{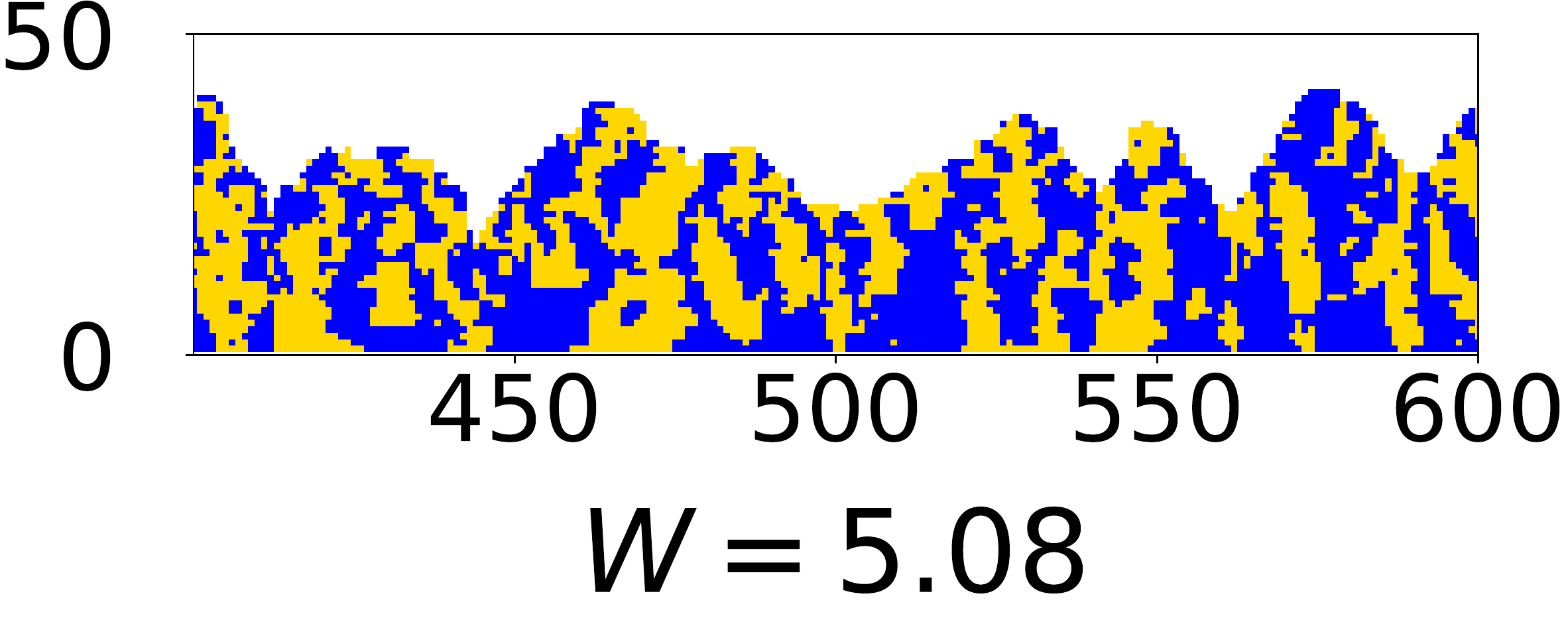}} \hspace{1.5cm}
    \subfloat{\includegraphics[width=0.35\textwidth]{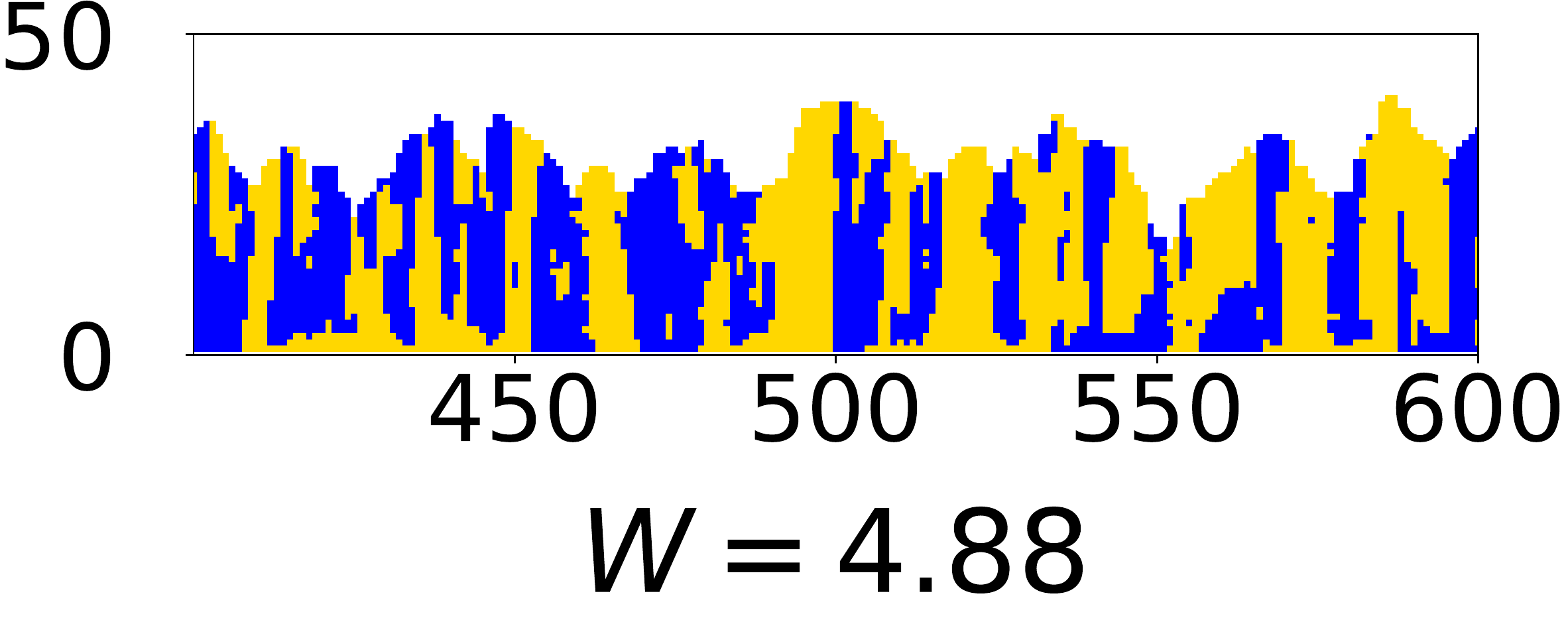}} \\
    \subfloat{\includegraphics[width=0.35\textwidth]{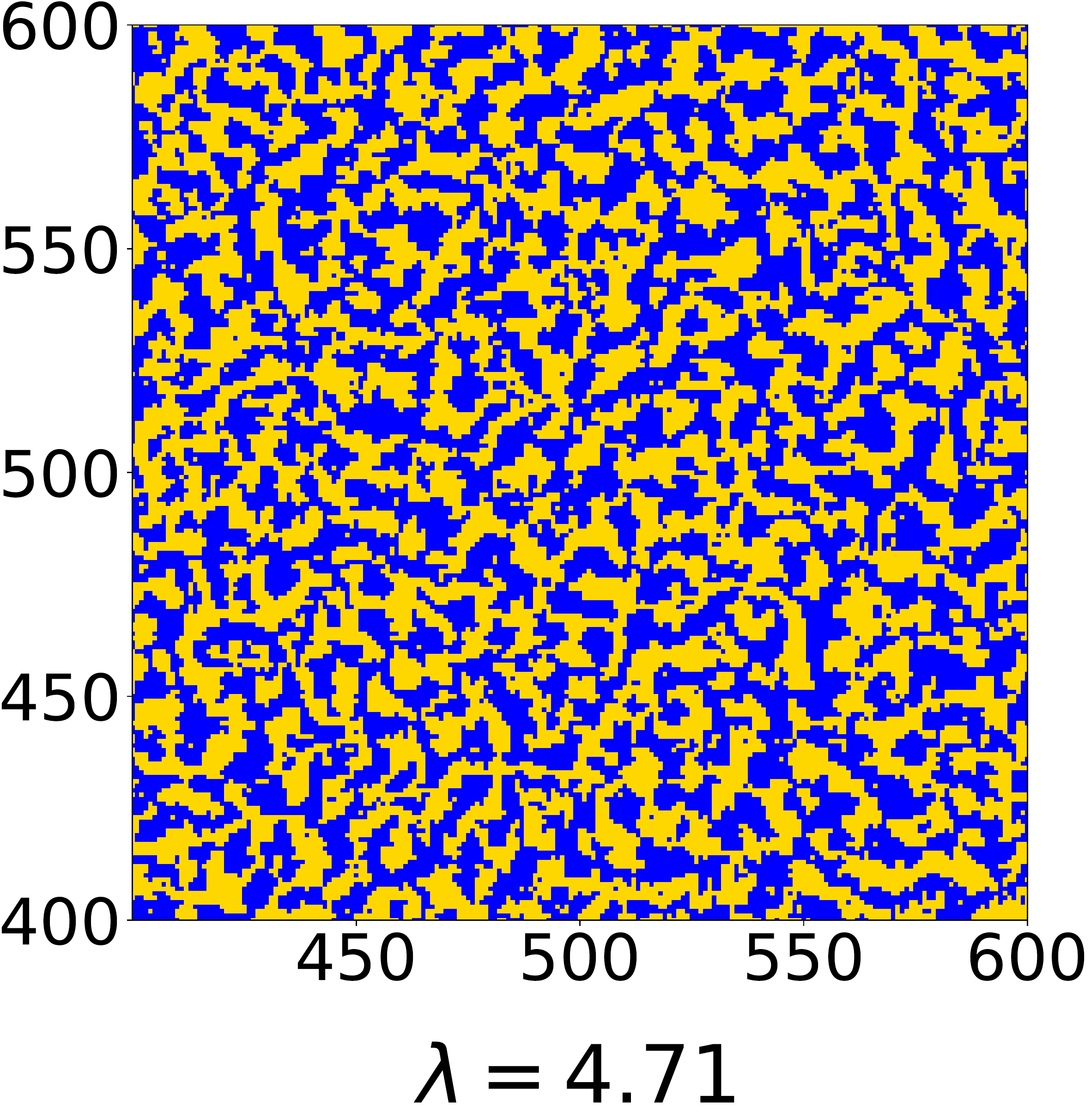}} \hspace{1.5cm}
    \subfloat{\includegraphics[width=0.35\textwidth]{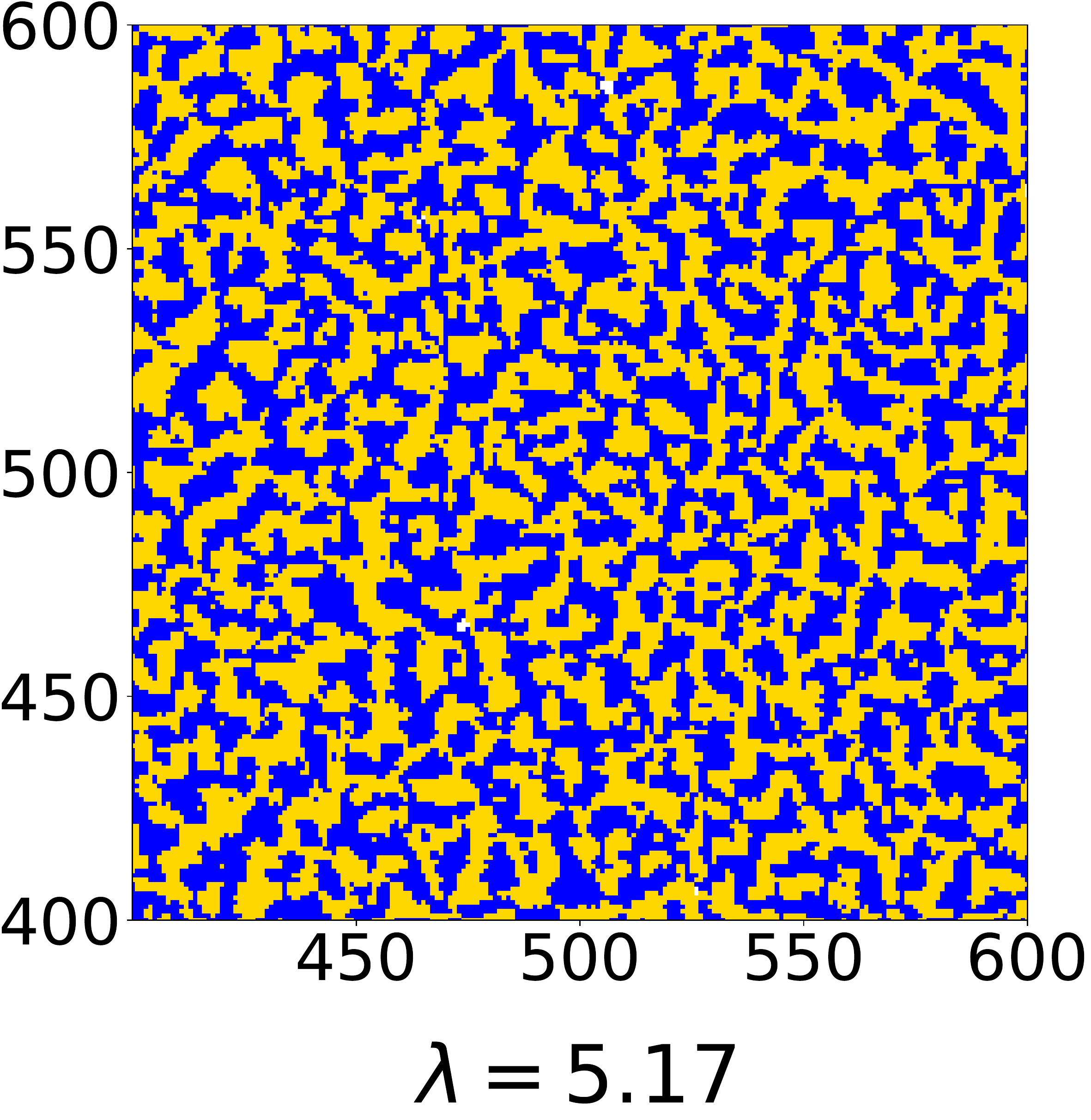}} \\
	\addtocounter{subfigure}{-4}
	\subfloat[]{\includegraphics[width=0.35\textwidth]{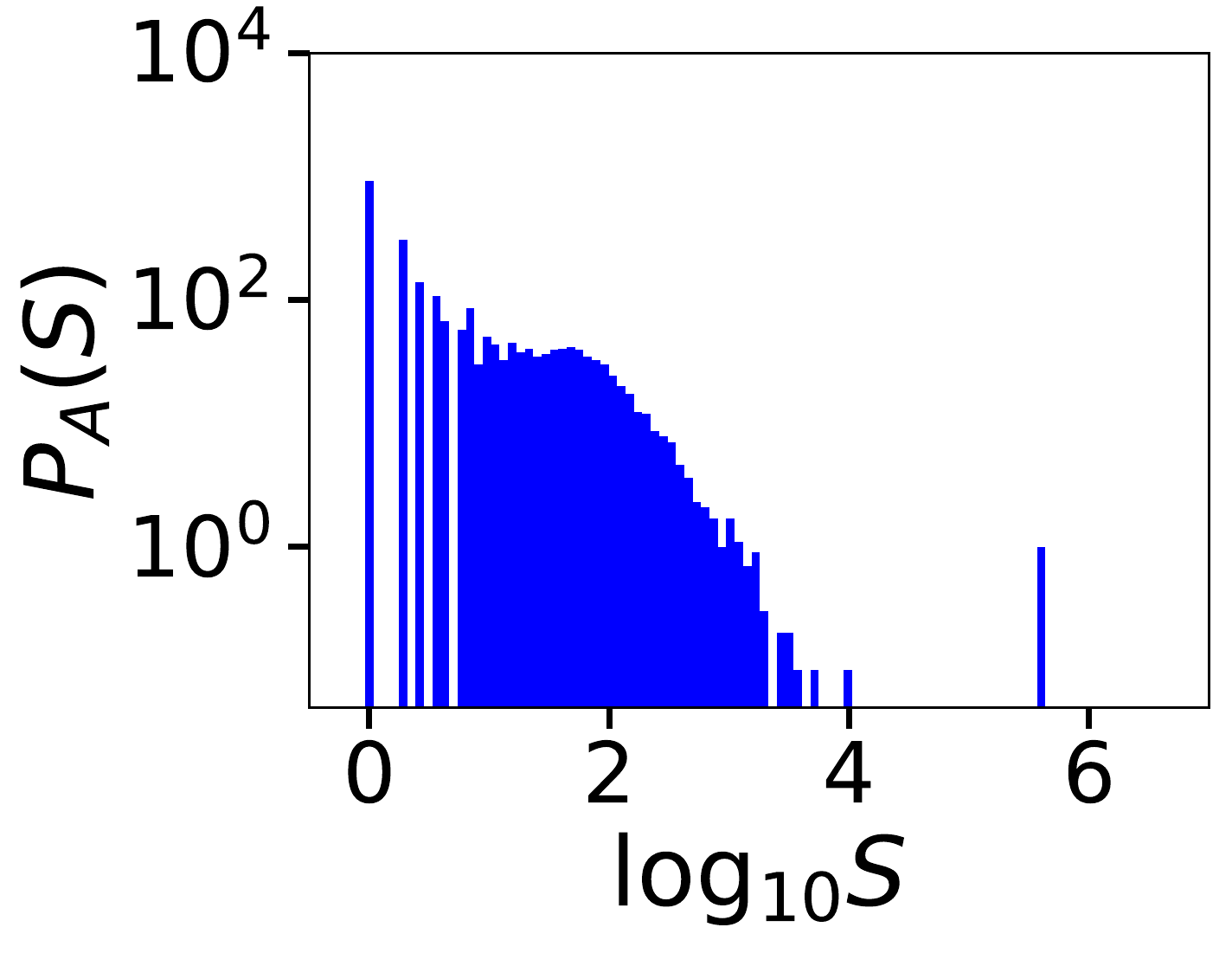}} \hspace{1.5cm}
    \subfloat[]{\includegraphics[width=0.35\textwidth]{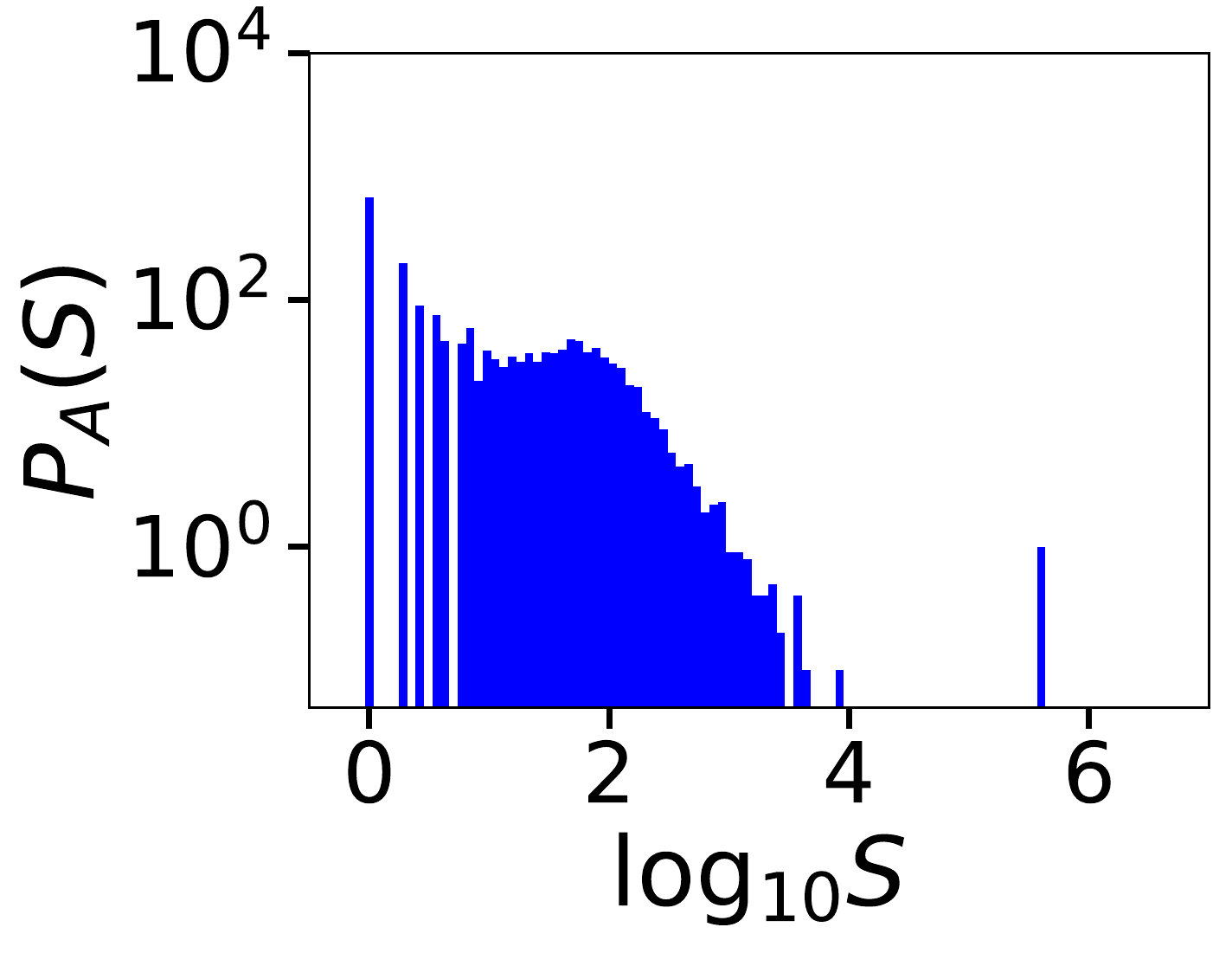}}
\caption{
Vertical cross sections, horizontal cross sections at $z=10$, and A domain size distributions
at $z=10$ in films deposited with
$R_{AA}={10}^5$, $\epsilon_{AA}={10}^{-2}$, $\epsilon_{AB}=0.1$, $P={10}^{-3}$, and:
(a) $R_{AB}={10}^6$; (b) $R_{AB}={10}^8$.
A particles are in blue, B in yellow.
The surface roughness and the average domain sizes in the cross sections at $z=10$ are indicated.
}
\label{symcompRAB}
\end{center}
\end{figure}

Fig. \ref{symcompRAB}(a) shows vertical and horizontal ($z=10$) cross sections and the domain size
distribution of films grown with $R_{AA}={10}^5$, $\epsilon_{AA}={10}^{-2}$, $P={10}^{-3}$,
$\epsilon_{AB}=0.1$, and $R_{AB}={10}^6$.
For comparison, Fig. \ref{symcompRAB}(b) shows cross sections and distributions of films grown
with the same parameters except $R_{AB}={10}^8$.
Domain size distributions obtained at $z=10$ and $z=15$ were similar in all cases.

Again we observe the formation of very large horizontal domains ($S\sim {10}^6$) in all films,
but with very small average widths $\lambda\sim 5$.
These results show that $D_{AB}$ has small effects on the domain size at layers $z\geq 10$.
The vertical cross sections in Figs. \ref{symcompRAB}(a),(b) show that the only significant effect
of $D_{AB}$ is to improve the vertical orientation of the domains, so that their meandering at
a given height is strongly correlated with the meandering at the layer below it.

\begin{figure}
\begin{center}
    \subfloat{\includegraphics[width=0.35\textwidth]{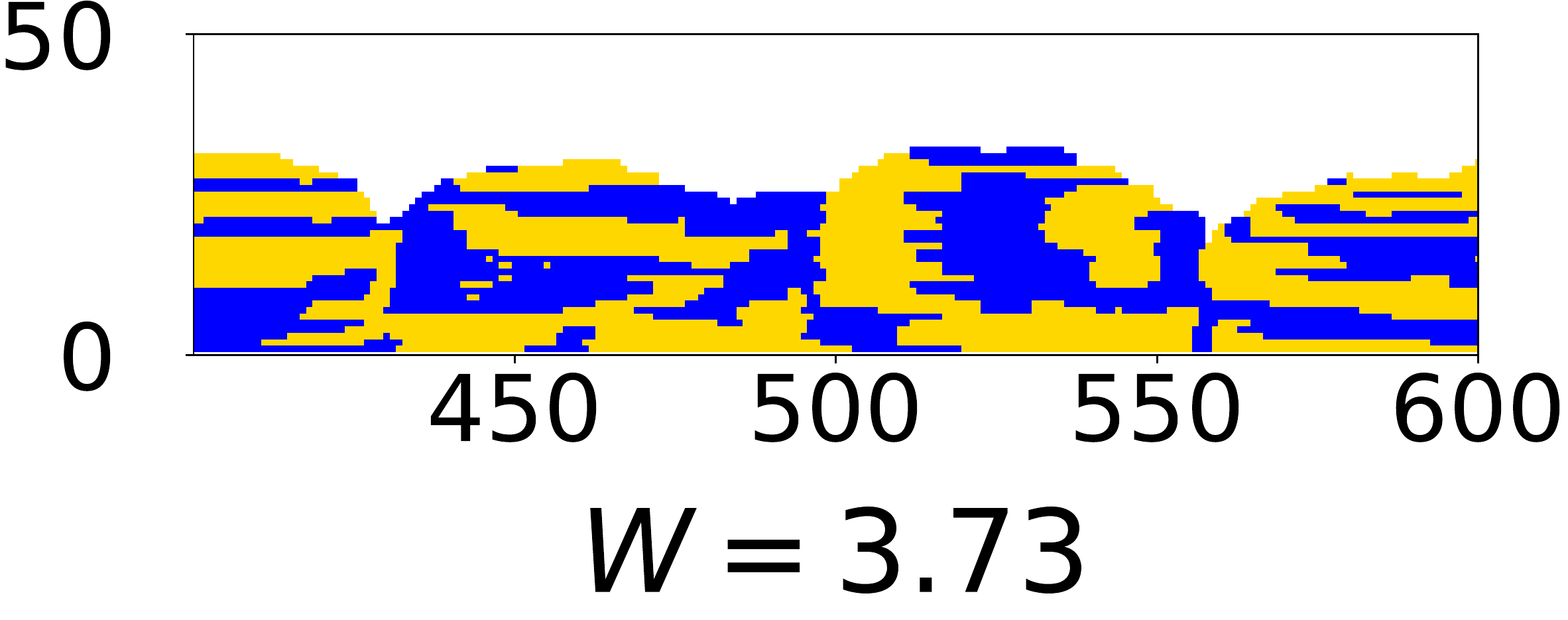}} \hspace{1.5cm}
    \subfloat{\includegraphics[width=0.35\textwidth]{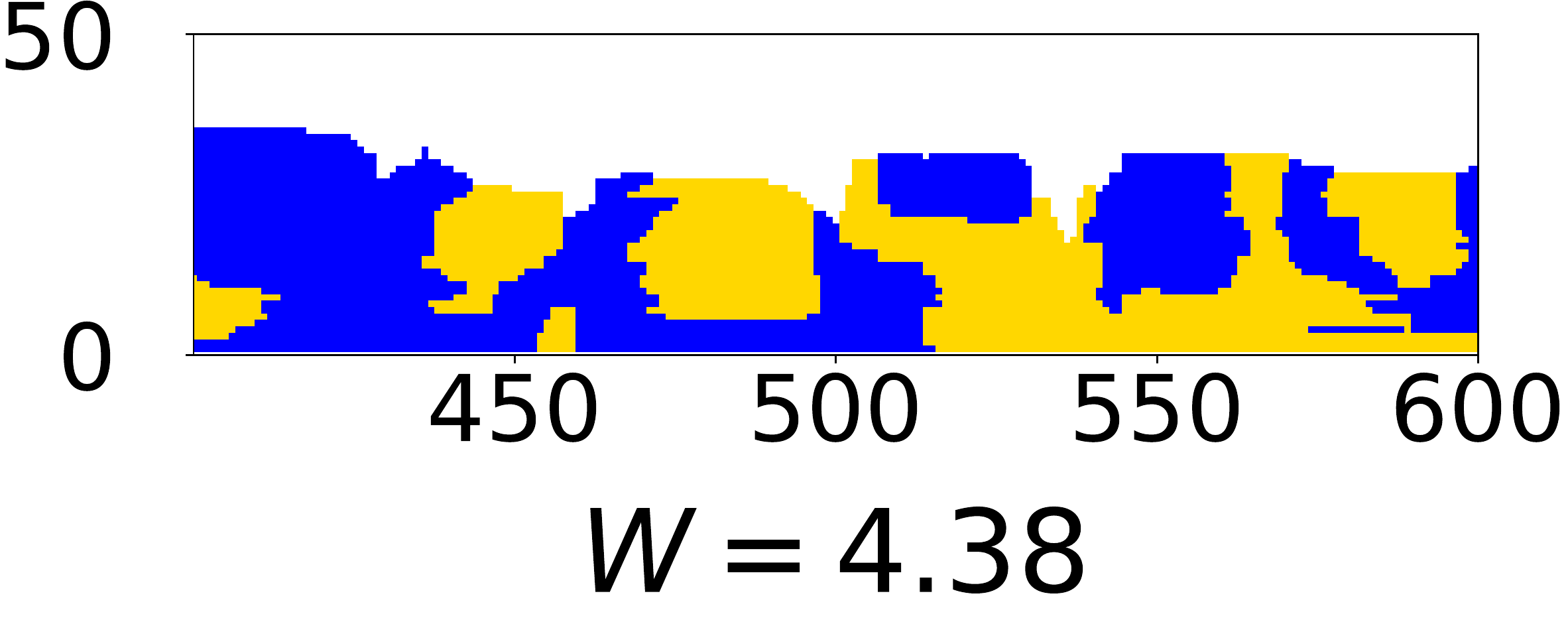}} \\
    \subfloat{\includegraphics[width=0.35\textwidth]{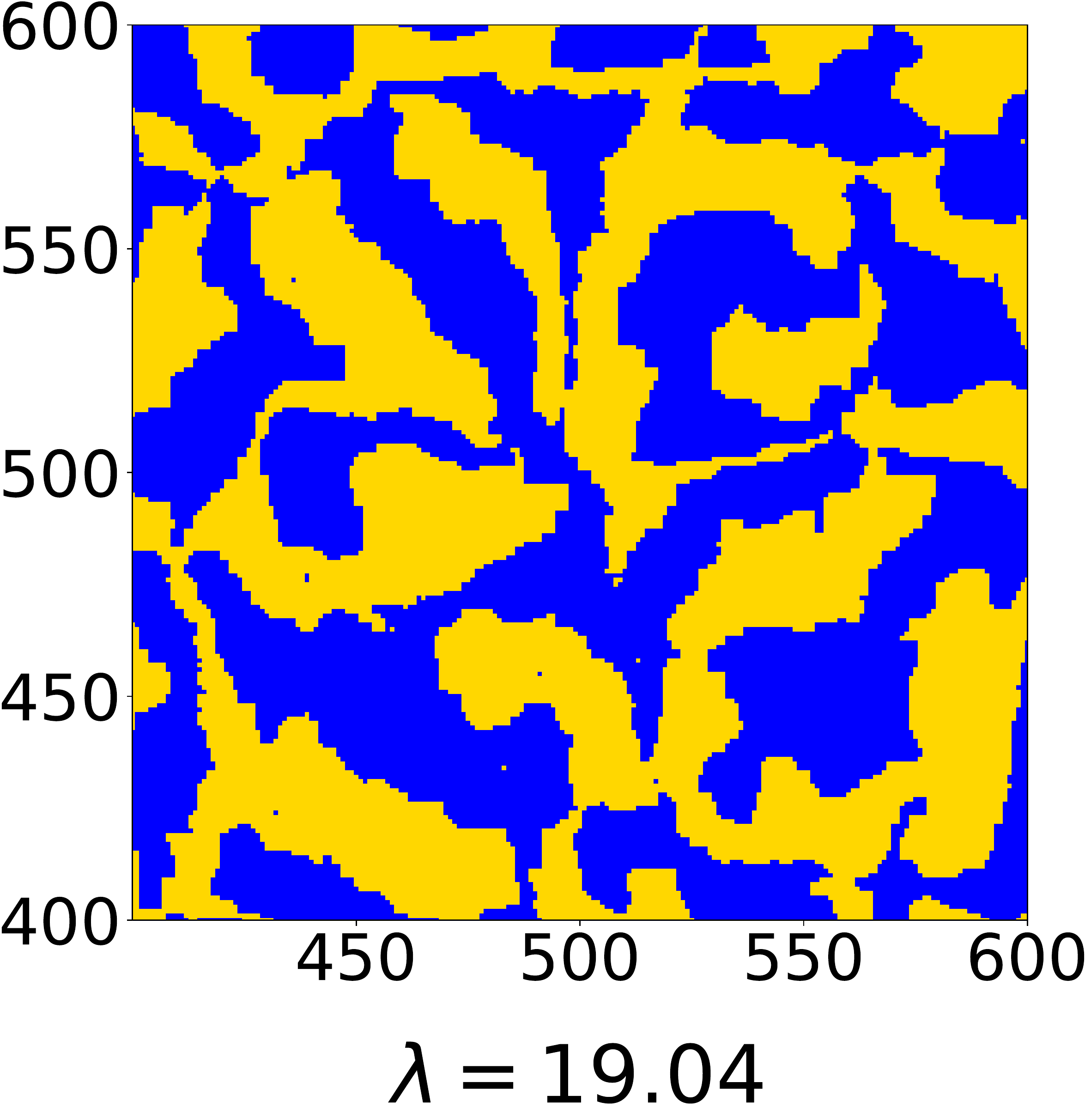}} \hspace{1.5cm}
    \subfloat{\includegraphics[width=0.35\textwidth]{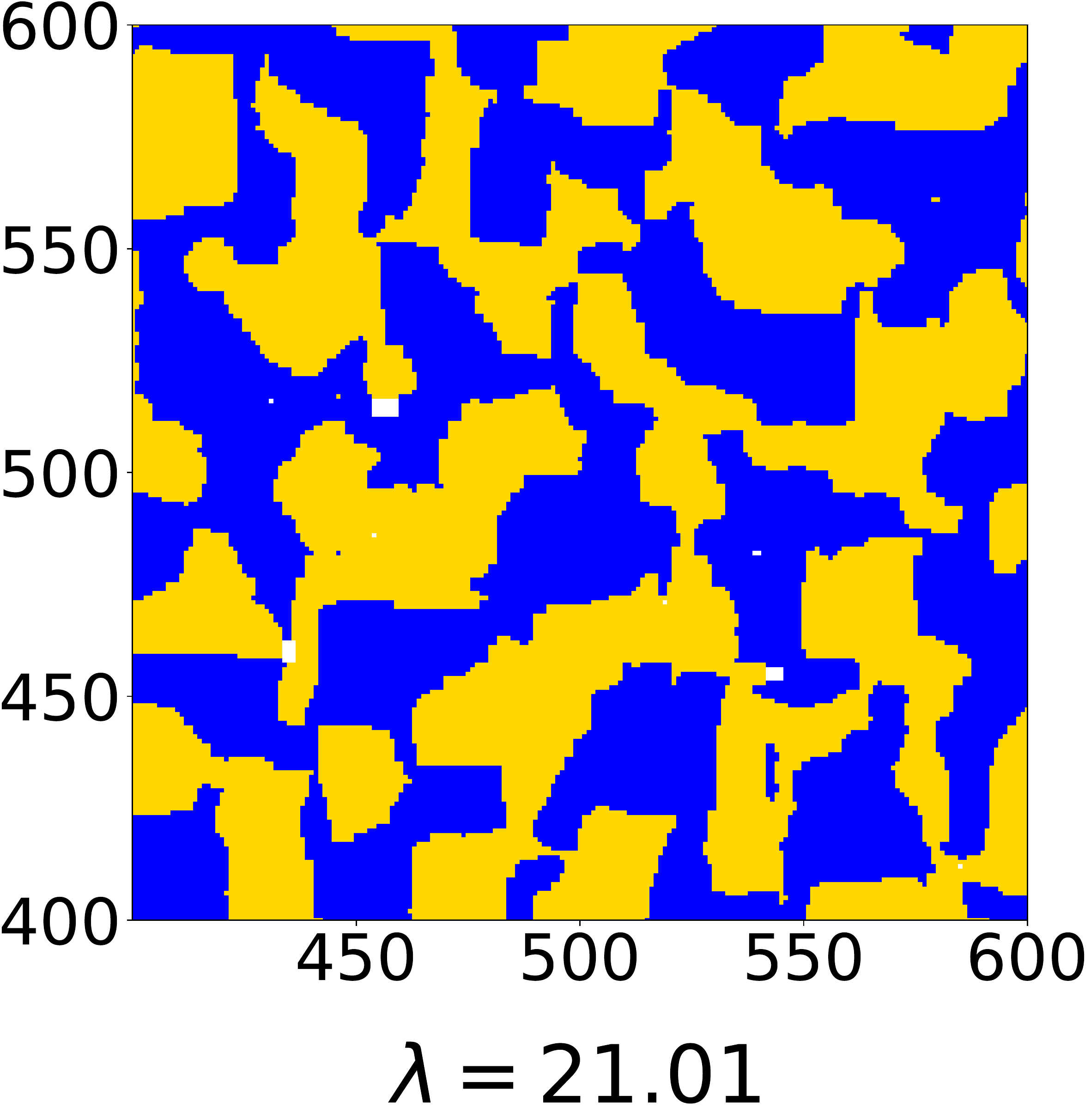}} \\
	\addtocounter{subfigure}{-4}
	\subfloat[]{\includegraphics[width=0.35\textwidth]{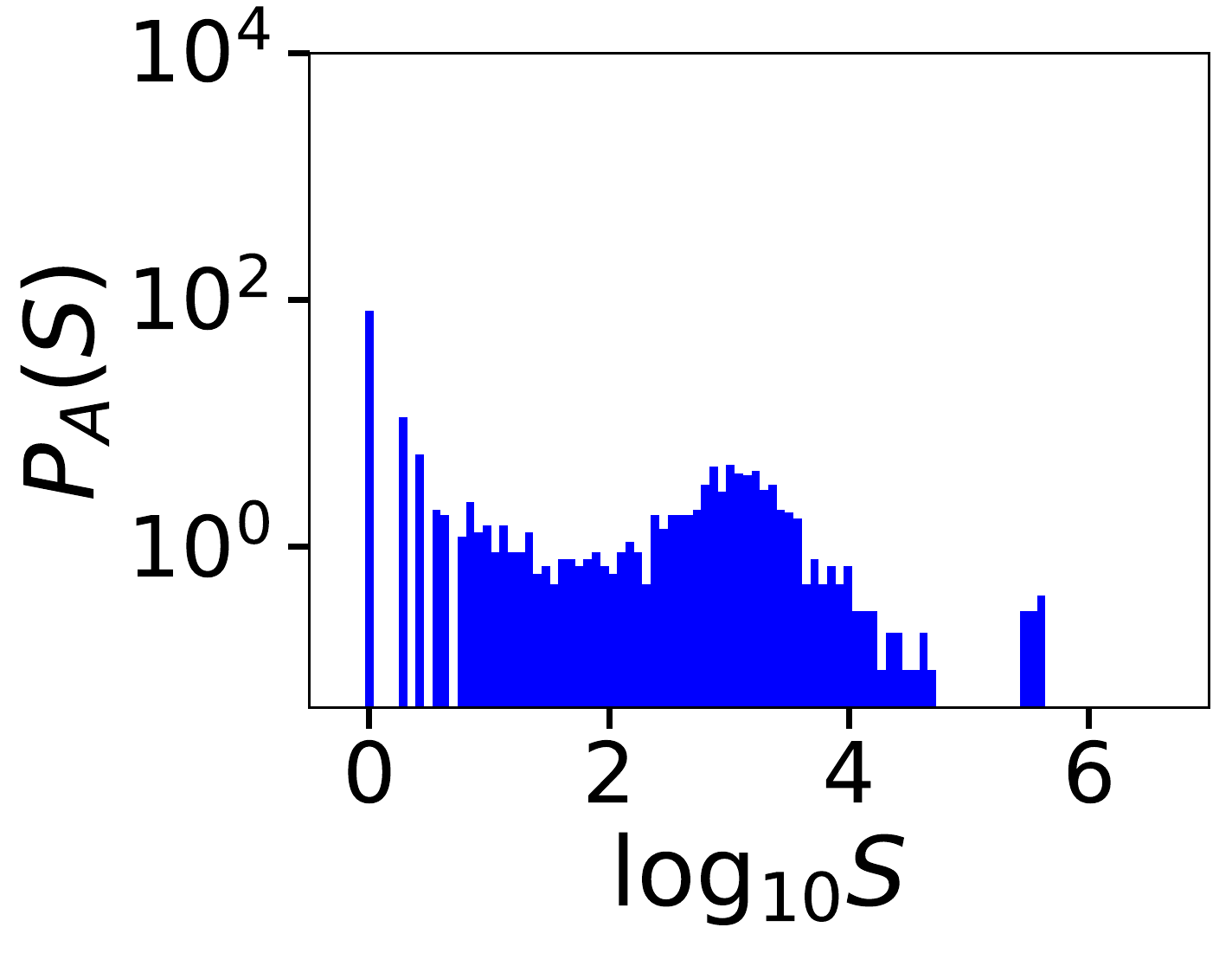}} \hspace{1.5cm}
    \subfloat[]{\includegraphics[width=0.35\textwidth]{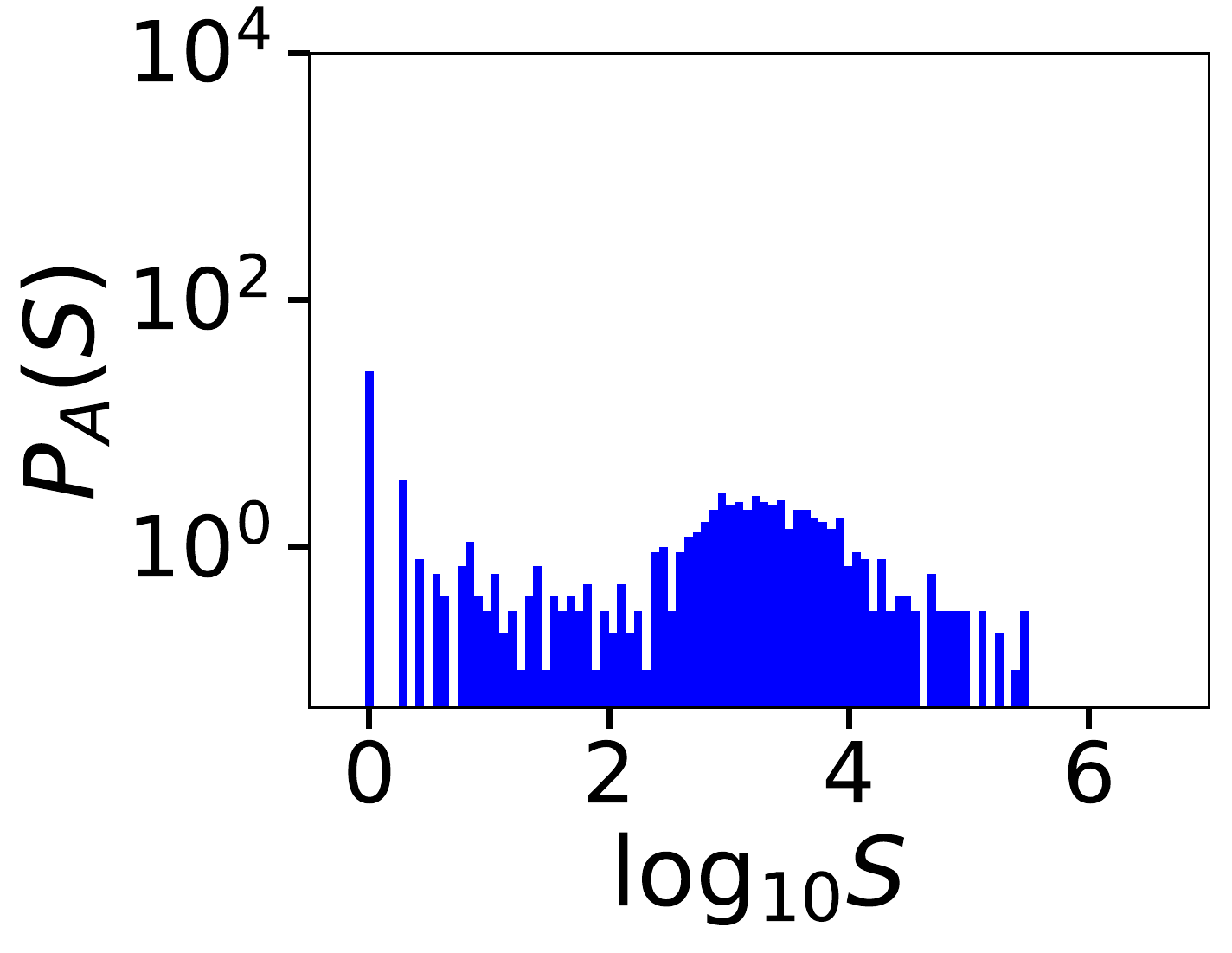}}
\caption{
Vertical cross sections, horizontal cross sections at $z=10$, and A domain size distributions
at $z=10$ in films deposited with
$R_{AA}={10}^7$, $\epsilon_{AA}={10}^{-3}$, $\epsilon_{AB}=0.1$, $P={10}^{-2}$, and:
(a) $R_{AB}={10}^8$; (b) $R_{AB}={10}^9$.
A particles are in blue, B in yellow.
The surface roughness and the average domain sizes in the cross sections at $z=10$ are indicated.
}
\label{symcompRAB1}
\end{center}
\end{figure}

Similar conclusions are obtained by comparing films whose data are shown in 
Figs. \ref{symcompRAB1}(a) ($R_{AB}={10}^8$) and \ref{symcompRAB1}(b) ($R_{AB}={10}^9$),
in which the other parameters are
$R_{AA}={10}^7$, $\epsilon_{AA}={10}^{-3}$, $P={10}^{-2}$, and $\epsilon_{AB}=0.1$.
Note that the domain widths are much larger in this case if compared to Figs. 
\ref{symcompRAB}(a),(b).

The diffusion coefficients $D_{AB}=D_{BA}=D_{AS}=D_{BS}$ also have weak effect on the domain sizes
and widths in other films in which only these parameters were changed
and in which the energies of step barriers are large (small $P$).
However, in Sec. \ref{PAAsym}, we will show that $D_{AB}$ plays a role in cases with $P=0.1$,
which is important for high temperature deposition.

We recall that the island density in submonolayer island growth with two species is predominantly
affected by the diffusion coefficients of those species on the substrate \citep{einax2007,einax2009}.
Thus, the results of this section show that the effects of the short range interactions
with the substrate propagate only to a small number of atomic layers in conditions of low
to intermediate temperatures.

\subsubsection{Effects of the diffusion coefficient on terraces of the same species}
\label{DAAsym}

Here we analyze the role of $R_{AA}=D_{AA}/F$ on the film structure.

\begin{figure}
\begin{center}
    \subfloat{\includegraphics[width=0.35\textwidth]{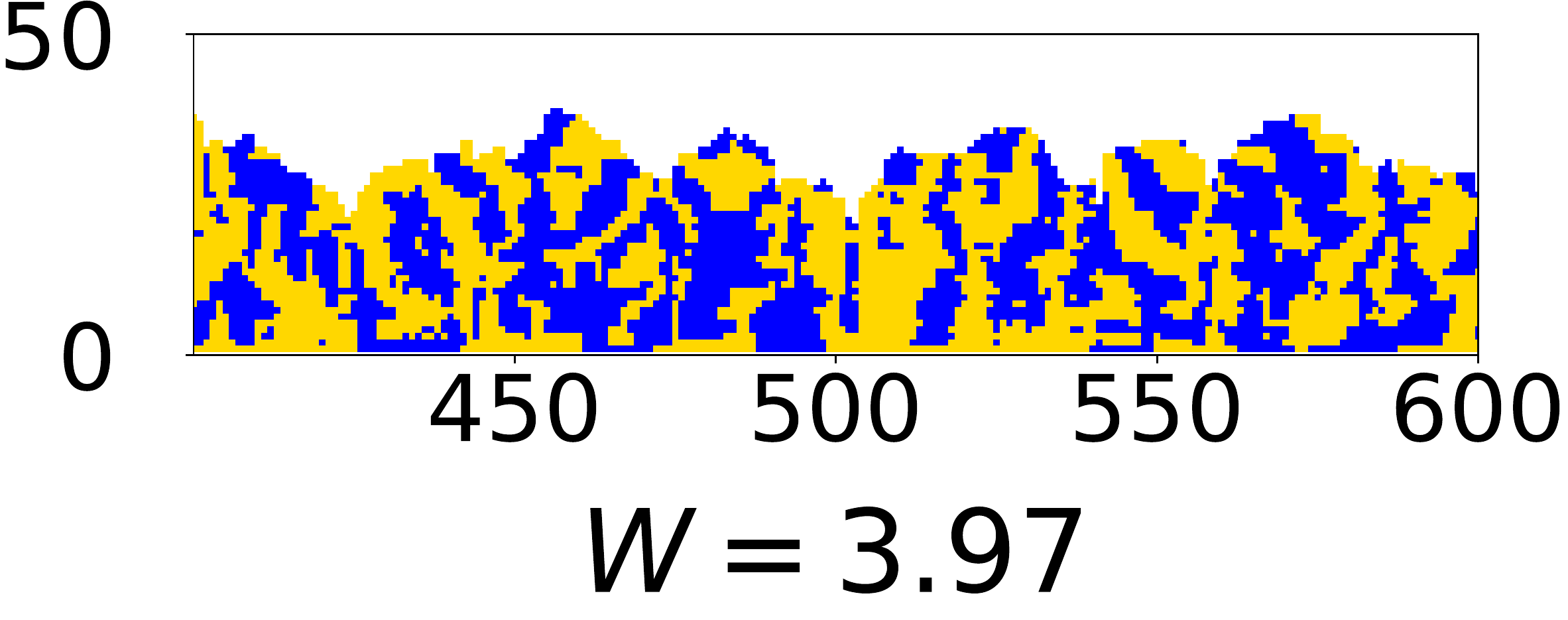}} \hspace{1cm}
    \subfloat{\includegraphics[width=0.35\textwidth]{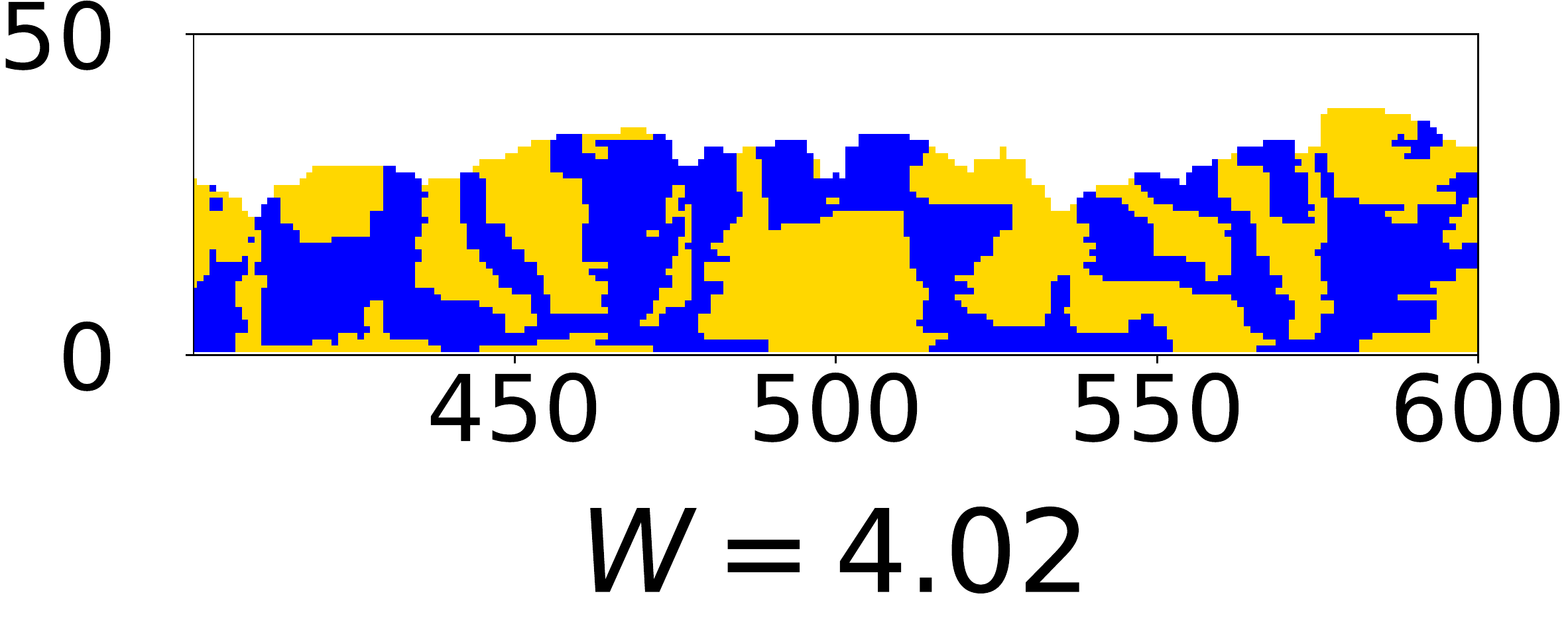}} \\
    \subfloat{\includegraphics[width=0.35\textwidth]{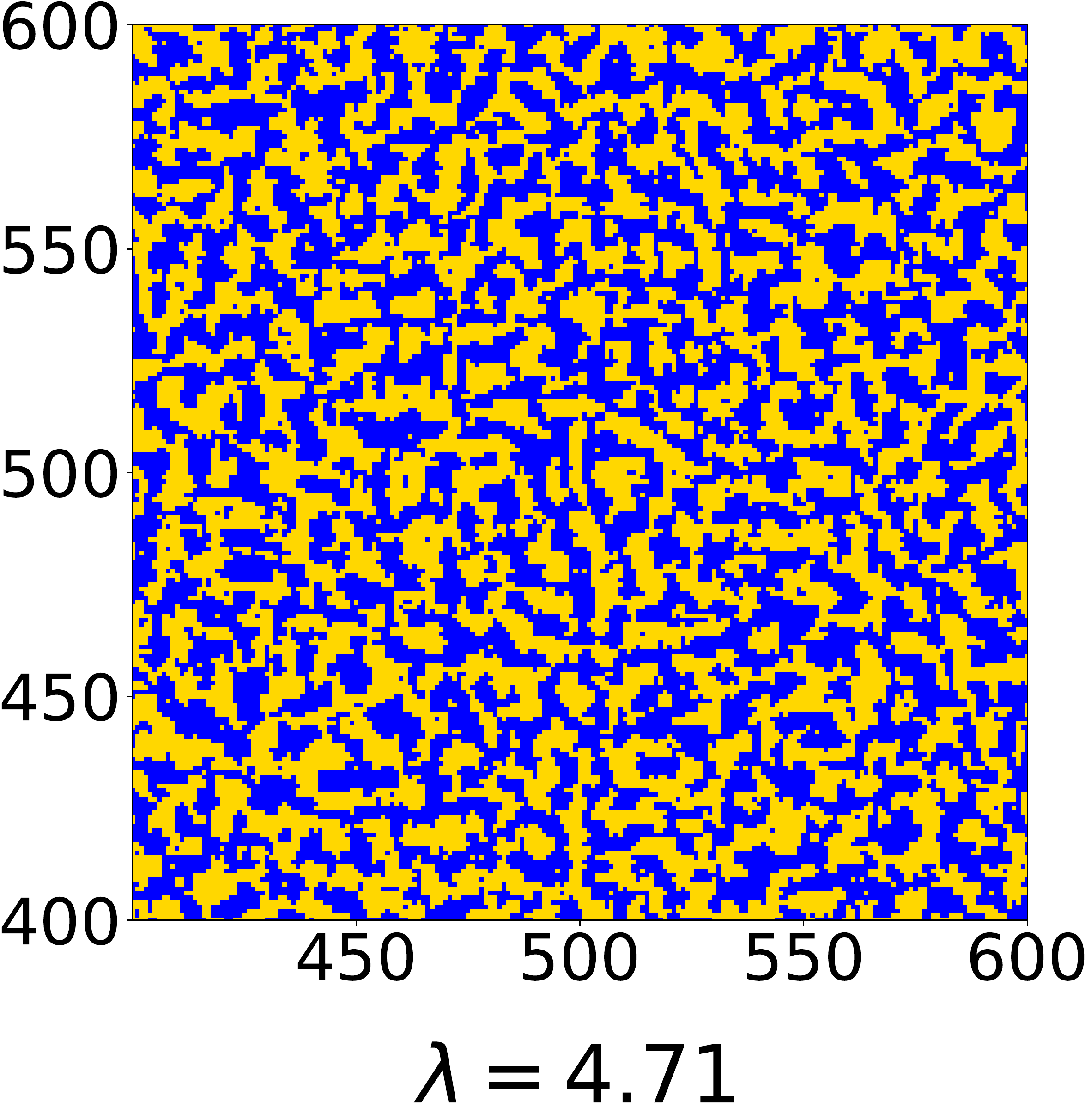}} \hspace{1cm}
    \subfloat{\includegraphics[width=0.35\textwidth]{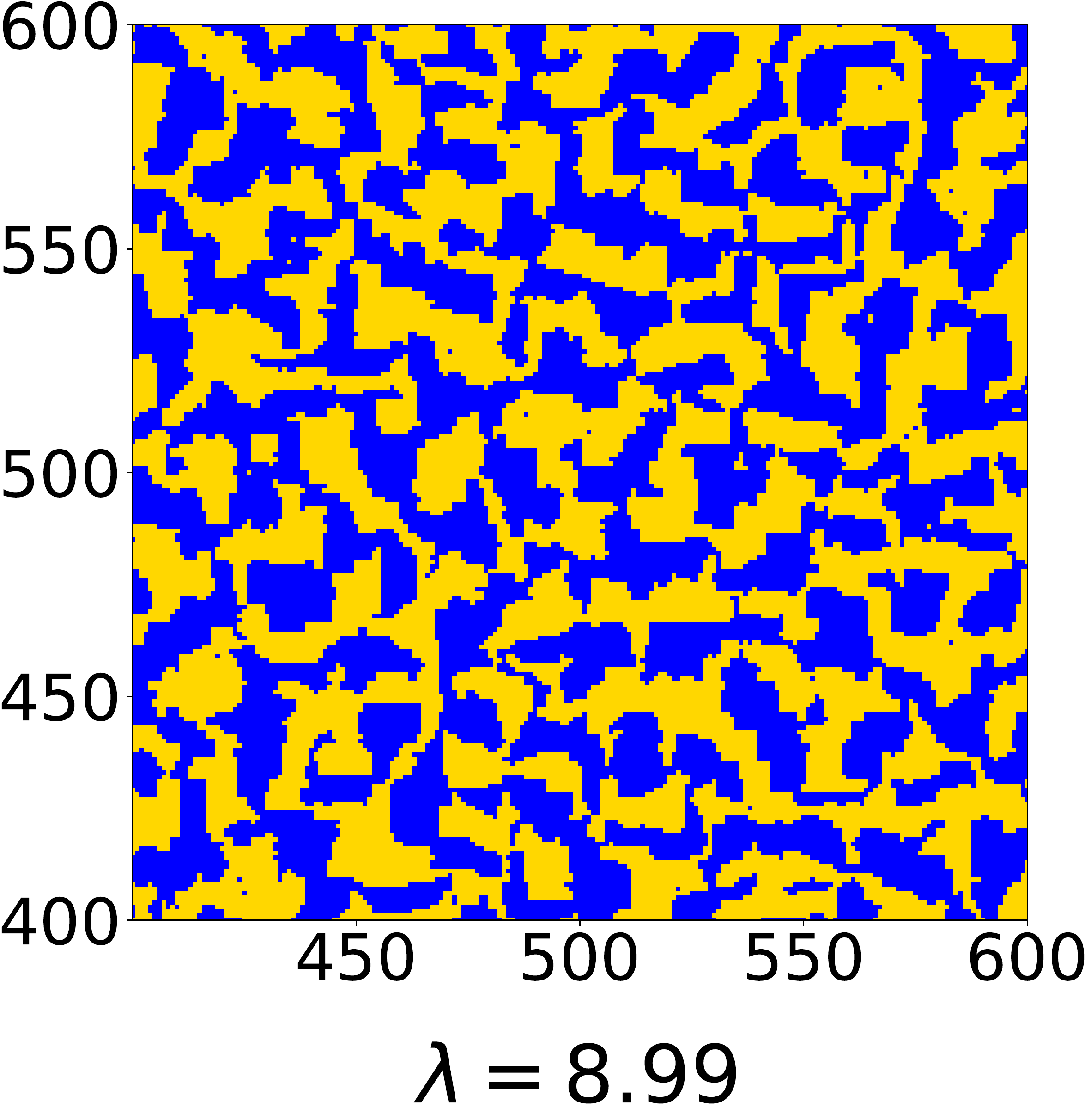}} \\
	\addtocounter{subfigure}{-4}
	\subfloat[]{\includegraphics[width=0.35\textwidth]{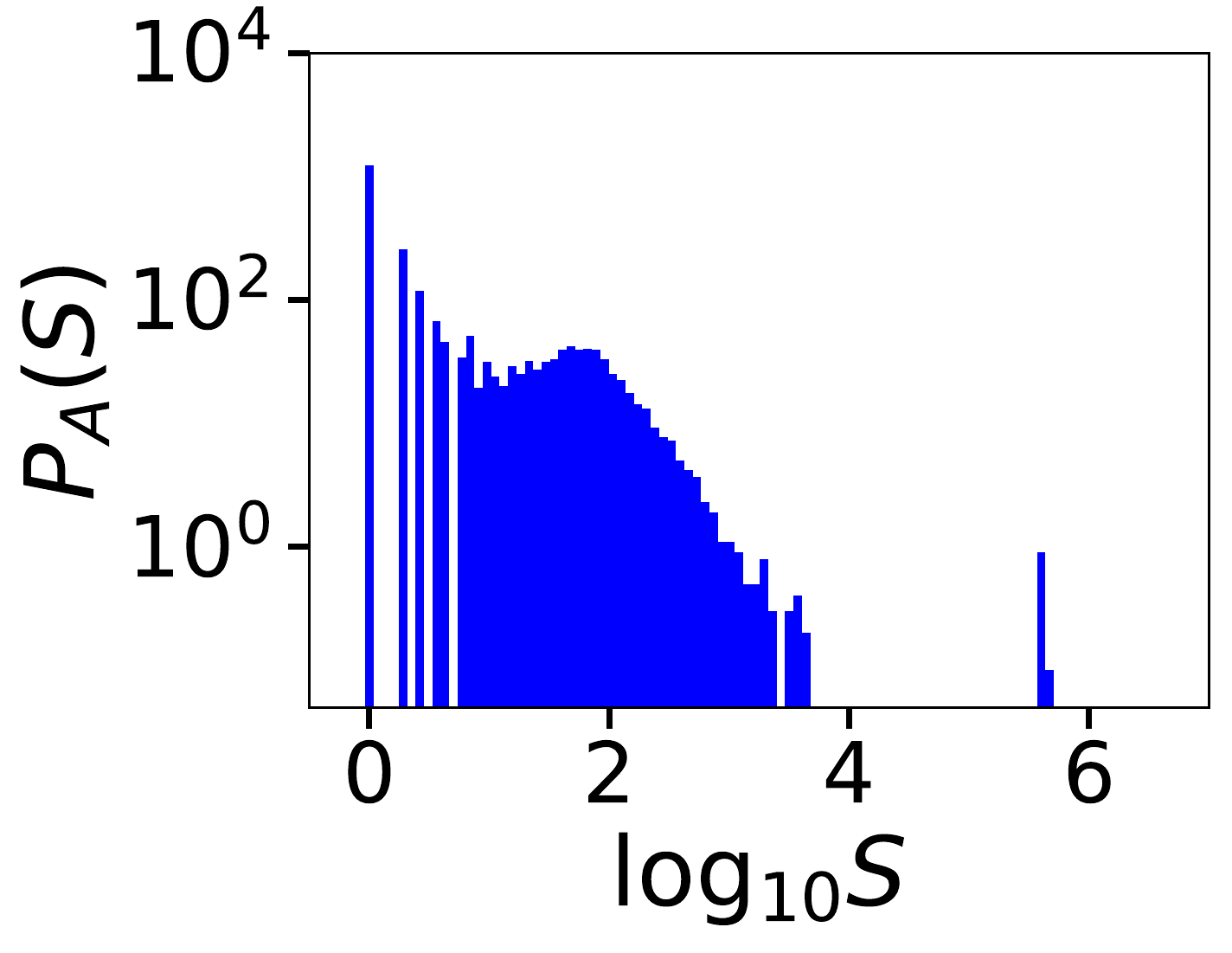}} \hspace{1cm}
    \subfloat[]{\includegraphics[width=0.35\textwidth]{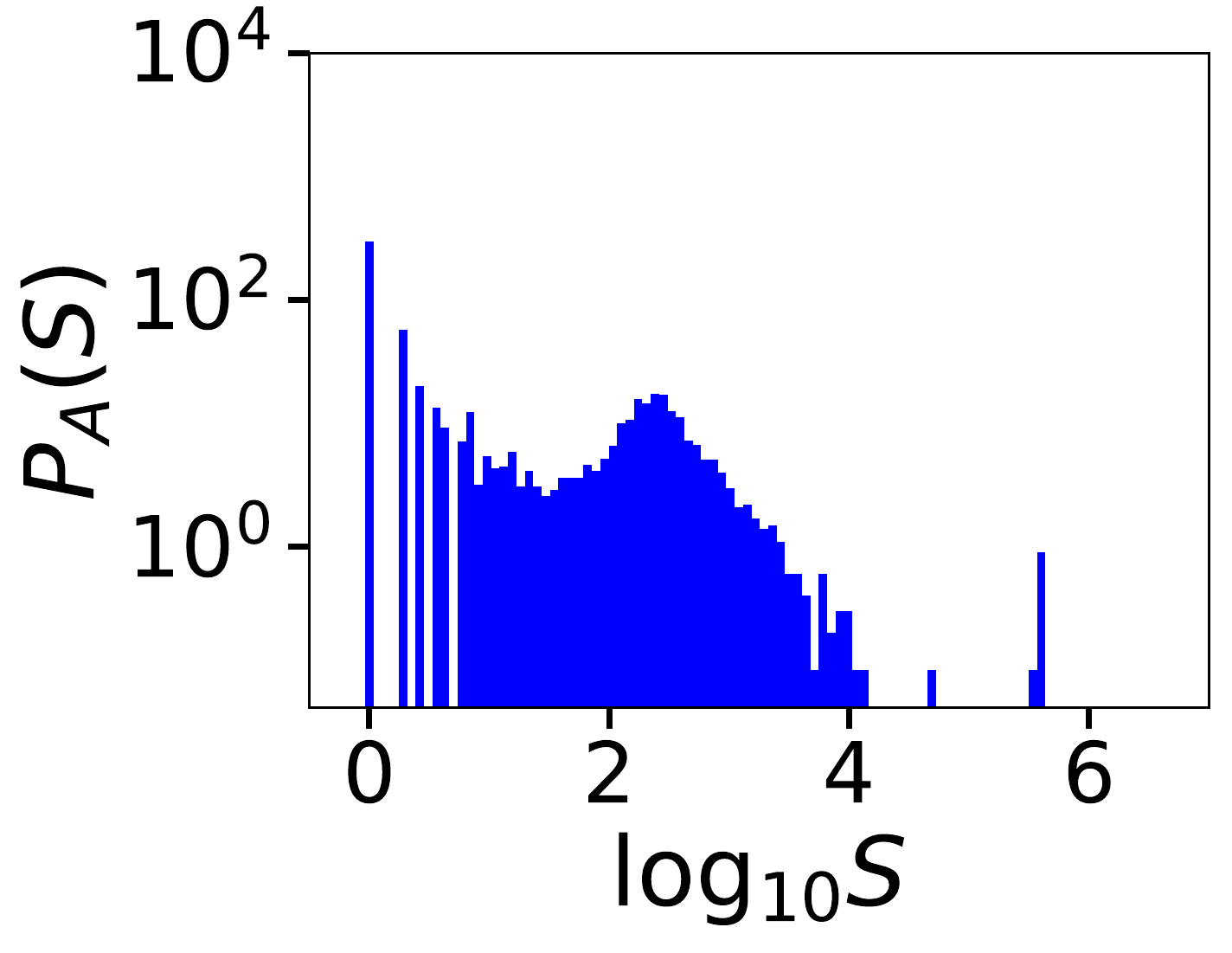}}
\caption{
Vertical cross sections, horizontal cross sections at $z=10$, and A domain size distributions
at $z=10$ in films deposited with
$\epsilon_{AA}={10}^{-3}$, $\epsilon_{AB}=0.1$, $P={10}^{-2}$, and:
(a) $R_{AA}={10}^5$, $R_{AB}={10}^7$; (b) $R_{AA}={10}^6$, $R_{AB}={10}^8$.
A particles are in blue, B in yellow.
The surface roughness and the average domain sizes in the cross sections at $z=10$ are indicated.
}
\label{symcompRAA}
\end{center}
\end{figure}

Fig. \ref{symcompRAA}(a) shows cross sections and domain size distributions
of films grown with $R_{AA}={10}^5$ and $R_{AB}={10}^7$;
Fig. \ref{symcompRAA}(b) shows cross sections and domain size distributions of films grown with
$R_{AA}={10}^6$ and $R_{AB}={10}^8$.
The remaining parameters in Figs. \ref{symcompRAB1}(a) and \ref{symcompRAB1}(b) are the same:
$\epsilon_{AA}={10}^{-3}$, $P={10}^{-2}$, and $\epsilon_{AB}=0.1$.

The change of two orders of magnitude in $R_{AA}$ (from ${10}^5$ to ${10}^7$) does not lead to
a significant change in the surface roughness $W$.
In deposition of films with a single species and no step barrier, $W<1$ at short times for
${10}^5\leq R\leq {10}^7$, i.e. the films have very smooth surfaces \citep{cv2015}.
Consequently, we actually do not expect that the relatively larger values of $W$ shown in
Figs. \ref{symcompRAB1} and \ref{symcompRAA} are related to the values of $R_{AA}$.

On the other hand, the average width of the domains increases with $R_{AA}$:
as this ratio varies from ${10}^5$
[Fig. \ref{symcompRAA}(a)] to ${10}^7$ [Fig. \ref{symcompRAB1}(a),(b)], that width increases by a
factor between $4$ and $5$.
Indeed, a larger mobility on the terrace of a domain facilitates its lateral expansion.
Figs. \ref{symcompRAA}(a),(b) shows that this large mobility works against the vertical
alignment of the domains, which is consistent with the interpretation of a facilitated lateral
propagation.

The cluster size distributions for $R_{AA}\geq {10}^5$ differ from those of low temperatures
(Sec. \ref{lowtempsym}) because they have a peak at a finite size, which moves from $S\sim {10}^2$
in $R_{AA}={10}^5$ to $S\sim {10}^3$ in $R_{AA}={10}^7$.
This means that the small domains have a characteristic size;
some of those domains can be observed in the horizontal cross sections of Figs. \ref{symcompRAB1}(a) and
\ref{symcompRAB1}(b), in which $R_{AA}$ is large.
The sizes of the largest domains decrease as $R_{AA}$ increases:
for $R_{AA}\leq {10}^6$ [Figs. \ref{symcompRAA}(a),(b)], the largest domains have $S\sim {10}^6$,
but for $R_{AA}={10}^7$ [Fig. \ref{symcompRAB1}(b)], they have 
$S\sim {10}^5$.
This indicates that the increase of the size of finite domains, which is a signature of enhanced
phase separation, is accompanied by a decrease in the long distance connectivity of the
largest domains across the film.

\subsubsection{Effects of lateral interactions of the same species}
\label{epsilonAAsym}

Now we analyze the role of the detachment probability $\epsilon_{AA}$ on the film structure.
We compare the structures of films grown with constant values $R_{AA}={10}^6$,
$P={10}^{-2}$, and $\epsilon_{AB}=0.1$.
The values of $R_{AB}$ are in the range ${10}^7$-${10}^8$, but the effect of this change
is negligible, as discussed in Sec. \ref{DABsym}.

\begin{figure}
\begin{center}
    \subfloat{\includegraphics[width=0.35\textwidth]{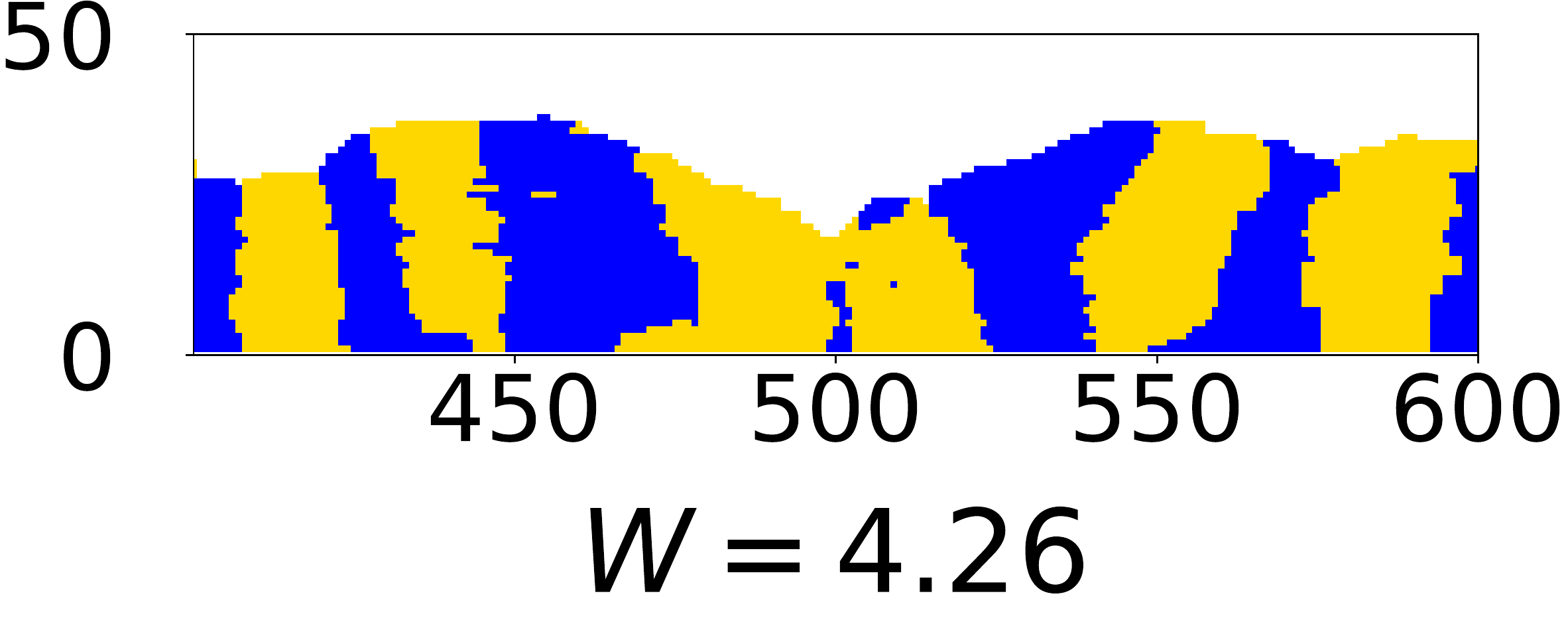}} \hspace{1cm}
    \subfloat{\includegraphics[width=0.35\textwidth]{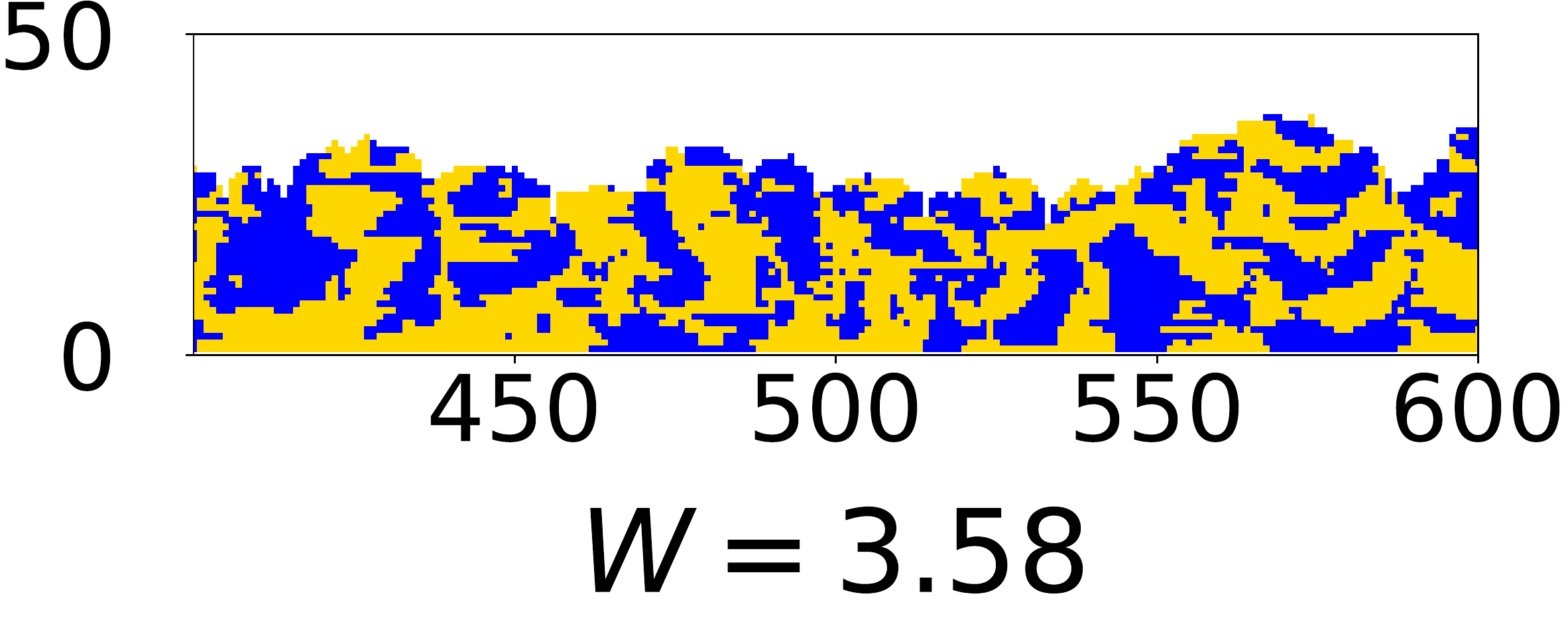}} \\
    \subfloat{\includegraphics[width=0.35\textwidth]{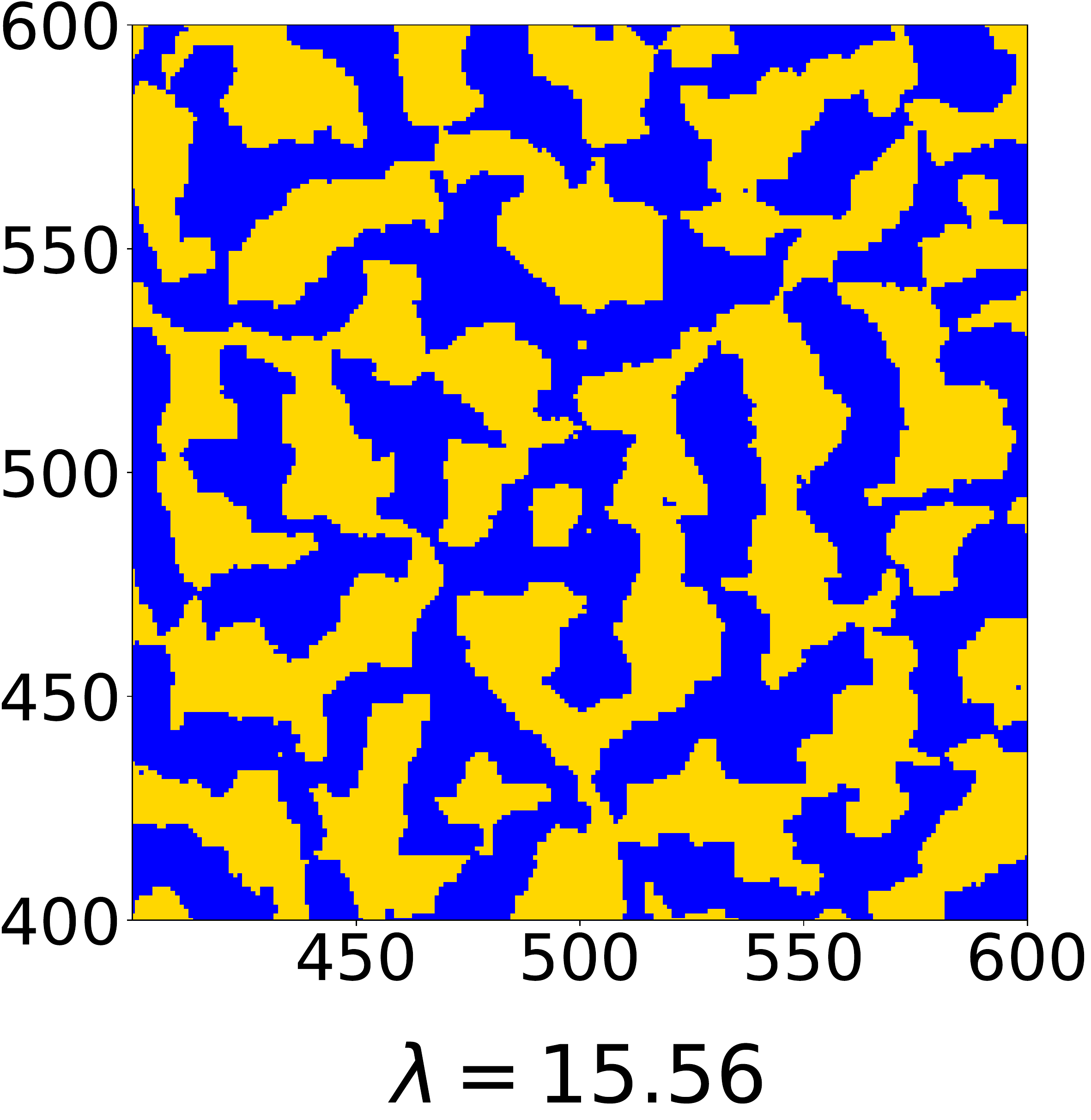}} \hspace{1cm}
    \subfloat{\includegraphics[width=0.35\textwidth]{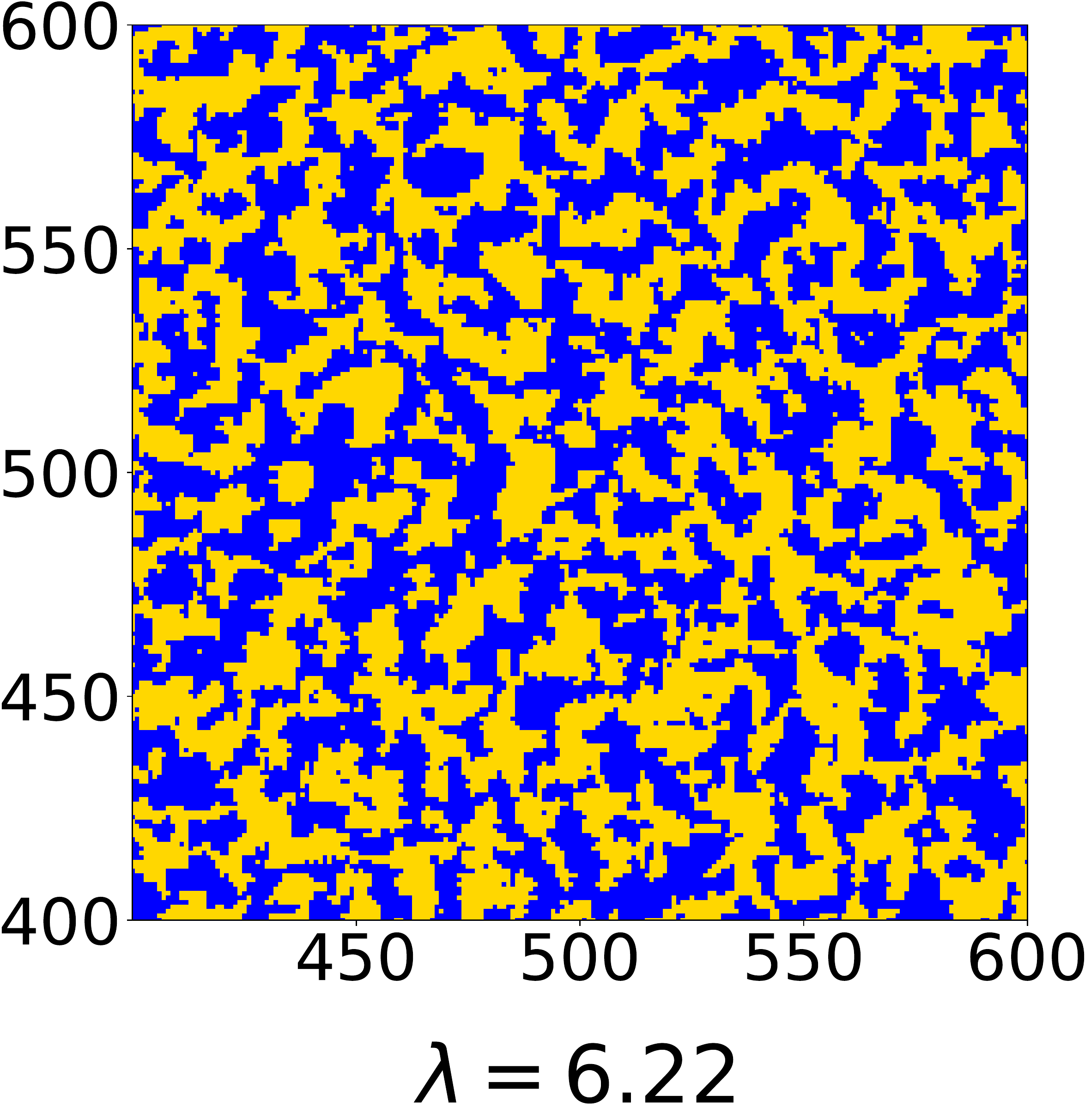}} \\
	\addtocounter{subfigure}{-4}
	\subfloat[]{\includegraphics[width=0.35\textwidth]{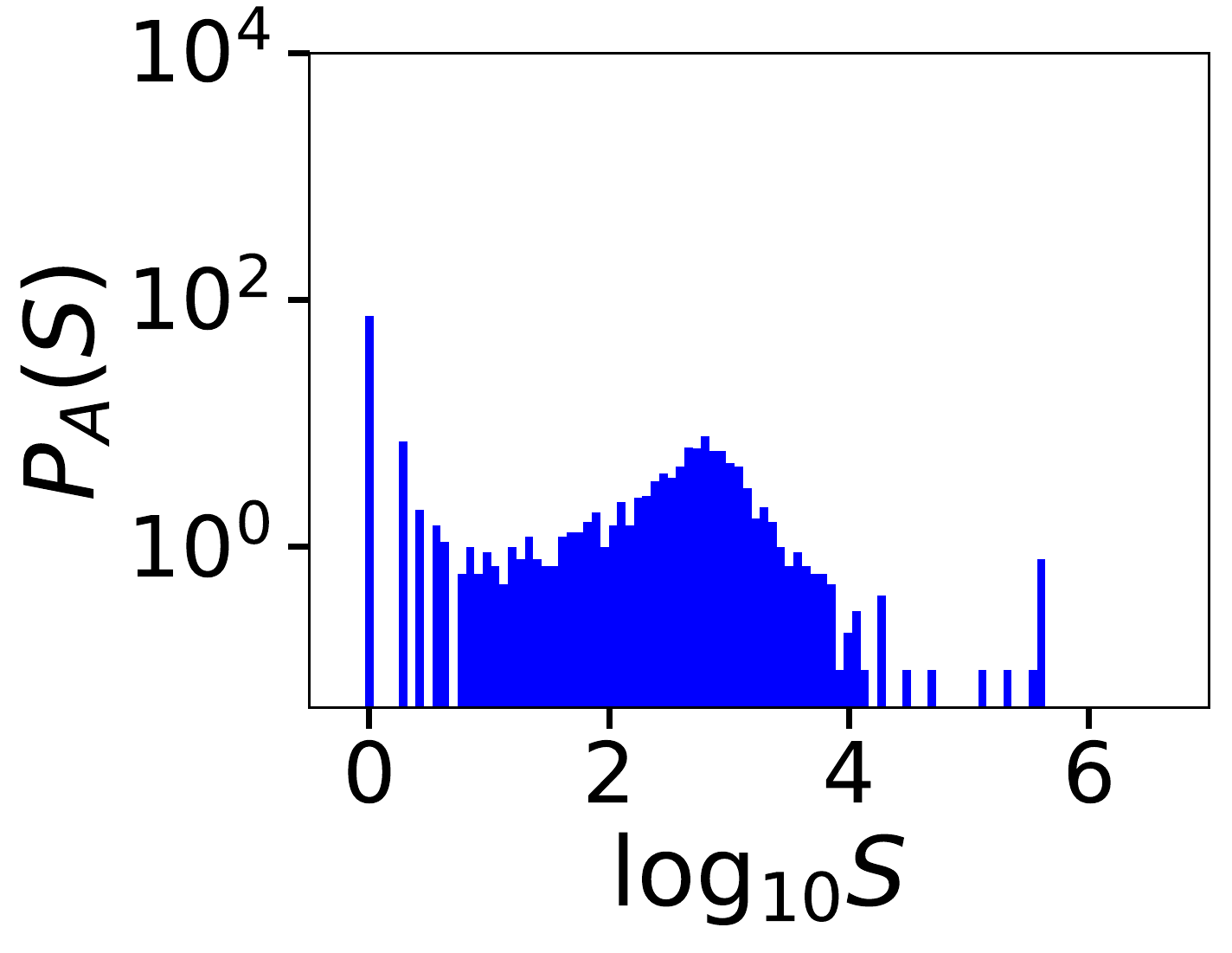}} \hspace{1cm}
    \subfloat[]{\includegraphics[width=0.35\textwidth]{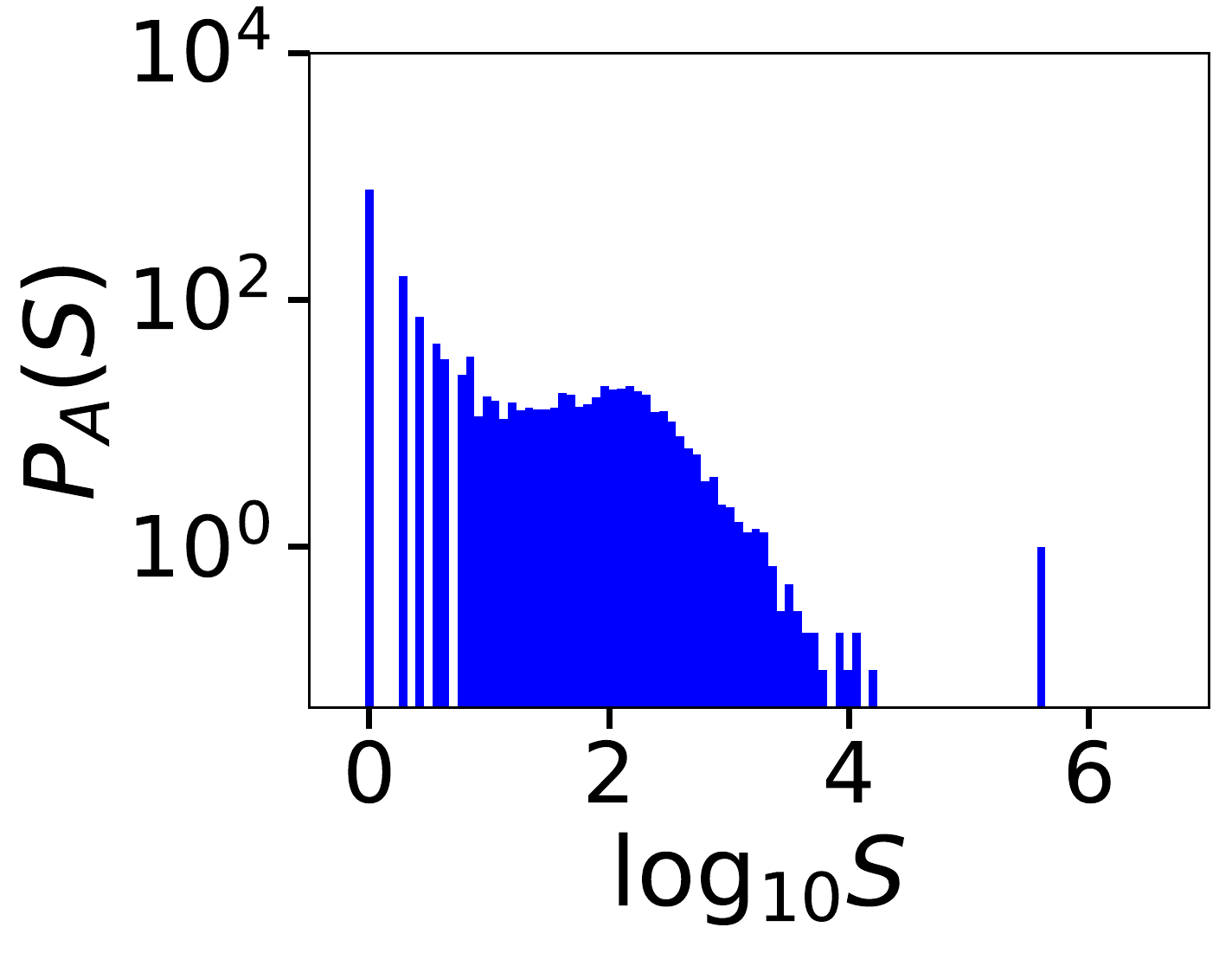}}
\caption{
Vertical cross sections, horizontal cross sections at $z=10$, and A domain size distributions
at $z=10$ in films deposited with
$R_{AA}={10}^6$, $\epsilon_{AB}=0.1$, $P={10}^{-2}$, and:
(a) $\epsilon_{AA}={10}^{-2}$, $R_{AB}={10}^7$; (b) $\epsilon_{AA}={10}^{-4}$, $R_{AB}={10}^8$.
A particles are in blue, B in yellow.
The surface roughness and the average domain sizes in the cross sections at $z=10$ are indicated.
}
\label{symcompeAA}
\end{center}
\end{figure}

Figs. \ref{symcompeAA}(a) and \ref{symcompeAA}(b) show cross sections and domain size distributions
of films grown with $\epsilon_{AA}={10}^{-2}$ ($R_{AB}={10}^7$) and 
$\epsilon_{AA}={10}^{-4}$ ($R_{AB}={10}^8$).
For comparison, consider the data shown in Fig. \ref{symcompRAA}(b) for an intermediate 
probability $\epsilon_{AA}={10}^{-3}$ ($R_{AB}={10}^8$).

The effect of $\epsilon_{AA}$ on the surface roughness is small; this is also observed
in films with a single species \citep{cv2015}.
However, the vertical and horizontal cross sections show that the domain width significantly
increases as $\epsilon_{AA}$ increases: when this parameter varies from ${10}^{-4}$ to ${10}^{-2}$,
that width changes by a factor near $2.5$.
The increase of $\epsilon_{AA}$ also leads to an increase in the average size of the finite domains,
which are of order $S\sim {10}^2$ for $\epsilon_{AA}={10}^{-4}$ and 
$S\sim {10}^3$ for $\epsilon_{AA}={10}^{-2}$.

The vertical cross sections also show that the increase of $\epsilon_{AA}$ improves the
vertical domain orientation.
This is also observed with other values of $R_{AA}$, from ${10}^5$ to
${10}^7$, and with other values of $P$, from ${10}^{-1}$ to ${10}^{-3}$.

Work on deposition of films with a single species showed that the increase of the detachment probability
from lateral neighbors ($\epsilon$) leads to the formation of more compact islands and
smoother terrace borders \citep{tosousareis2018}.
Conversely, decreasing $\epsilon$ leads to more branched islands and terraces because the branches
typically have atoms with low coordination.
With two species, the branching mechanism facilitates the lateral propagation of a domain over
a domain of the other species; consequently, decreasing $\epsilon_{AA}$ contributes to destroy
the vertical orientation, as observed in the simulations.

\subsubsection{Effects of step energy barriers}
\label{PAAsym}

The increase of $P$ means that it becomes easier for a particle to jump between different heights,
upward or downward.
For a given step energy barrier, $P$ increases with the temperature.

\begin{figure}
\begin{center}
    \subfloat{\includegraphics[width=0.35\textwidth]{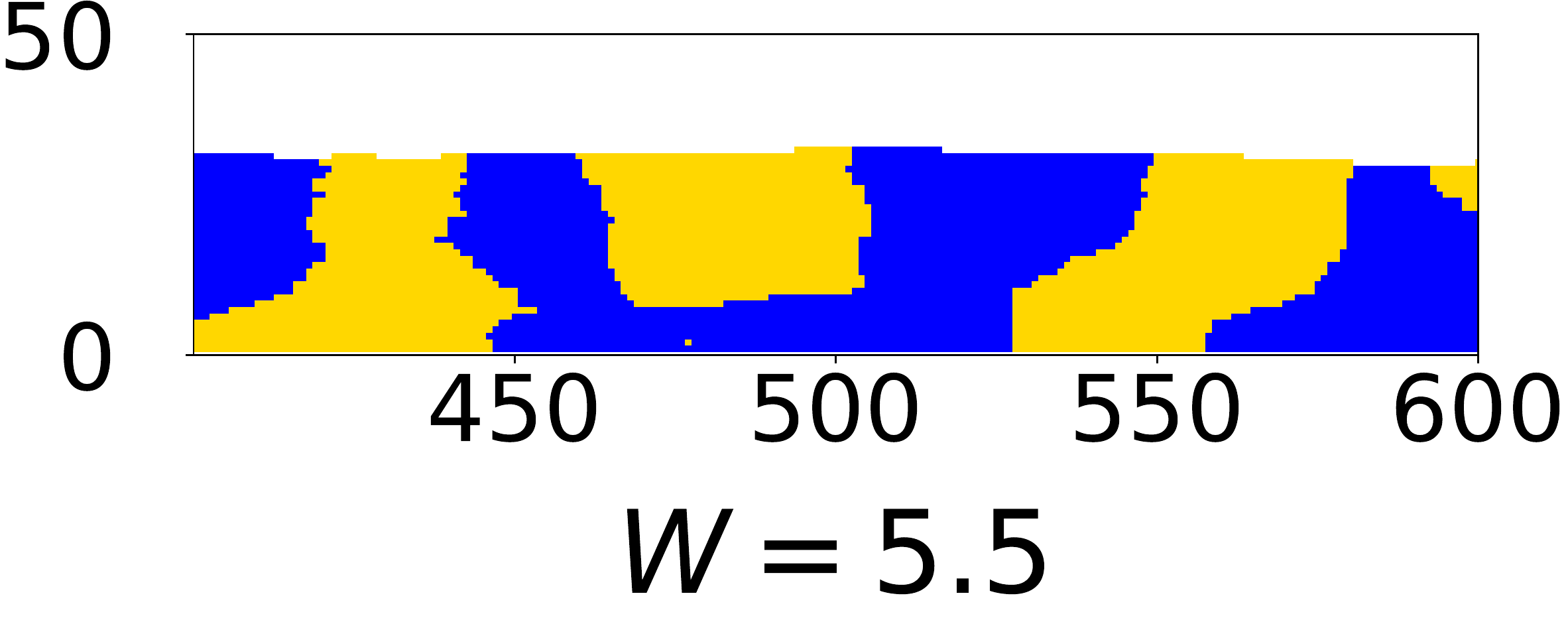}} \hspace{1cm}
    \subfloat{\includegraphics[width=0.35\textwidth]{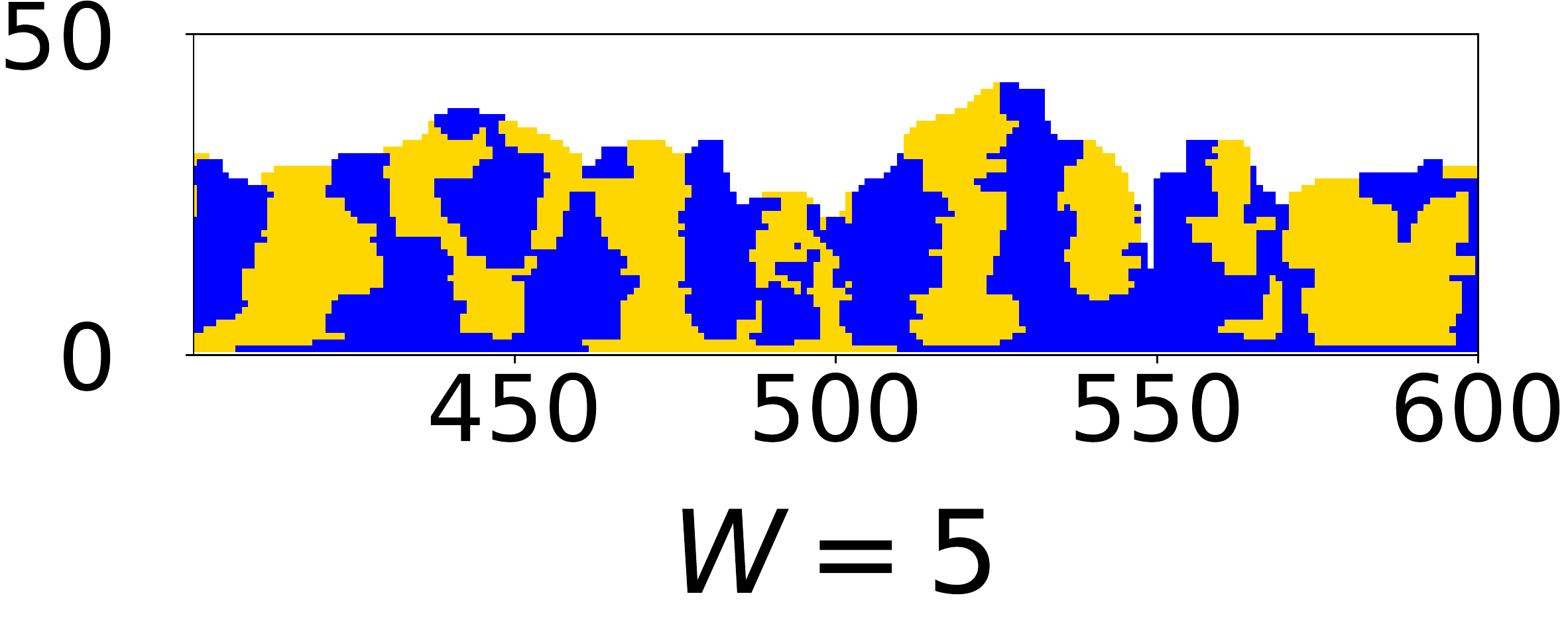}} \\
    \subfloat{\includegraphics[width=0.35\textwidth]{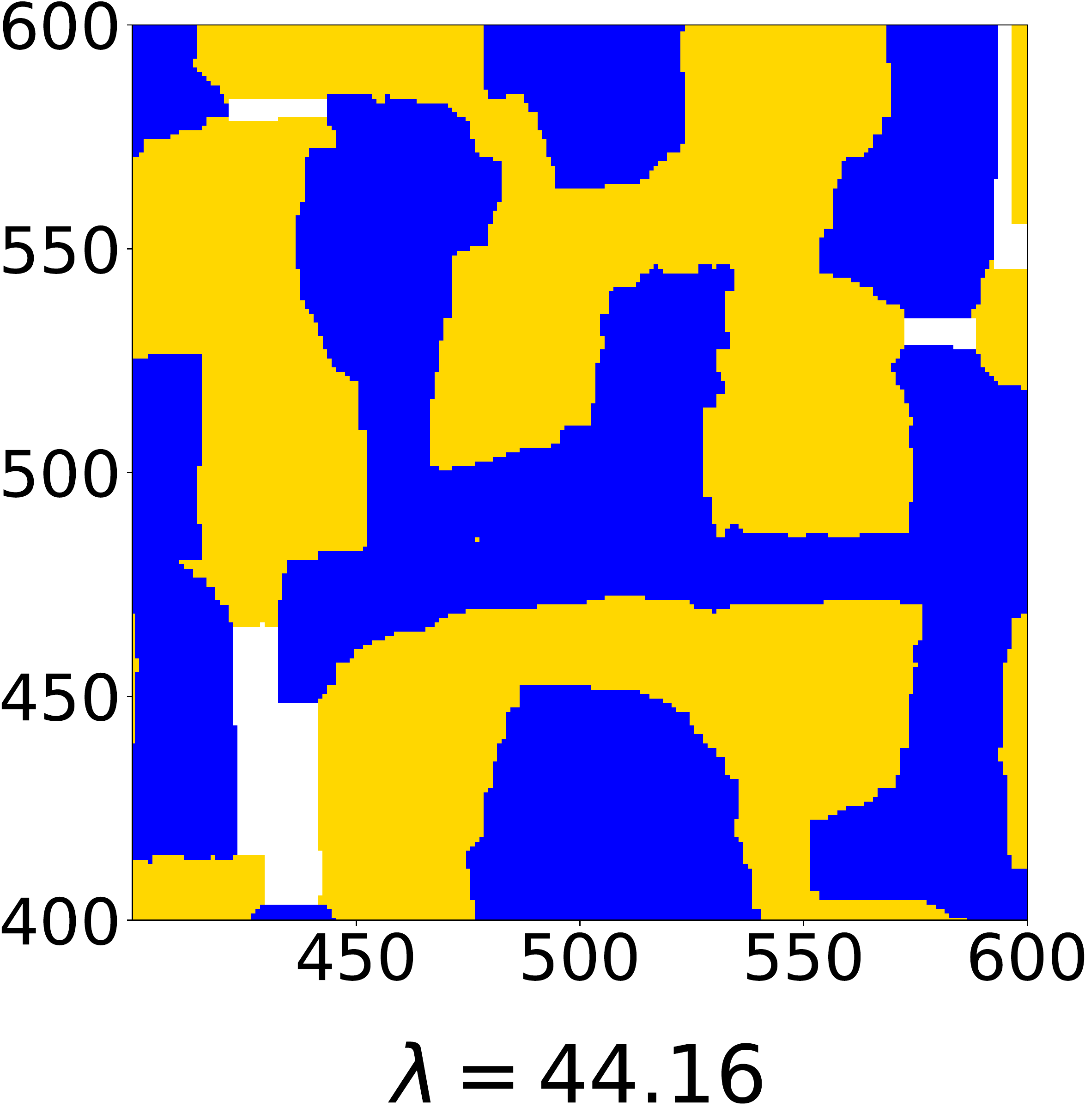}} \hspace{1cm}
    \subfloat{\includegraphics[width=0.35\textwidth]{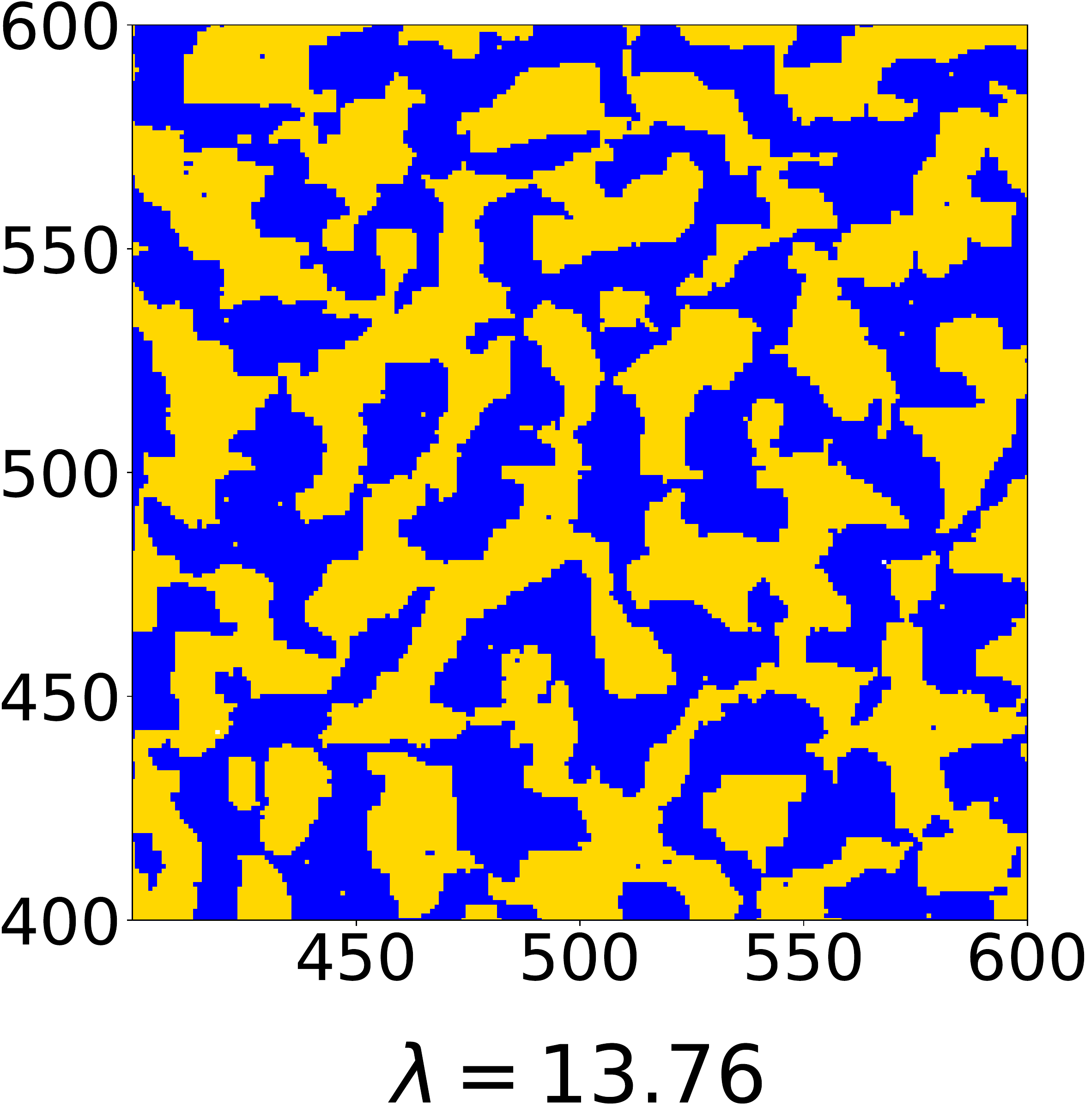}} \\
	\addtocounter{subfigure}{-4}
	\subfloat[]{\includegraphics[width=0.35\textwidth]{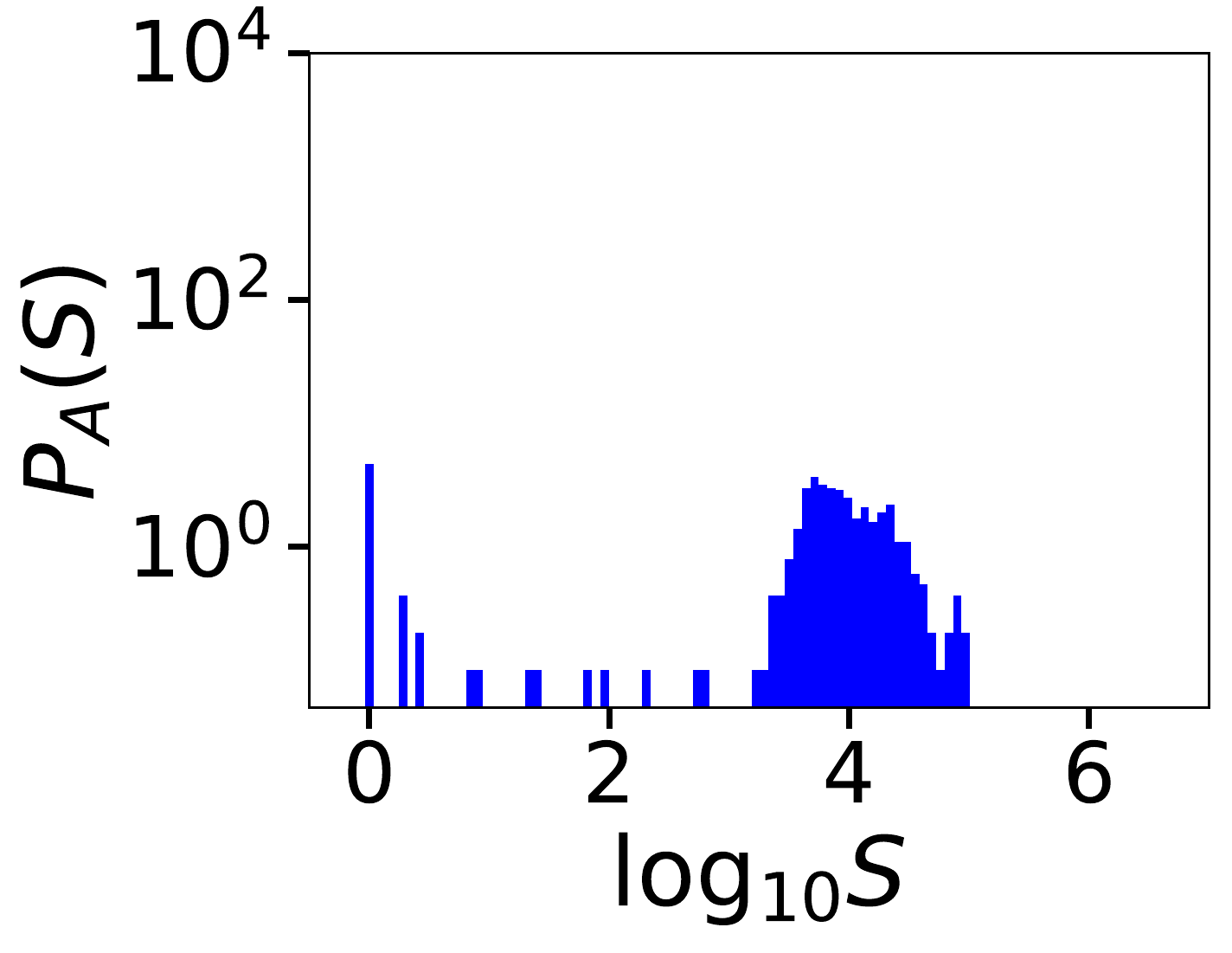}} \hspace{1cm}
    \subfloat[]{\includegraphics[width=0.35\textwidth]{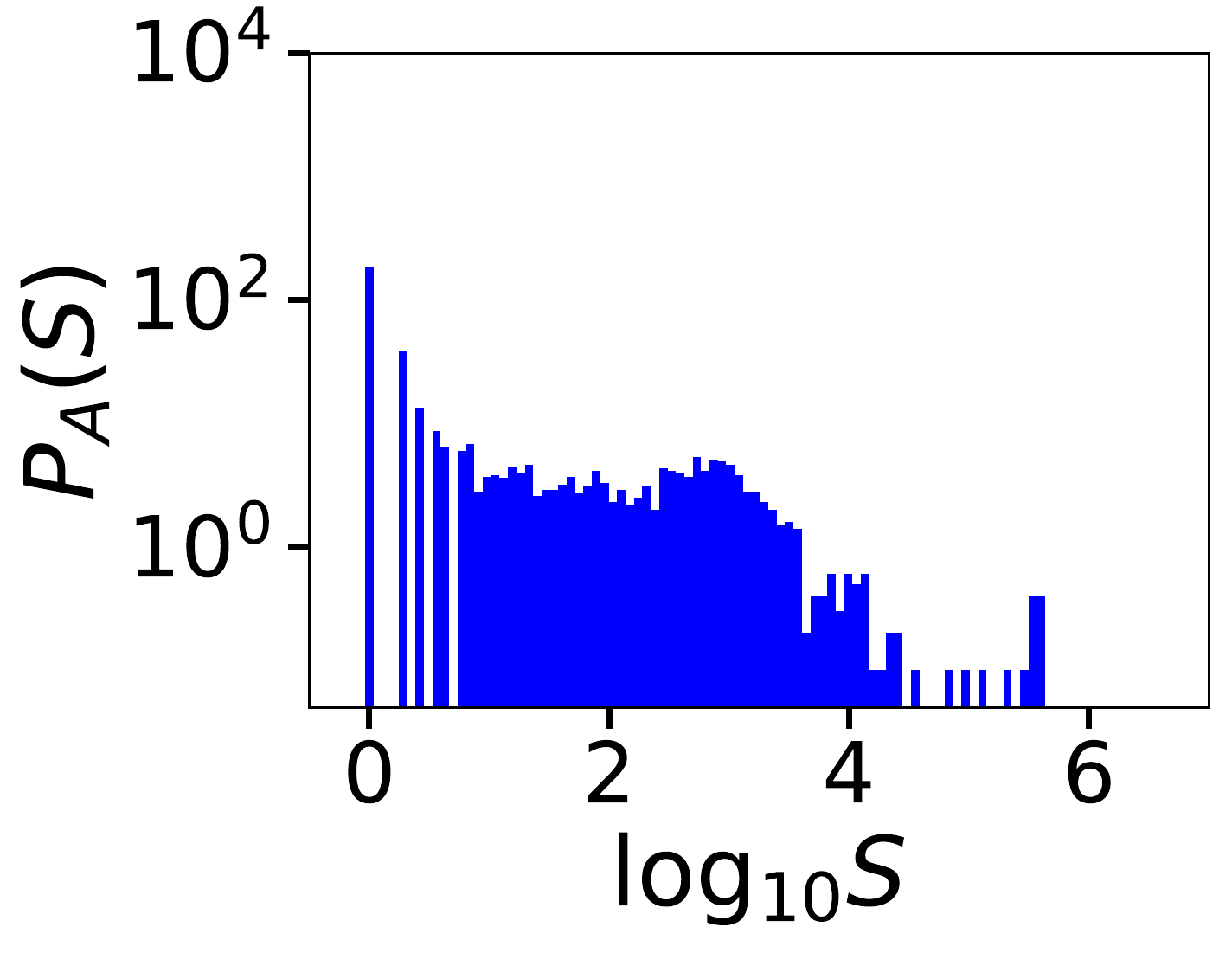}}
\caption{
Vertical cross sections, horizontal cross sections at $z=10$, and A domain size distributions
at $z=10$ in films deposited with
$R_{AA}={10}^{7}$, $R_{AB}={10}^9$, $\epsilon_{AA}={10}^{-3}$, $\epsilon_{AB}={10}^{-1}$, and:
(a) $P={10}^{-1}$; (b) $P={10}^{-3}$.
A particles are in blue, B in yellow.
The surface roughness and the average domain sizes in the cross sections at $z=10$ are indicated.
}
\label{symcompP}
\end{center}
\end{figure}

Figs. \ref{symcompP}(a) and \ref{symcompP}(b) show cross sections and domain size distributions
of films grown with $P={10}^{-1}$ and $P={10}^{-3}$, respectively, in both cases with
$R_{AA}={10}^{7}$, $R_{AB}={10}^9$, $\epsilon_{AB}={10}^{-1}$, and $\epsilon_{AA}={10}^{-3}$.
They can be compared with the data in Fig. \ref{symcompRAB1}(b), in which $P={10}^{-2}$ and
the other parameters have the same values as here.

The vertical cross sections show that the film surface is locally smoother as $P$ increases.
However, the large values of the roughness $W$ in Figs. \ref{symcompP}(a),(b)
do not seem to be consistent with this result.
The reason is that $W$ is affected by the large height differences between some domains when $P={10}^{-1}$;
this leads to the formation of the gaps observed in the horizontal cross sections of
Fig. \ref{symcompP}(a).
Increasing $P$ also leads to the increase of the width of the domains and to the enhancement of the vertical
orientation, which are effects similar to those of the detachment probability $\epsilon_{AA}$.

The distributions of domain sizes show that, as $P$ increases, the peak corresponding to finite
domains shifts from $S\sim {10}^{3}$ ($P={10}^{-3}$) to $S\sim {10}^{4}$ ($P={10}^{-1}$).
The increase in the size of finite domains is accompanied by a decrease in the size of
the largest domain.
Careful inspection of Fig. \ref{symcompP}(a) ($P={10}^{-1}$) shows that A and B domains do not connect
the opposite boundaries of the image; at some points, they are separated by gaps because the
layer $z=10$ is not completely filled.
This indicates the loss of long range connectivity of this largest cluster in conditions
of high temperatures or in cases where the energy barriers at edges are very small (large $P$).

\subsubsection{High temperature deposition}
\label{hightempsym}

The separate analysis of the effects of each model parameter in the previous sections
can be used to provide a consistent picture of the structure of the binary films
deposited at high temperatures.
This regime corresponds to large values of the diffusion coefficients, detachment probabilities,
and step jump probability.
The results in Fig. \ref{symcompP}(a) are helpful at this point.
The general trend is that the film surfaces become locally smoother and that the domains have larger
widths.
However, the size of the largest clusters decreases and the long range connectivity may be lost.

The diffusion coefficients on terraces of different species are also large at high temperatures.
This contributes to the formation of large islands on the substrate, which subsequently leads to
the formation of large compact domains with gaps between them.

Our simulations for $P={10}^{-1}$ also show that the average domain width has a non-negligible
increase as $R_{AB}$ increases.
This contrasts with the results of Sec. \ref{DABsym} for small $P$.

We conclude that the formation of long connected domains with 
non-negligible width is possible in a narrow range of parameters.
In the present model, this range is represented by most of our results with
$R_{AA}={10}^{5}$-${10}^{6}$ (except those with very small $P$ or $\epsilon_{AA}$),
and the results with $R_{AA}={10}^{7}$, $P\leq {10}^{-2}$, and $\epsilon_{AA}\leq {10}^{-2}$.
Thus, the appropriate tuning of temperature and flux is necessary for production of films with
long enough domains that cross it in the horizontal directions with a nanometer size width.

\subsubsection{Scaling of the average domain width}
\label{scalingsym}

Considering the observed effects of model parameters on the average domain width, we searched
for a quantitative relation between them.
A simple approximate relation is proposed with the form
\begin{equation}
\lambda \approx A {\left[ R_{AA} {\epsilon_{AA}}^x P^y \right]}^\gamma ,
\label{scalinglambda}
\end{equation}
in which the prefactor $A$ and the exponents $x$, $y$, and $\gamma$ are to be determined.
Here we will show data obtained in $z=10$, but the values obtained in $z=15$ are similar.

\begin{figure}[!h]
\begin{center}
    \subfloat[]{\includegraphics[width=0.45\textwidth]{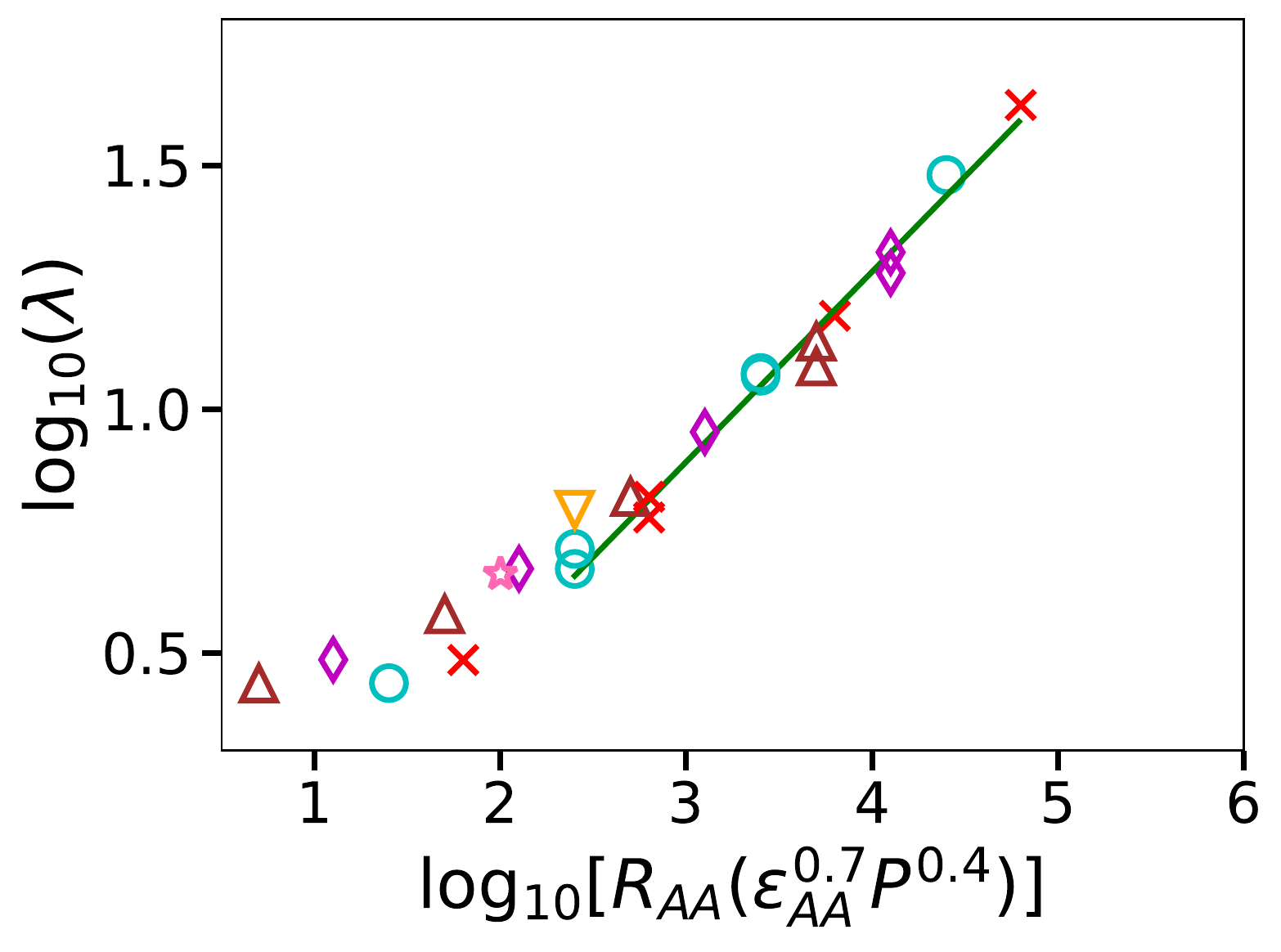}}
    \subfloat[]{\includegraphics[width=0.45\textwidth]{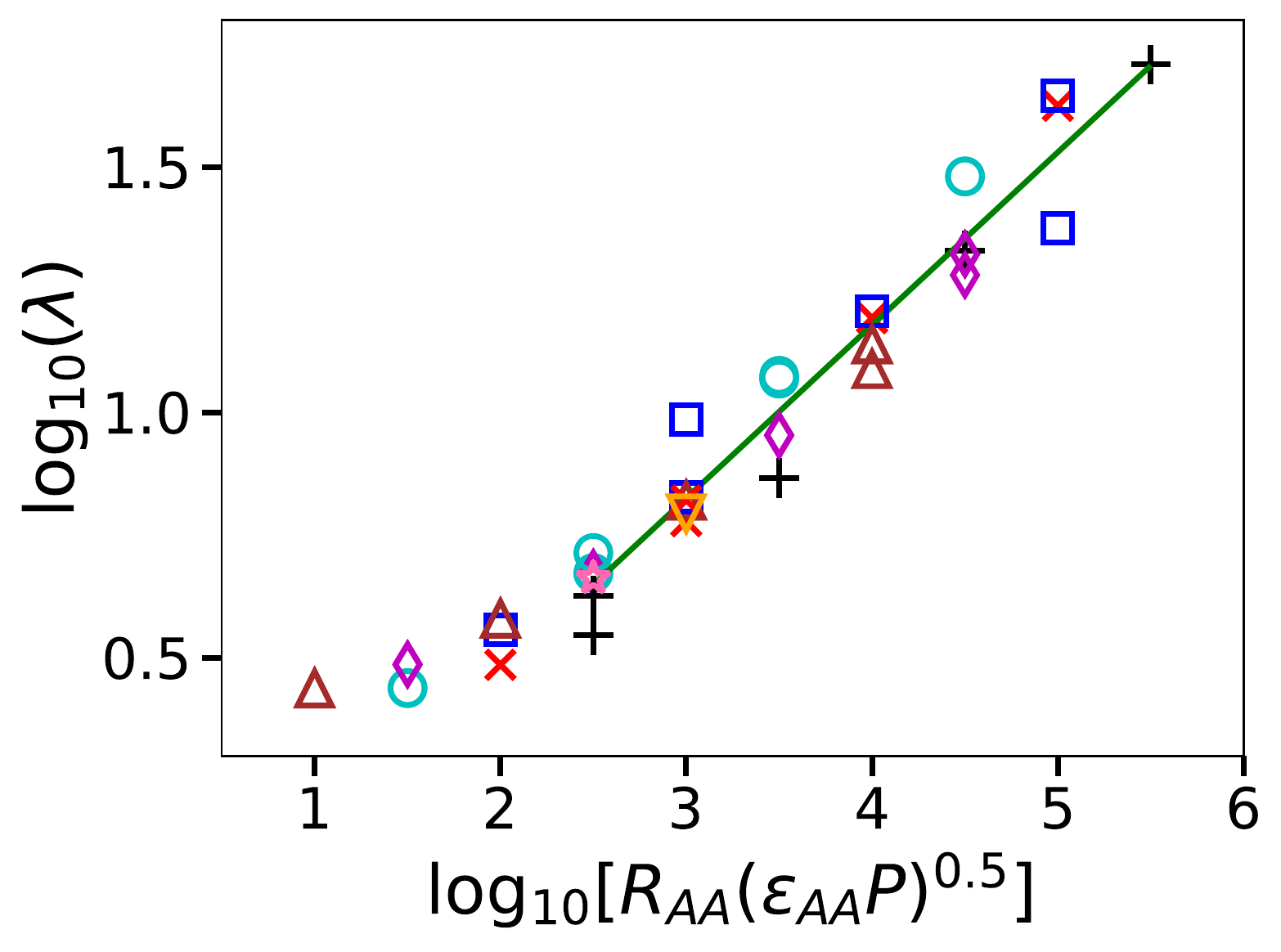}}
\caption{
Scaling plots of the average domain width at $z=10$ with (a) $P\leq {10}^{-2}$ and
(b) all simulated values of $P$.
Different symbols correspond to different combinations of the parameters.
The values of ($\epsilon_{AA}$,$P$) for each symbol are:
plus sign ($10^{-2}$, $10^{-1}$); crosses ($10^{-2}$, $10^{-2}$); circles ($10^{-2}$, $10^{-3}$);
squares ($10^{-3}$, $10^{-1}$); diamonds ($10^{-3}$, $10^{-2}$); triangles ($10^{-3}$, $10^{-3}$);
inverted triangles ($10^{-4}$, $10^{-2}$); stars ($10^{-4}$, $10^{-3}$).
The solid lines are linear fits of the data with $\lambda\geq 5$ with slopes (a) 0.39 and (b) 0.35.
}
\label{lambdasym}
\end{center}
\end{figure}

We first consider data with non-negligible terrace edge barriers, $P\leq {10}^{-2}$.
Plots of $\log{\lambda}$ as a function of $\log{\left[ R_{AA} {\epsilon_{AA}}^x P^y \right]}$
were constructed with several choices of $x$ and $y$.
The best data collapse is obtained with $x=0.7$ and $y=0.4$, as shown in Fig. \ref{lambdasym}(a).
It also shows a linear fit of the data with large values in the abscissa,
which gives $\gamma\approx 0.39$ and $A\approx 0.5$.

If we include data with small terrace edge barriers ($P={10}^{-1}$), the dispersion of the data
is larger for any choice of $x$ and $y$.
The best data collapse is now obtained with $x=0.5$ and $y=0.5$, as shown in Fig. \ref{lambdasym}(b).
However, the linear fit of the data with $\lambda\geq 5$ has a slope $\gamma\approx 0.35$
and the prefactor is $A\approx 0.6$, which are not very different from the previous estimates.

The values of $D_{AA}$, $\epsilon_{AA}$, and $P$ for various metals and semiconductors with
different surface orientations are given in the literature; see e.g. Ref. \protect\cite{etb}.
For a given binary mixture, those data and the fits shown in Figs. \ref{lambdasym}(a),(b) may
help to predict the order of magnitude of the size of nanoscale domains formed in films of
that mixture.

\subsection{Two species with different surface mobilities}
\label{assymmetric}

Now we consider cases with very different mobilities of A and B in terraces of the same
species, which are represented by $R_{AA}=100 R_{BB}$.
The probabilities of detachment from lateral neighbors follow the same relation.
The diffusion coefficients in terraces of a different species are much larger than both 
$D_{AA}$ and $D_{BB}$, which contributes to the domain separation.

The inspection of several data sets shows some qualitative trends that are similar to those
of the case with equal mobilities of the two species.
First, the roughness decreases as $R_{AA}$ or $R_{BB}$ increase and as $P$ increases.
Second, the average domain width increases with the increase of all the main model parameters:
$R_{AA}$, $R_{BB}$, $\epsilon_{AA}$, $\epsilon_{BB}$, and $P$.

\begin{figure}
\begin{center}
    \subfloat{\includegraphics[width=0.35\textwidth]{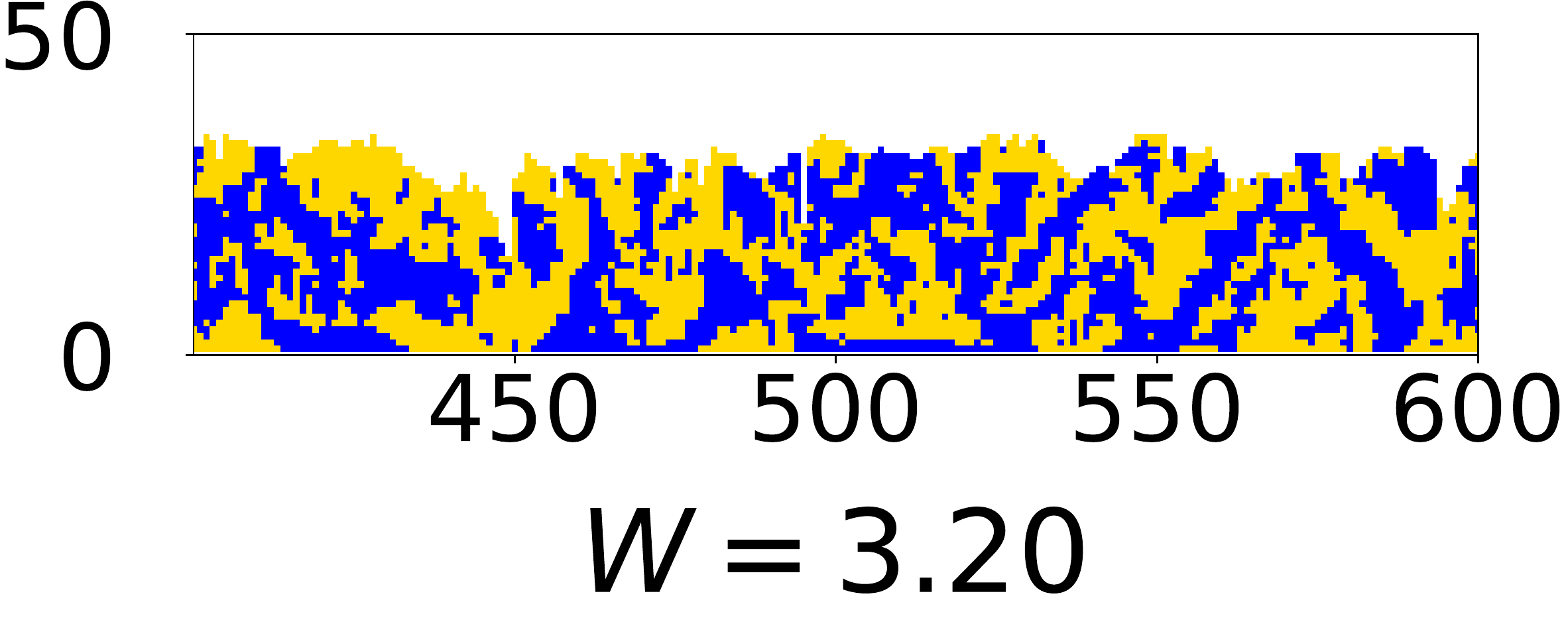}} \\
    \subfloat{\includegraphics[width=0.35\textwidth]{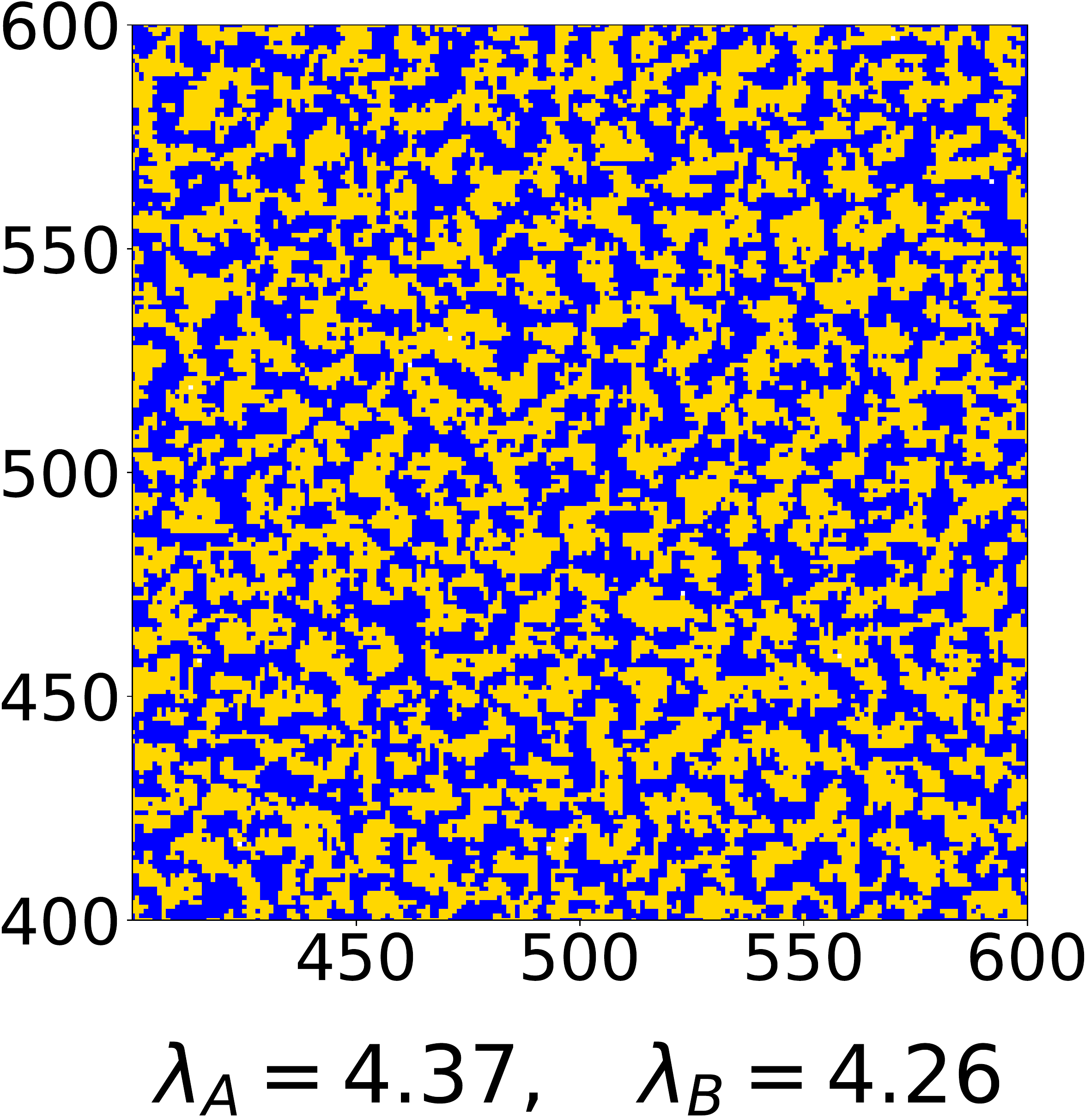}} \\
	\subfloat{\includegraphics[width=0.35\textwidth]{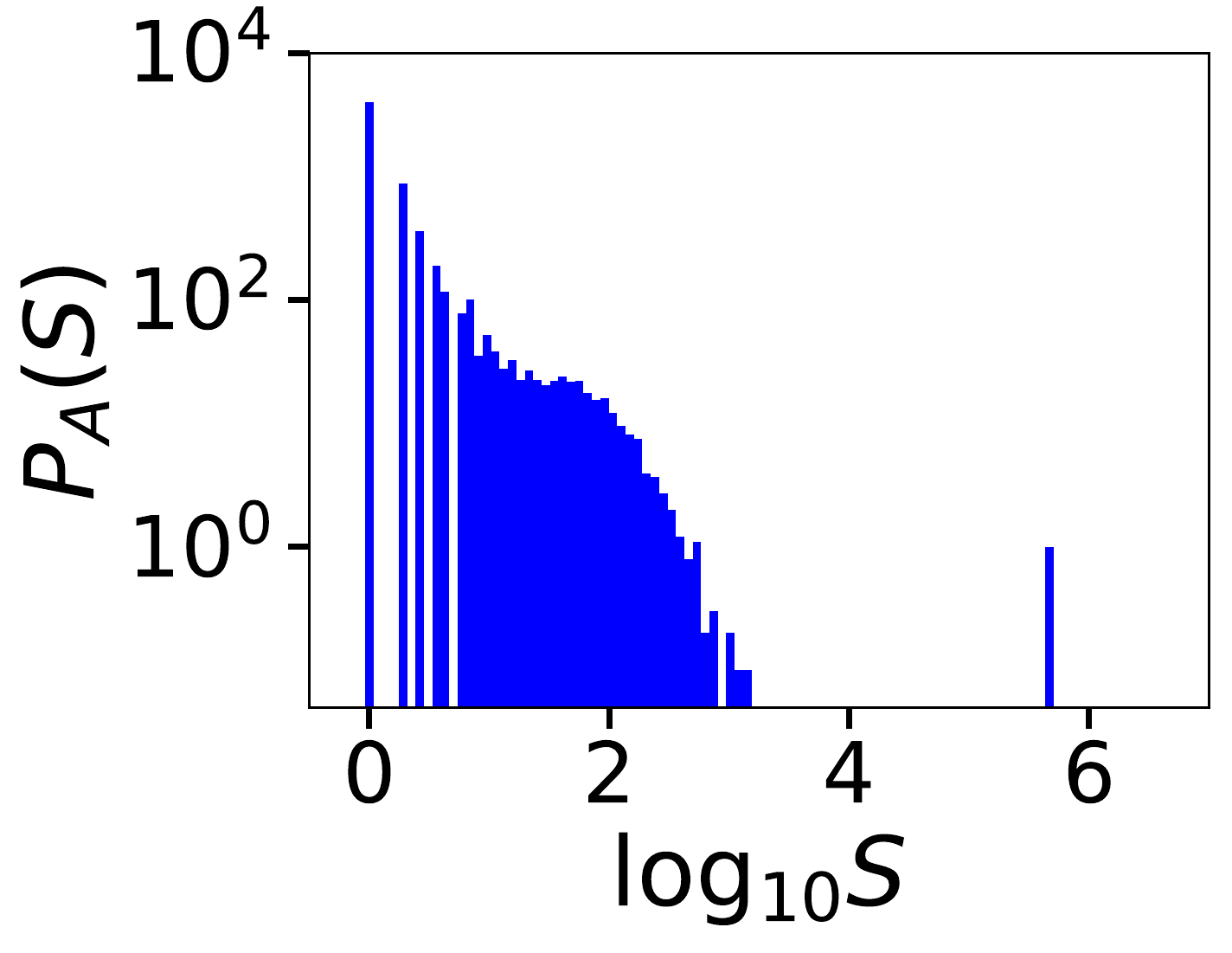}} \hspace{1cm}
    \subfloat{\includegraphics[width=0.35\textwidth]{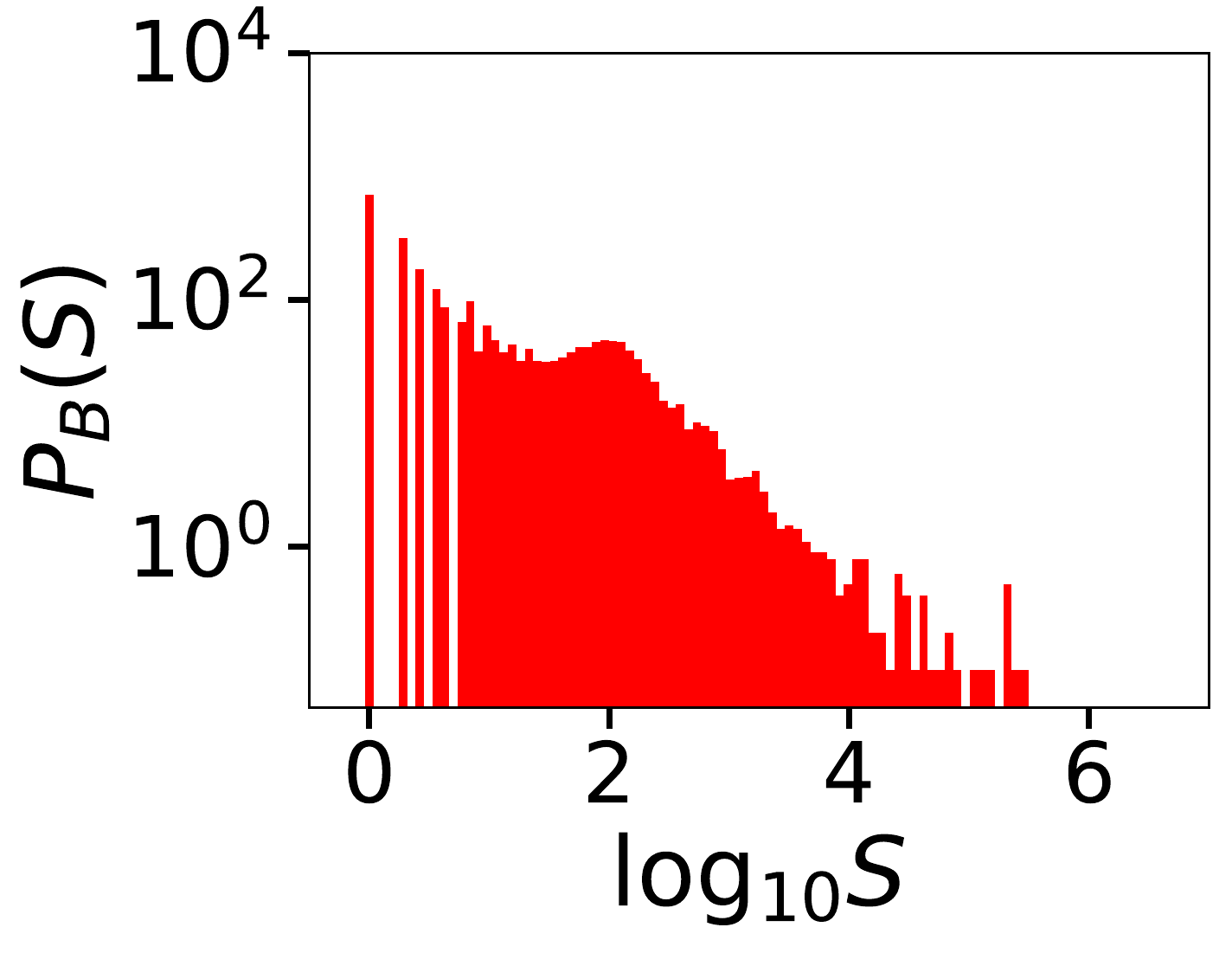}}
\caption{
Vertical cross section, horizontal cross section at $z=15$, and domain size distributions
of A and B at $z=15$ in films deposited with
$R_{AA}={10}^{6}$, $R_{BB}={10}^{4}$, $R_{AB}={10}^8$, $\epsilon_{AA}={10}^{-3}$,
$\epsilon_{BB}={10}^{-5}$, $\epsilon_{AB}={10}^{-1}$, $P={10}^{-2}$.
A particles are in blue, B in yellow.
The surface roughness and the average domain sizes in the horizontal cross section are indicated.
}
\label{asymcomplow}
\end{center}
\end{figure}

To investigate the case of low to intermediate temperatures, in Fig. \ref{asymcomplow} we show
vertical and horizontal cross sections of the deposits
and the distributions of sizes of domains A and B at $z=15$, for films grown with:
$R_{AA}={10}^{6}$, $R_{BB}={10}^{4}$, $R_{AB}={10}^8$, $\epsilon_{AA}={10}^{-3}$,
$\epsilon_{BB}={10}^{-5}$, $\epsilon_{AB}={10}^{-1}$, and $P={10}^{-2}$.
The distribution of A shows a large domain ($S\sim {10}^6$) at $z=15$ and at other heights.
The largest domains of B have $S\sim {10}^5$; at $z=10$, their areas are smaller than ${10}^5$.
The size distributions of both A and B are superpositions of a monotonically decreasing part
and a peaked part; the peak of A is located at a size smaller than ${10}^2$ and the peak of
B is near size ${10}^2$.
This indicates a trend that the most mobile species (A) forms the largest domain, which connects
long distances, while the less mobile species (B) forms finite domains with areas
typically larger than the other one.

The domain size distributions of A and B are different from those of the symmetric case with
$R_{AA}=R_{BB}={10}^{4}$ [Figs. \ref{sym4}(e)-(f)]
and with $R_{AA}=R_{BB}={10}^{6}$ [Figs. \ref{symcompRAA}(b), \ref{symcompeAA}(a)-(b)].
Here, the distribution of A is similar to the distribution obtained in the symmetric case
with $R_{AA}=R_{BB}={10}^{5}$ [Fig. \ref{symcompRAA}(a)], i.e. with a diffusion-to-deposition ratio
intermediate between those of A and B in the asymmetric case.
The comparison of horizontal cross sections in Figs. \ref{symcompRAA}(a)
and \ref{asymcomplow} visually confirms that the domain morphologies are similar.

\begin{figure}
\begin{center}
    \subfloat{\includegraphics[width=0.35\textwidth]{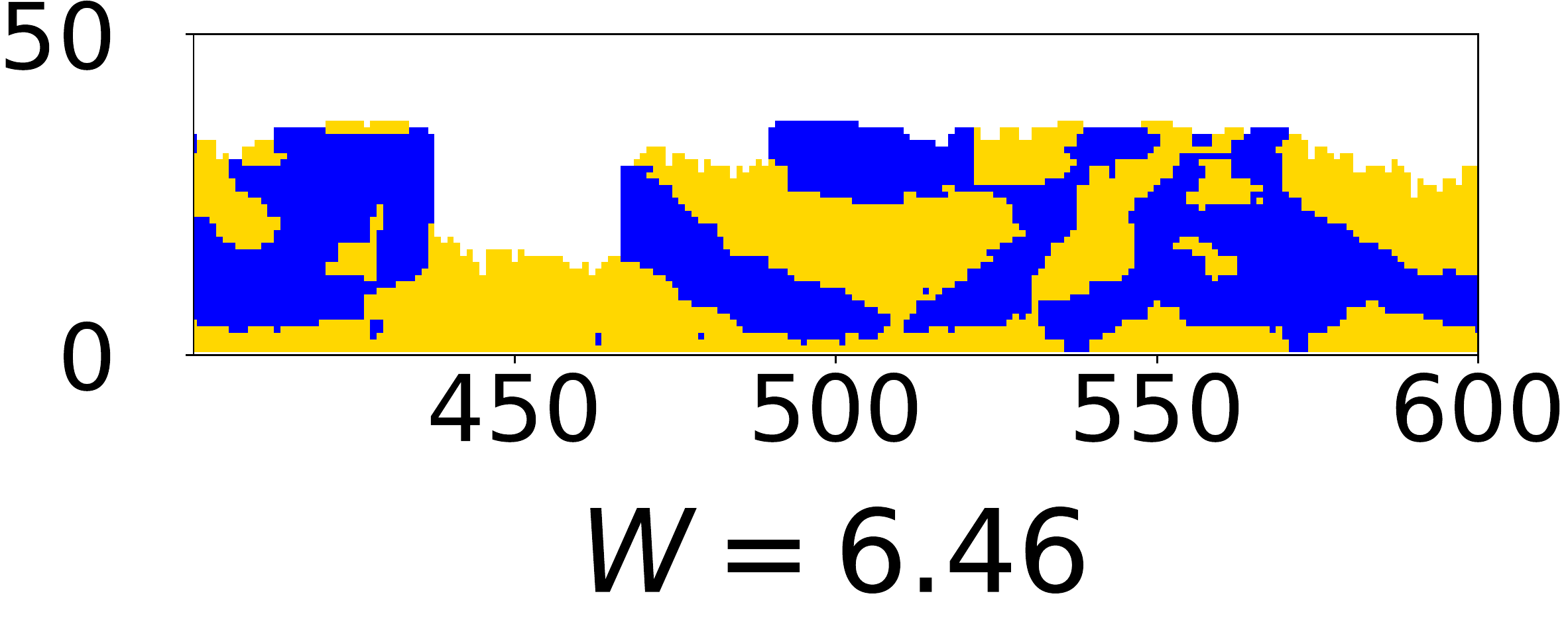}} \\
    \subfloat{\includegraphics[width=0.35\textwidth]{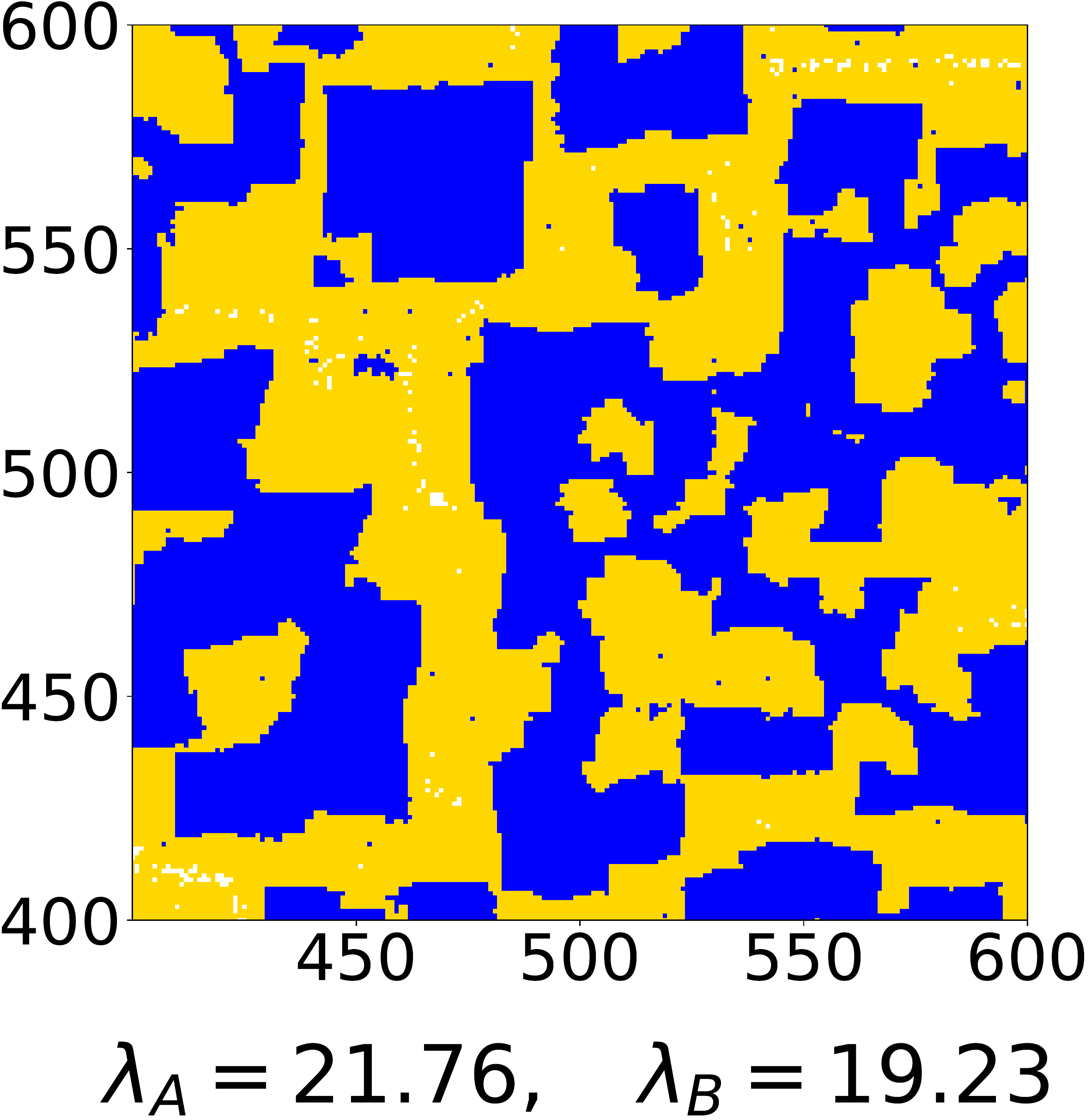}} \hspace{1cm}
	\subfloat{\includegraphics[width=0.35\textwidth]{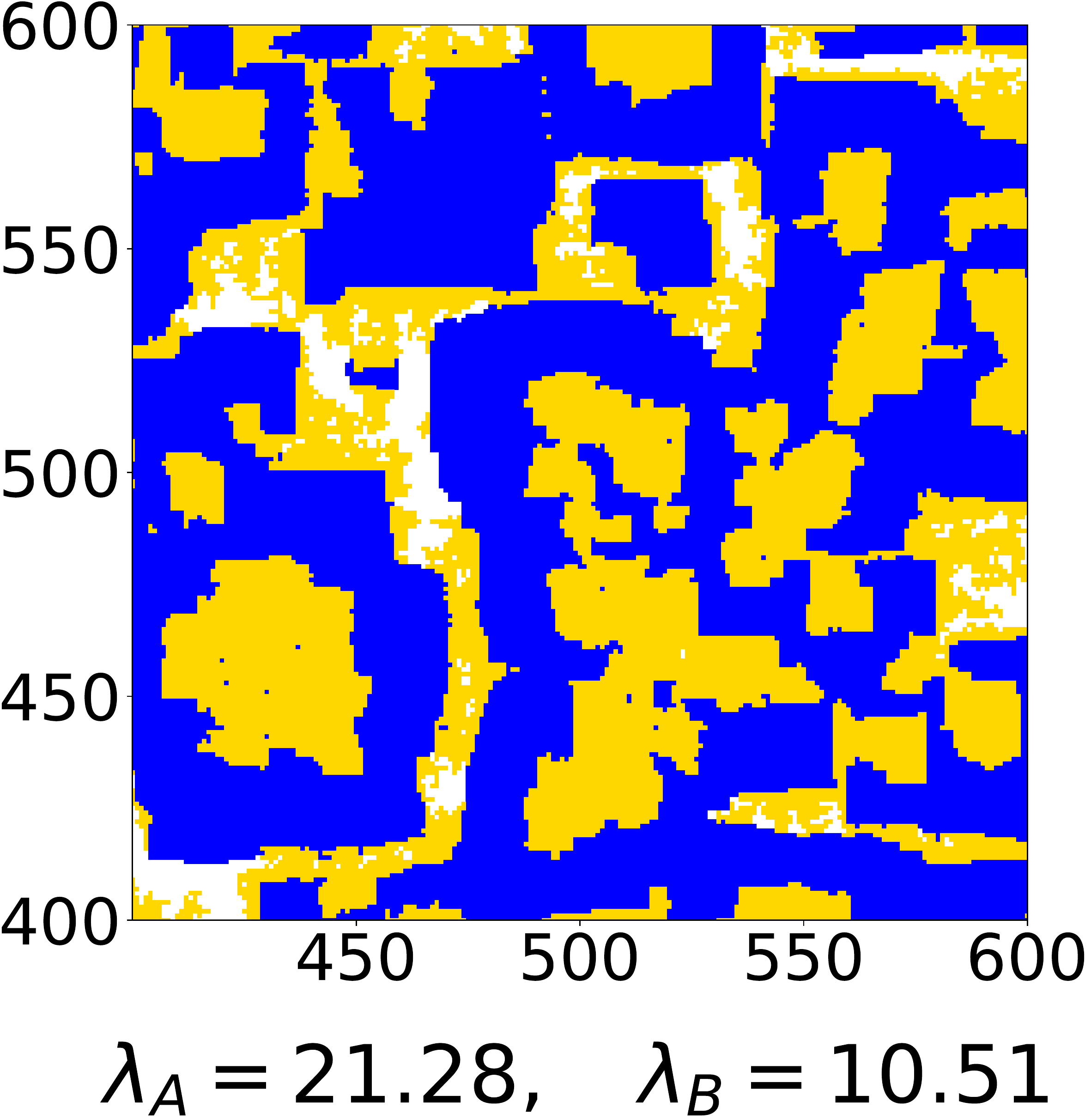}} \\
    \subfloat{\includegraphics[width=0.35\textwidth]{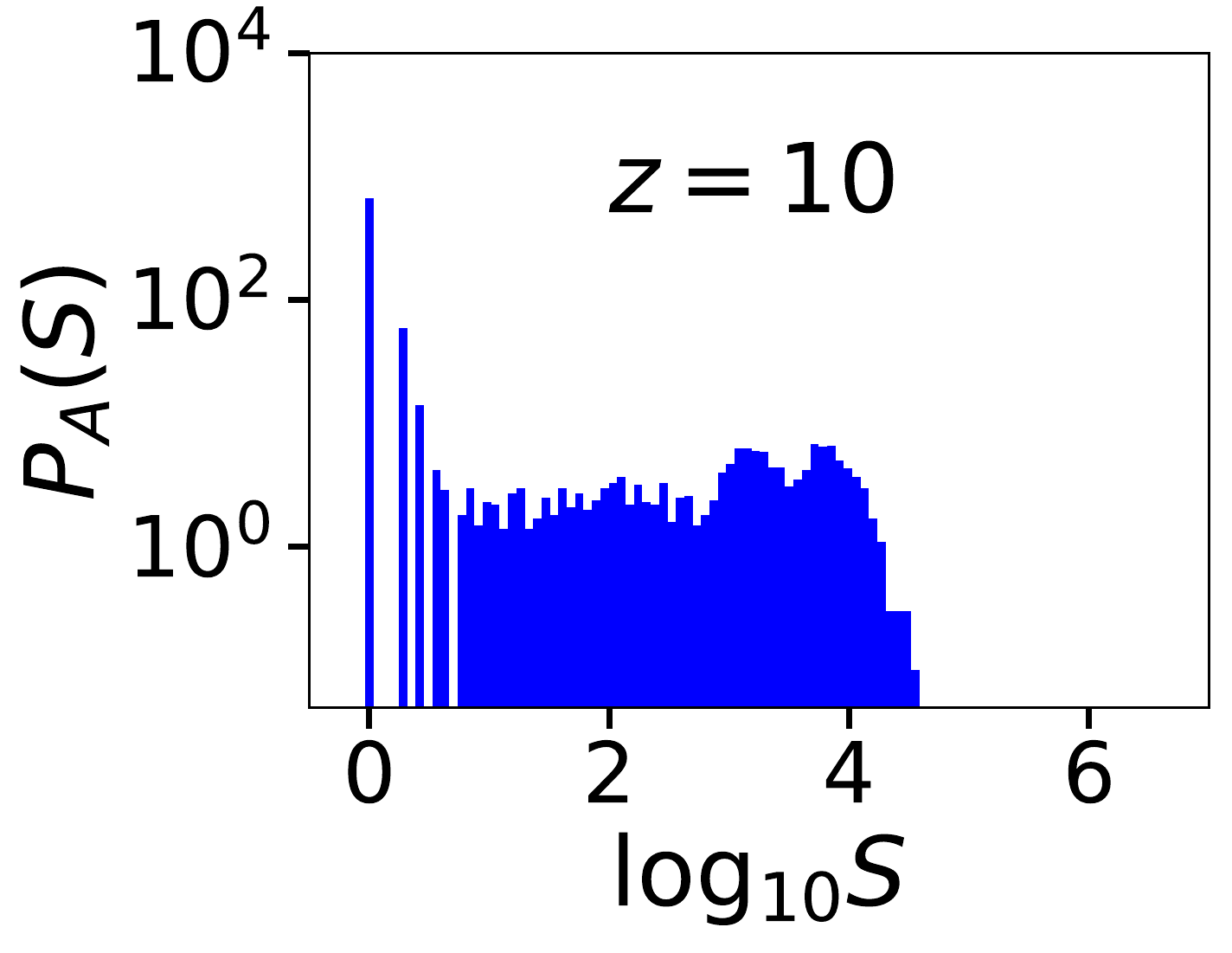}} \hspace{1cm}
	\subfloat{\includegraphics[width=0.35\textwidth]{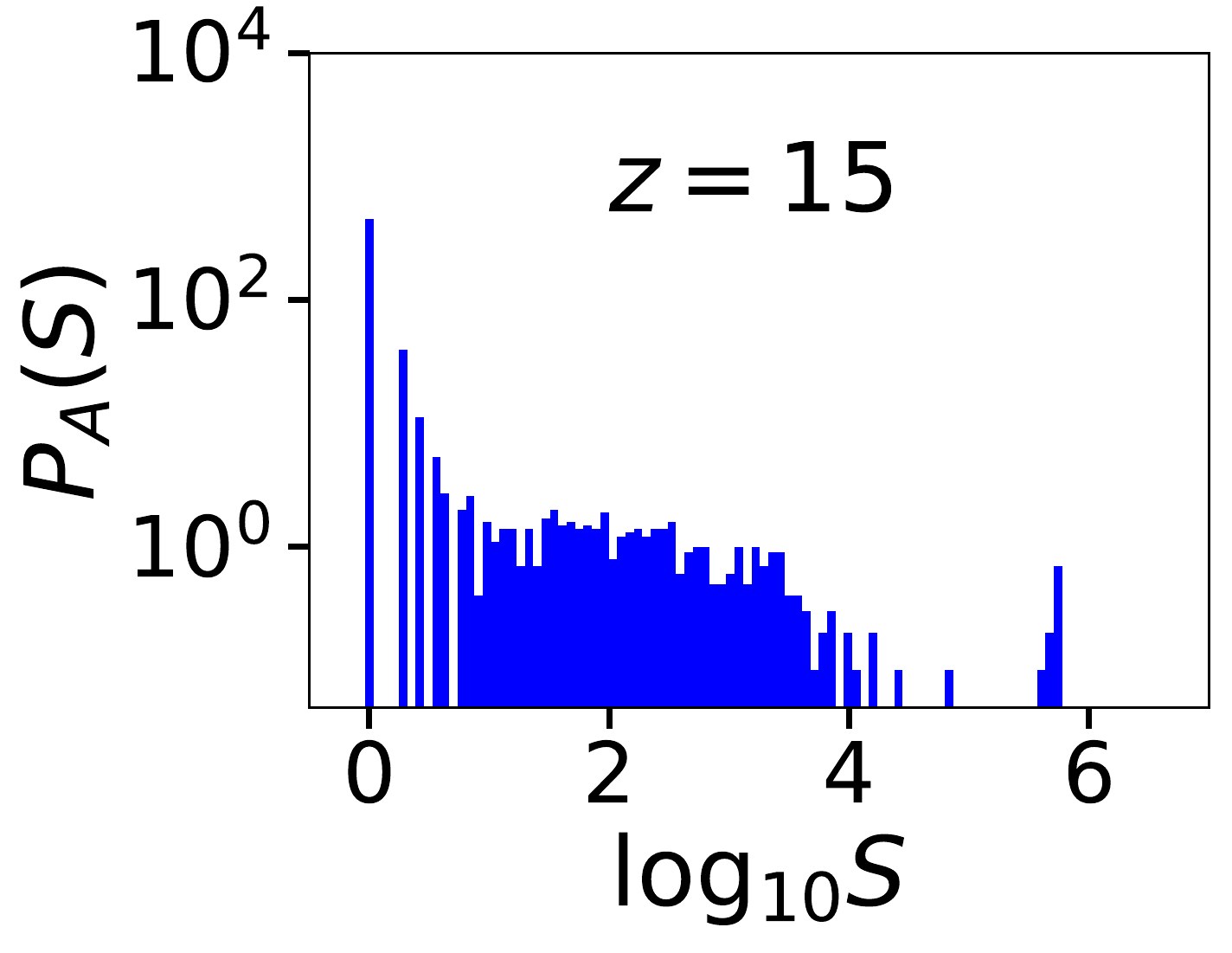}} \\
	\addtocounter{subfigure}{-5}
	\subfloat[$\mathbf{z=10}$]{\includegraphics[width=0.35\textwidth]{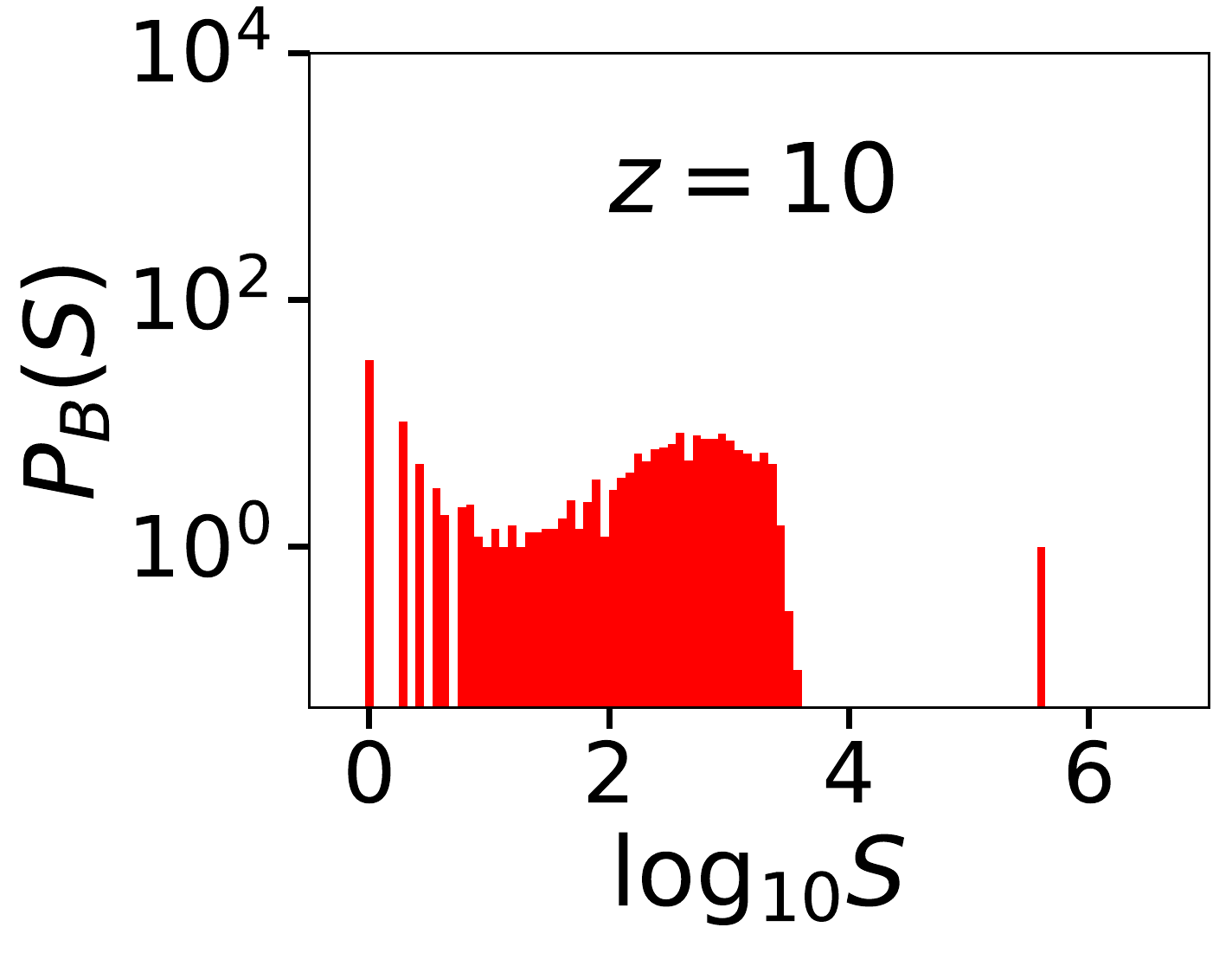}} \hspace{1cm}
	\subfloat[$\mathbf{z=15}$]{\includegraphics[width=0.35\textwidth]{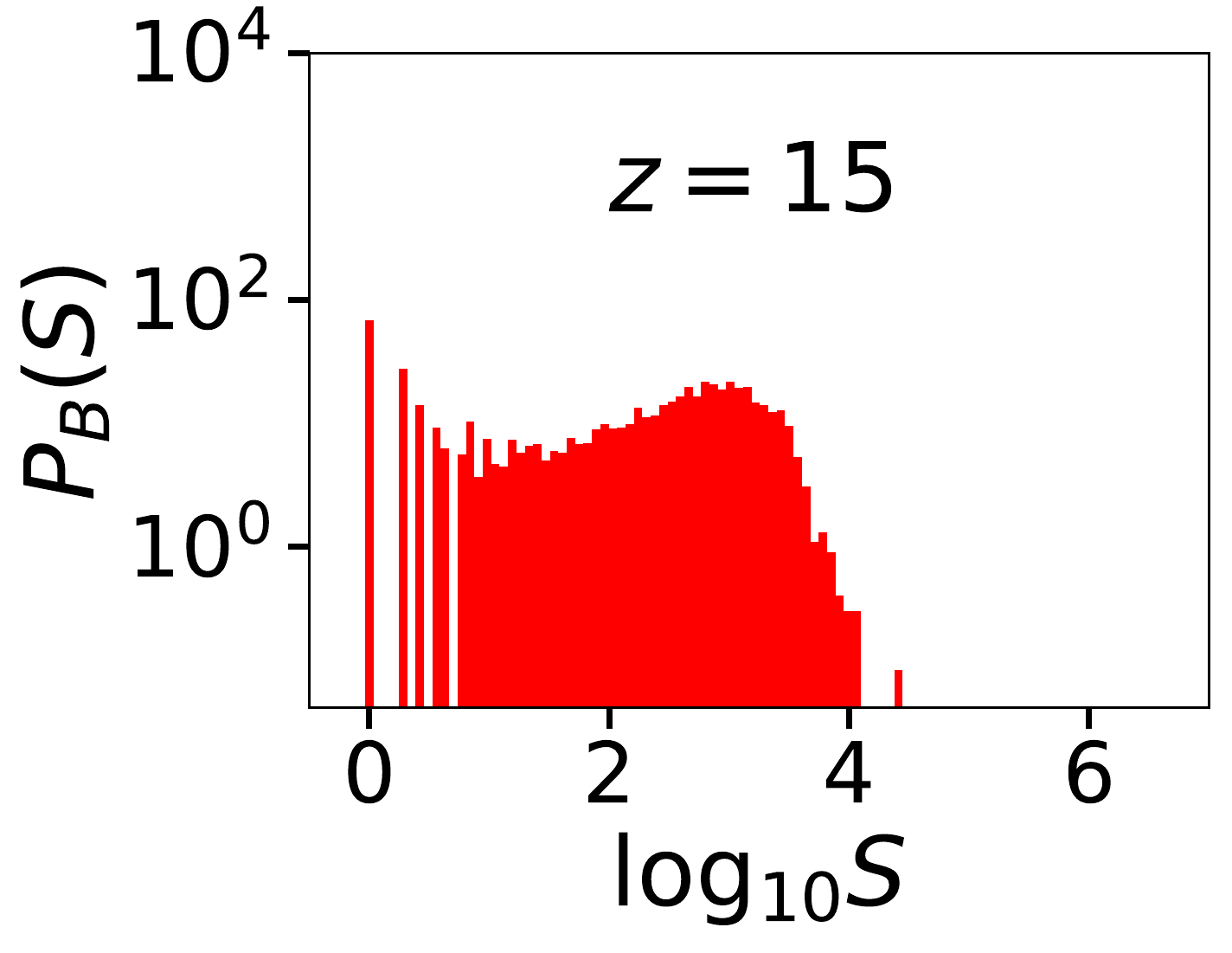}}
\caption{
Vertical cross section of a film deposited with
$R_{AA}={10}^{7}$, $R_{BB}={10}^{5}$, $R_{AB}={10}^9$, $\epsilon_{AA}={10}^{-3}$,
$\epsilon_{BB}={10}^{-5}$, $\epsilon_{AB}={10}^{-1}$, and $P={10}^{-1}$.
(a) and (b) show horizontal cross section and domain size distributions at $z=10$ and $z=15$,
respectively.
A particles are in blue, B in yellow.
The surface roughness and the average domain sizes in the horizontal cross section are indicated.
}
\label{asymcomphigh}
\end{center}
\end{figure}

A case of intermediate to high temperature is shown in Fig. \ref{asymcomphigh}.
At or below $z=10$, a very large domain of species B is formed, but not of A;
the situation changes at or above $z=15$, in which there is a large domain of A but not of B.
The horizontal cross section at $z=10$ confirms that finite domains of A are surrounded
by a narrower domain of B which connects the external borders of that image; the inverse
occurs at $z\geq 15$.

The film structure in Fig. \ref{asymcomphigh} resembles that of the symmetric case in
Fig. \ref{symcompP}(a),
in which $R_{AA}=R_{BB}={10}^{7}$, $\epsilon_{AA}=\epsilon_{BB}={10}^{-3}$, and $P=0.1$.
Thus, in intermediate to high temperatures, the structure is similar to that obtained with
two similar species with the largest mobility.
This suggests that the most mobile species is responsible for the loss of long range domain
connectivity at high temperatures.

\begin{figure}[!h]
\begin{center}
    \subfloat[]{\includegraphics[width=0.45\textwidth]{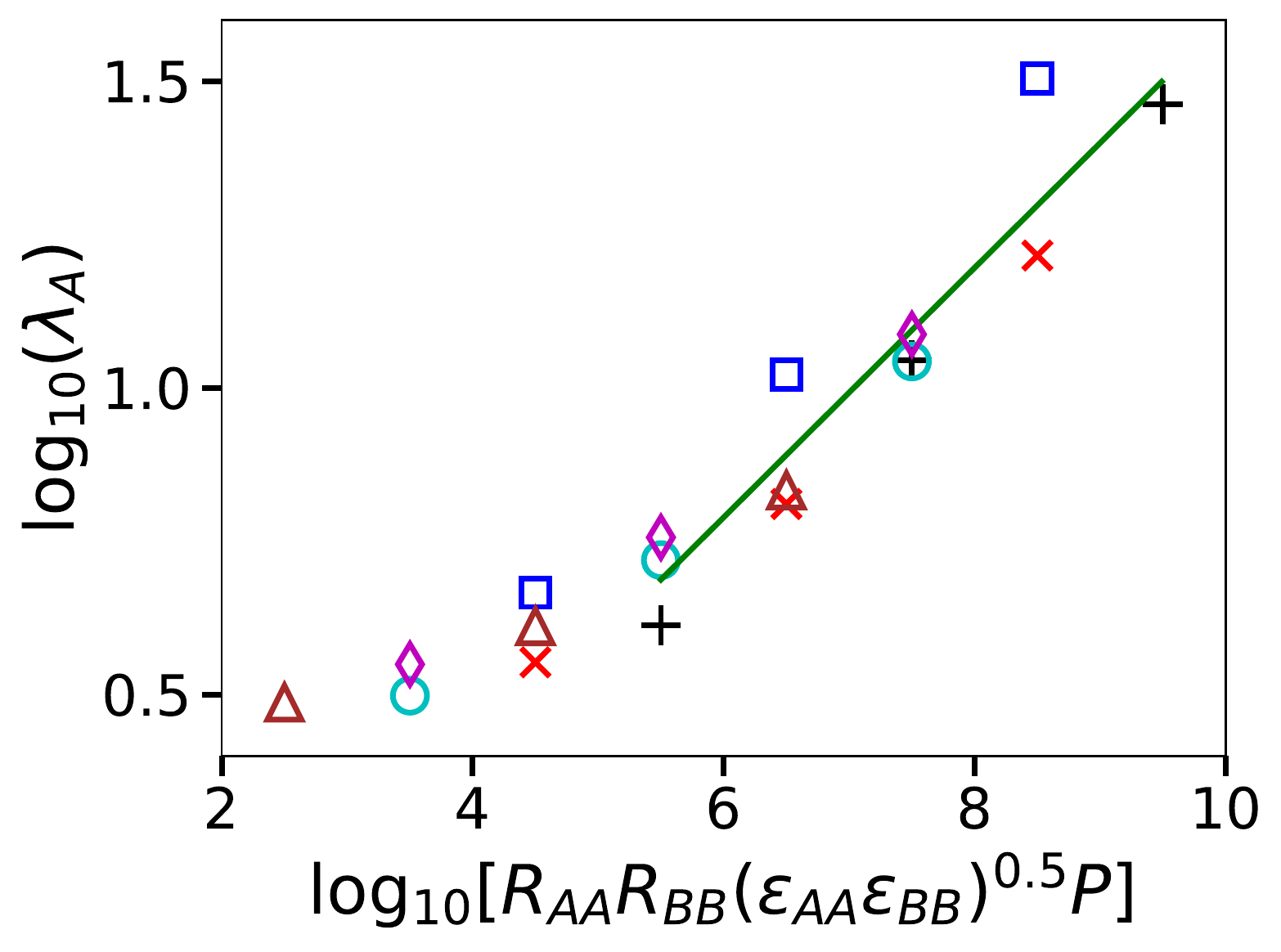}}
    \subfloat[]{\includegraphics[width=0.45\textwidth]{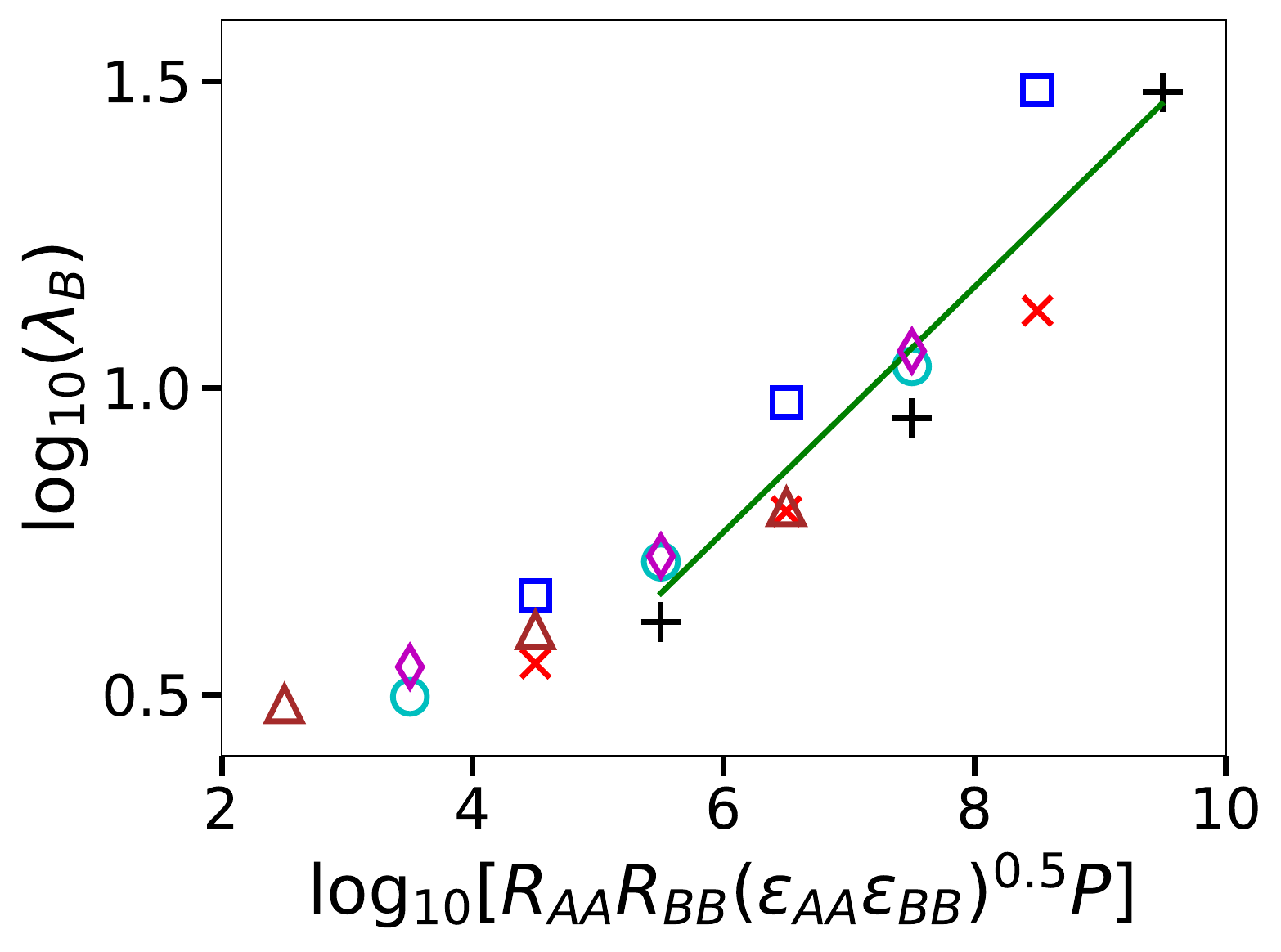}} \\
	\subfloat[]{\includegraphics[width=0.45\textwidth]{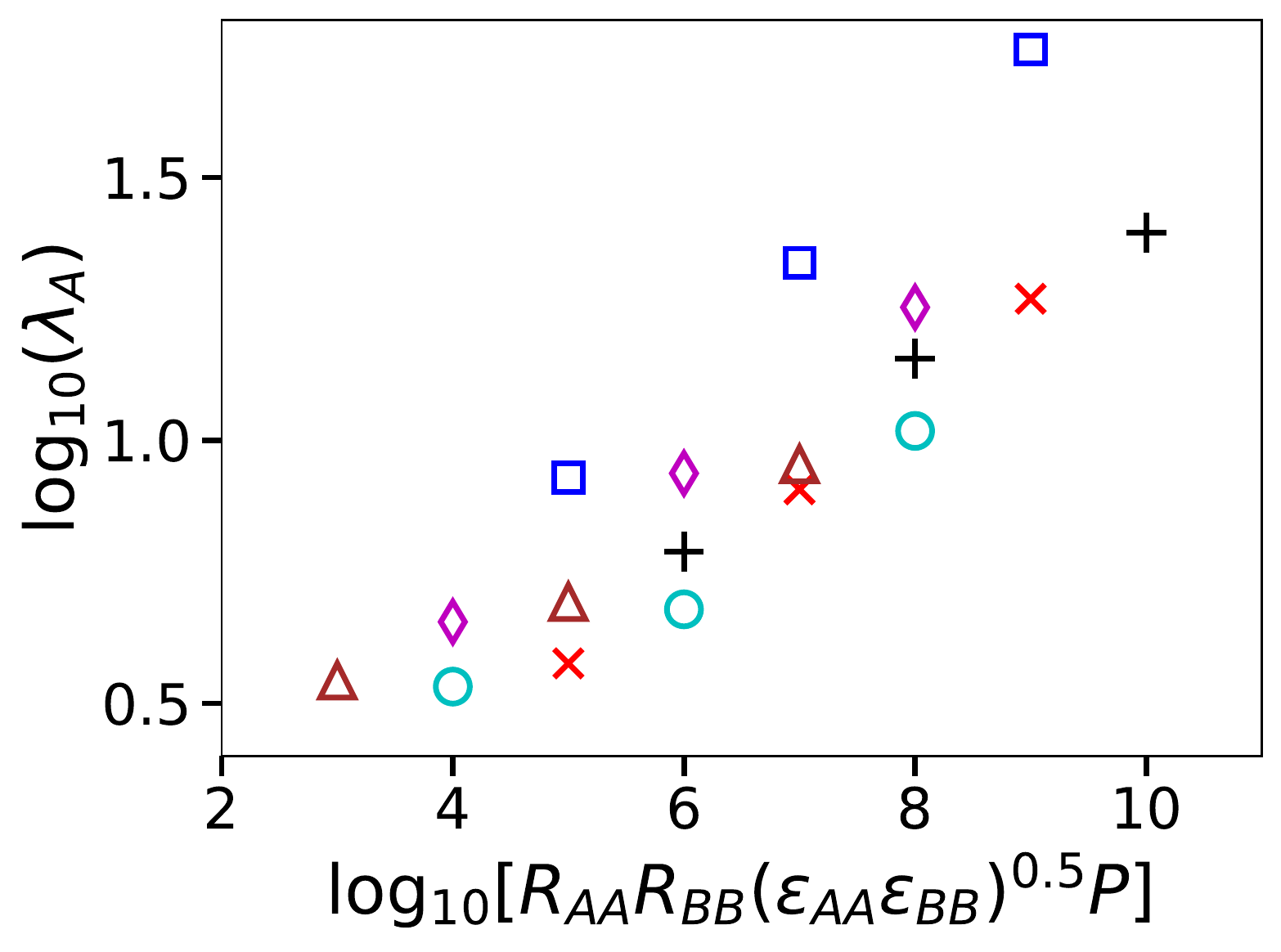}}
    \subfloat[]{\includegraphics[width=0.45\textwidth]{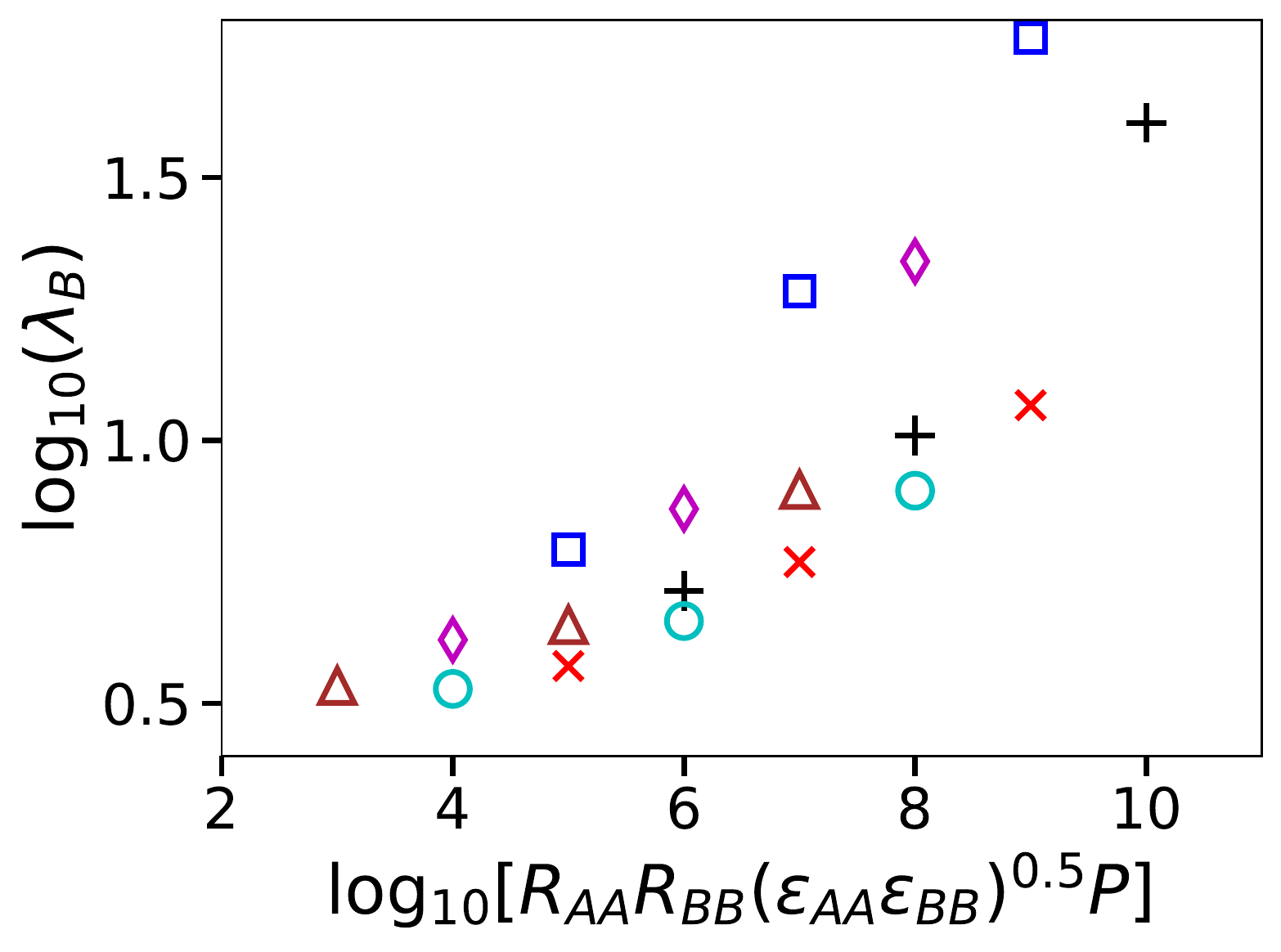}}
\caption{
Average domain widths of species A and B for: (a),(b) $R_{AA}/R_{BB}=10$; (c),(d) $R_{AA}/R_{BB}=100$.
Each symbol corresponds to a different parameter set.
The solid lines in (a) and (b) are linear fits of the data with $\lambda\geq 5$.
}
\label{lambdaasym}
\end{center}
\end{figure}

We also tried to fit the data for this case with a relation similar to Eq. (\ref{scalinglambda}),
but using a ratio $R$ and a probability $\epsilon$ equal to the geometric means of their
values for A and B species.
Figs. \ref{lambdaasym}(a) and \ref{lambdaasym}(b) show domain widths $\lambda_A$ and $\lambda_B$
obtained at $z=10$, respectively, as a function of the appropriate scaling variable, in cases
with $R_{AA}/R_{BB}=10$ (small difference between the surface diffusivities of A and B).
Figs. \ref{lambdaasym}(c) and \ref{lambdaasym}(d) show the same quantities in cases with
$R_{AA}/R_{BB}=100$.

We observe that the scatter of the data is smaller for $R_{AA}/R_{BB}=10$.
The fits in Figs. \ref{lambdaasym}(a),(b) give
\begin{equation}
\lambda_A \approx A_1 {\left[ R_{AA} R_{BB} {\left( \epsilon_{AA}\epsilon_{BB}\right)}^{0.5} P \right]}^\delta
\qquad , \qquad
\lambda_B \approx B_1 {\left[ R_{AA} R_{BB} {\left( \epsilon_{AA}\epsilon_{BB}\right)}^{0.5} P \right]}^\delta ,
\label{scalinglambdaasym}
\end{equation}
with $\delta\approx 0.2$, $A_1\approx 0.4$, and $B_1\approx 0.4$.
This shows that a difference of one order of magnitude in surface diffusion coefficients and
lateral detachment probabilities do not change the main conclusions obtained in the symmetric
case of Sec. \ref{symmetric}.
However, for $R_{AA}/R_{BB}=100$, we have already observed that
characteristic sizes of the domains of the two species are very different, which
explains why it is difficult to collapse data of different species with a single scaling variable.

\section{Discussion}
\label{discussion}

The deposition of thin films with two-component mixtures was simulated and the
effects of the terrace diffusion coefficients, the probabilities of detachment from island
and terrace borders, and the barriers for crossing terrace edges were analyzed.
Some features of the model are similar to those of previous models of submonolayer growth
of binary mixtures \citep{einax2007,einax2009}; however, the main difference here is that
the surface diffusion coefficient of a given species also depends on the terrace where it is moving.
In all cases, we considered that the surface mobility of a given species is larger (possibly much larger)
in terraces of the other species in comparison with the mobility in terraces of the same species,
which is beneficial to the separation of domains.

Most of our simulations considered the case of two species with approximately the same activation
energies for terrace diffusion and for detachment from lateral neighbors.
When terrace diffusion coefficients differ by a factor $10$, the results are similar.
The effects of increasing temperature summarized below must be understood as effects of increasing
the diffusion-to-deposition ratios, so that similar effects can be obtained by decreasing the
deposition rate.

For low and intermediate temperatures, we observe the formation of very narrow domains
with highly disordered shape; for metals and semiconductors, the domain widths are of some nanometers.
Those domains occupy a large fraction of horizontal cross sections, so that they connect very
distant points across the substrate directions.
The scaling of the average width obtained here involves only parameters describing the interactions
of a single species, which depend on activation energies for surface diffusion and on the temperature.
The leading contribution comes from terrace diffusion.
Since the values of activation energies of this process were already obtained by experimental and ab initio
calculations for several materials and deposition conditions \citep{etb}, they may be used to predict the
domain widths of real binary mixtures.
Note that most metals and semiconductors have lattice structure different from the simple cubic
model considered here, but the present scaling relations may be reasonable to predict
the orders of magnitude of domain widths.

The long range domain connectivity may be interesting for the production of multifunctional materials
if the two deposited species have different thermal, electric, or magnetic properties.
In alloys where previous works showed the formation of lamellar domains in the nanoscale,
the investigation of their connectivities would be interesting
\citep{elofsson2016,derby2018,ghafoor2018}.
The long connectivity of domains may also help the production of nanoporous materials by dealloying
of the less noble component \citep{ding2004,fujitaSTAM2017,atwater2018}.
One of the technical difficulties of dealloying processes is to warrant that the
remaining component forms very long connected clusters, but the present models suggests that
such clusters are already present in the precursor films.
Indeed, recent works on dealloying of sputter-deposited NiAl films \citep{luo2019} and
AgAu films \citep{huPhysB2019} suggest that the connectivity of the final porous materials
(Ni and Au films, respectively) evolved from the connectivity of their components in the precursors.

At high temperatures, we showed an enhanced separation of domains, which parallels the observations
of previous models \citep{adams1993,mouton2014,zhu2014,minkowski2016}.
The increase in the width of the finite, nanosized domains is usually accompanied by a loss of
the long range connectivity, with possible formation of gaps between domains of different species.
The vertical orientation of the domains may also be impaired because the large surface mobility
facilitates their lateral propagation during the formation of each atomic layer.
Thus, the production of films with long connected clusters of the same species requires
a careful control of temperature and deposition flux, which have to be set according to
the energetics of both species.

We also observe that local surface fluctuations depend on the model parameters in
a similar way of films with a single species.
The main parameter controlling those fluctuations is the probability of adatom hops across steps:
as that rate increases, the surface becomes locally smoother.
This smoothening is also obtained with the increase of the terrace diffusion coefficients.
Thus, the local smoothening is generally a consequence of the increase of temperature or of the decrease
of the deposition flux.
However, at high temperatures, the value of the roughness $W$ measured in the whole surface may be large
because there are large height differences in the gaps between domains, although the top surfaces
of individual domains are smooth.

We also investigated some cases in which the terrace diffusion coefficients of the two species differ
by two orders of magnitude.
The scaling of domain width with the model parameters is less accurate;
rough estimates of those widths can be obtained using geometric averages of
diffusion-to-deposition ratios (which are obtained from averages of the terrace diffusion coefficients
of the two species), of probabilities of detachment from island borders,
and of probabilities of hops across terrace steps.
However, the species with the largest mobility determines the maximal temperature in which
long connected domains can be obtained.

Note that the increase of a characteristic domain size with the increase
of the temperature or with the decrease of the deposition rate was already observed in some previous
models on deposition of binary alloy films
\citep{adams1993,kairaitis2014,leonard1997,he2006,mouton2014,hennes2018}.
However, those models did not present the domain morphology described here, in which the
different sizes that characterize a single domain may range from a few nanometers (the widths)
to hundreds of nanometers or micrometers (the connected lengths in longitudinal directions).
Thus, despite the simple features of our model, the results presented here may be
useful to search for suitable conditions for producing two-component thin films with such nontrivial
morphology.

\section{Conclusion}
\label{conclusion}

We analyzed results of kinetic Monte Carlo simulations of deposition of thin films
with two components, A and B, in conditions that favor domain formation: energy barriers
for adatom diffusion are larger (smaller) when represent the interactions of adatoms
of the same (different) species.
The film morphology is determined in terms of the ratios between terrace
diffusivity and atomic flux, of the probabilities of detachment from lateral neighbors,
and of probabilities of crossing step barriers.

For low and intermediate temperatures, we observe the formation of narrow domains that meander
in the layers parallel to the substrate.
Their widths are small (up to $\sim 30$ lattice constants), but they are connected through
very long distances along the substrate directions.
Scaling relations for the domain widths were obtained.
Using tabulated values of the energy barriers in single species film deposition,
those relations may be used to predict the orders of magnitude of domains in alloy films.
At high temperatures, the domains are thicker, but their long range connectivity is lost.
In cases of significantly different A-A and B-B interactions, the species with the largest
terrace mobility is the one that constrains the temperature range in which the long domains are found.

\section*{Acknowledgements}

FDAAR acknowledges support from the Brazilian agencies
CNPq (grant number 305391/2018-6), CAPES (project number 88881.068506/2014-01), and
FAPERJ (project numbers E-26/110.129/2013 and E-26/202.881/2018).
TBTT acknowledges support from CAPES (grant number PNPD20130933 - 31003010002P7).

\section*{Conflict of interest}

The authors declared that there is no conflict of interest.

\bibliographystyle{unsrt}
\bibliography{mixture}

\end{document}